\documentclass[openany,oneside]{scrbook} 
\usepackage[titletoc]{appendix}  
\usepackage[T1]{fontenc} 
\usepackage[utf8]{inputenc} 
\usepackage{graphicx}
\usepackage{hyperref} 
\usepackage{multirow} 
\usepackage{tabularx} 
\usepackage{color} 
\usepackage{textcomp} 
\usepackage{tipa}
\usepackage{amsmath} 
\usepackage{amssymb} 
\usepackage{amsfonts} 
\usepackage{amsxtra} 
\usepackage{wasysym} 
\usepackage{isomath} 
\usepackage{mathtools} 
\usepackage{txfonts} 
\usepackage{upgreek} 
\usepackage{enumerate} 
\usepackage{tensor} 
\usepackage{pifont} 
\usepackage{ulem} 
\usepackage{xfrac} 
\usepackage{arydshln} 
\usepackage[english]{babel}
\usepackage{tikz}
\usetikzlibrary{shapes.geometric, arrows.meta}
\usepackage{float}
\definecolor{yellow}{rgb}{1,1,0}
\definecolor{white}{rgb}{1,1,1}
\definecolor{color-2E74B5}{rgb}{0.18,0.45,0.71}
\definecolor{color-00B0F0}{rgb}{0,0.69,0.94}
\definecolor{color-404040}{rgb}{0.25,0.25,0.25}
\definecolor{color-4472C4}{rgb}{0.27,0.45,0.77}
\definecolor{color-A6A6A6}{rgb}{0.65,0.65,0.65}
\definecolor{gray}{rgb}{0.5,0.5,0.5}
\definecolor{color-AEAAAA}{rgb}{0.68,0.67,0.67}

\begin{document}

\begin{titlepage}
    \centering
    \vspace*{2cm}
    
    {\LARGE \textbf{Ouroboros AutoSyn: Time Based Permissionless Synchrony Model for PoS}}
    
    \vspace{2cm}
    
    {\Large Joshua Shen}
    
    \vspace{2cm}
    
    {\Large \textbf{Abstract}}
    
    \vspace{0.5cm}
    
    \begin{minipage}{0.85\textwidth}
        \normalsize
        \noindent Blockchain as a promising technology is gaining its popularity ever since proof-of-work based Bitcoin came to the world.
                Nevertheless, Bitcoin achieves consensus at an expensive cost of energy.
                Proof-of-stake is one of the solutions for such a problem.
                Participants of PoS protocols achieve dynamic-availability in permissionless settings.
                Parties can join and leave the protocol at their will without notifying others.
                However, such protocol relies heavily on a central clock, providing the function of synchrony by collecting the finish status of every honest participant. 
        
        \vspace{0.3cm}
        
        \noindent In our protocol, the global function maintaining the round information for each participant is no longer needed.
                We analyze and modify the round into a real-time based round model.
                Message delivery delay is also taken into consideration in the round length.
                However, participants need the connection of a real-world time global clock which is crucial to calculate the current round.
                And round length is also adjusted due to the changing network situation at the start of every new epoch.
    \end{minipage}
    
    \vfill
\end{titlepage}

\clearpage
\chapter{ Introduction }

In the distributed protocol, synchrony is one of the most important concepts. 
Participants of the distributed protocol run the protocol in rounds.
        In the synchronous protocol, every participant needs to maintain a common clock keeping track of round time and make every participant run at the same pace.
        The round does not advance until every participant completes its task.
        During every round, participants deal with messages received from the previous round and prepare new messages to deliver them to other participants at the beginning of the next round. 

On the contrary, asynchronous protocols do not need such a strong setting of common rounds for every participant.
        Instead, each party runs the protocol at their own pace.
        The running time of the protocol and the message delivery order are different for each party.
        If an adversary intercepts the information delivery on behalf of corrupted parties, the protocol execution can be stalled.
        Correctness guarantees in the asynchronous model would be affected {[}2{]}.
        To deal with such a problem, Miller et al. proposed a model known as ``guaranteed termination'' in {[}3{]}.
        They assumed that the messages are eventually delivered to the correct nodes in the asynchronous network.
        Also, nodes have no notion of ``real time'' clocks and only act on the order of messages received.
        Under the assumptions that the message from honest parties will be eventually delivered, the problems above can be solved.
        Participants advance to the following round without the need for the acknowledgement from everyone, once ``enough'' messages have been received.
        The model ``guaranteed termination'' has been further exploited in {[}4{]},{[}5{]}, {[}6{]}. 

Although the protocol is simple without the need for synchronicity, the model would be at a high cost.
        The most famous asynchronous model is Bitcoin {[}7{]}, which uses a proof-of-work mechanism based on the asynchrony model, as does Ethereum {[}8{]}.
        There are many more participants and everyone can lead the protocol execution contributing to its security.
        They can join and leave the protocol at their own will, which increases the difficulty of the synchronization.
        In the Bitcoin model, participants running the protocol compete against each other for the right to issue the next block and get a reward achieving impressive agreement, but at the cost of energy and low efficiency.
        Though Bitcoin provides terrible performance compared with traditional distributed systems, it thrives in a highly adversarial environment when fault tolerance is taken into consideration.
        All well-prepared malicious attacks are expected to be encountered.
        In 2015, Garay et al. gave a thorough analysis of PoW-based protocol Bitcoin about the common prefix and chain-quality property {[}12{]}.
        To deal with the inherent flaw of Proof of Work, other permissionless models have been proposed such as proof of space {[}9{]}, proof of space time {[}10{]} and proof of stake {[}21{]} {[}22{]} {[}23{]}.
        In 2017, Badertscher et al. presented the first Ouroboros Proof of Stake protocol with rigorous security proof under the mild adversary with 0-round message delay condition {[}13{]}.
        In 2018, they adapted their model into a semi-synchronous setting by proposing Ouroboros Praos {[}14{]}, in which the adversary can control the message delivery by delaying them with at most $\Delta $-round.
        In order to solve the long range attack, in {[}15{]} they put forth Ouroboros Genesis with a new chain selection rule under the Global Universal Composable setting.
        In 2019, Ouroboros Crypsinous was proposed as the first privacy preserving PoS blockchain {[}17{]}.
        A novel clock synchronization mechanism was proposed in {[}16{]} to overcome the need for a time-providing global clock in Ouroboros Genesis {[}15{]}.
        The global clock function is replaced with ``tick'' function to inform nodes of the change of time.
        Participants maintain and rely on a local-clock for synchronization.

A global clock setting is essential, especially for the PoS model.
        Synchrony based on the global clock can prevent the adversary from exploiting the ``longest chain rule'' in PoS.
        However, the global clock function heavily relies on the agreement from every participant at the end of each round, which makes the goal of decentralization hard to achieve.
        Meanwhile, we cannot directly apply the permissioned model to the permissionless one.
        Take the Practical Byzantine Fault Tolerance model from {[}11{]} as an example, the synchronous model containing three periods is based on round structure and ends with the confirmation period.
        The protocol will require a great amount of communication if used under a permissionless setting, and it is almost impossible for the participant to acquire the round status of all other participants.

In this paper, we bring a solution to decentralize the clock synchronization and model the PoS protocol from {[}15{]} under no assumption of bounded delivery.
        In our model, the global clock function no longer maintains round status of each participant.
        Instead, it functions like a clock by maintaining an increase-only integer.
        Participants no longer get its round updated upon the command from the global clock.
        Instead, they calculate the current slot number from the time given by the clock.

The notion of round is defined by a period of time, and the length is denoted as $t_{\textit{round}}$ (for the shift of the clock function).
        Participants achieve synchrony by using the same amount of time for the same slot.
        They start at the same time and end at the same time read from the clock function.
        During each round, participants run the protocol and wait for their messages to be delivered to others.
        In the executing period, similar to {[}15{]}, participants process the chains sent from the network and issue a new block if elected as the round leader.
        After finishing execution, participants listen to the network for messages sent to them while waiting for this round to end.

The semi-synchronous network is an important assumption in the previous work.
        In our model, message delivery is no longer guaranteed, and message loss could happen.
        We no longer model the network with guaranteed delivery with delay of maximum $\Updelta $ rounds.
        Instead, message delivery is partially and probabilistically bounded, which means that message loss could happen.
        Initially, round length is a pre-defined number as $t_{{\textit{round}_{1}}}$.
        As the protocol executes (parties joining and leaving the protocol), this pre-defined number will no longer be suitable.
        Participants record the arrival time of each message and publish them on the blockchain.
        When it is a new epoch, participants use the information recorded on the block chains to calculate a new round length for this epoch.
        However, there is a lower bound on the message success ratio, so that part of the messages could be successfully sent to their destination before the corresponding time.

    The global clock function is maintained by the environment and the time parameter accumulates only.
            We stress the environment is rational, which means that the environment will not advance the clock so fast that none of the participants could finish its task within its round length.
            Also, we assume when constructing the genesis block information, the balance between the round length $t_{{\textit{round}_{1}}}$ and participants’ executing time is taking into consideration, so that most of the initial participants could finish their protocol execution within $t_{{\textit{round}_{1}}}$.

Newly joined parties do not have the information of current round length $t_{\textit{round}}$ and the beginning time of the next round $t_{next}$.
        They need to listen to the network for a certain amount of time and update the local chain, as to calculate $t_{\textit{round}}$ and $t_{next}$. 

\chapter{ Our Model}

\textbf{Basic idea.} In {[}15{]}, the protocol assumes a maximum message delay round of $\Delta $(not known to participants).
        And in analysis of the protocol, the slot is considered an honest one only if it is $\Delta $-right isolated.
        Here we think, if given enough time, more messages could be received.
        In the real world, messages spread in the network, could be received, delayed but received later, or simply just lost.
        So, if we extend each round with delivery time, the information of a new block can be delivered to more nodes within one round.
        The message delivery is considered successful, if it was delivered to the next slot leader within one round, and the leader can issue a block based on the message.

\textbf{Dynamic availability.} We adapt our model of participants from {[}15{]}.
        Dynamic availability will capture the notion that our parties are able to join and leave the protocol at will, which is decided by the environment.
        And same as {[}15{]}, we have our participants interact with the function with registration/de-registration commands, keeping track of the joining and leaving time of participants, and guarantee their flexibility.
        We refer to {[}15{]} for more details on this mechanism.

\textbf{The adversary and environment.} Similar to {[}15{]} and {[}12{]}, we assume a central adversary $A$, who is adaptive and corrupts miners to attack the protocol. 

\textbf{Time based synchrony.} A common assumption in the analysis of blockchain protocols in the permissionless model is the availability of a global clock that allows parties to directly acquire the current round index, and advance at the same speed.
        In {[}15{]}, a global central clock $\boldsymbol{G}_{\textit{\textbf{Clock}}}$ does not advance until every party sends an update signal to the clock.
        Also, the clock is accessible to participants and acts as a crucial part in the protocol. 

In this model, we no longer require the global clock function to maintain the current round number, instead it only maintains a counter of time, similar to the $\boldsymbol{G}_{\textit{\textbf{refClock}}}$ in {[}20{]}, detailed in \uline{Section A.2}.
        Parties no longer need to inform the global function to update the current round to keep all parties at the same pace.
        Different from {[}20{]}, we grant participants direct access to the clock function, by assuming each party has the same absolute notion of time in the real model.
        Every party who has registered to the global functionality $\boldsymbol{G}_{\textit{\textbf{AutoClock}}}$, is able to calculate the round number they are currently at.
        By doing so, every party running the protocol can achieve the same pace of round.

Every round has the same length during each epoch.
        Two important assumptions are that (1) the length of the first round $t_{{\textit{round}_{1}}}$ should be large enough so that initial participants can finish the execution of the protocol, and (2) the environment would not suddenly advance the clock so fast that none of the participants could finish its protocol execution making all the time related parameters e.g. $t_{\textit{round}}$ useless.
        Each round has two periods: execution period and waiting period.
        During the execution period, participants run the main protocol and prepare to deliver messages.
        And during the waiting period, after the messages have been sent to the network, participants wait for the end of the round and message delivery.
        For security reasons, the executing time of the protocol (within the execution period) should be no more than $t_{run}$ without the interference of deliberate delay from the adversary (the adversary responds to each message of the participant as fast as possible).
        The adversary can delay but needs to guarantee a certain amount of participants finishing the execution of the protocol and beginning the waiting period during $t_{run}$, so that minimum message delivery ratio could be satisfied.

During the first epoch, the round length is defined by Genesis block.
        But in case network situation changes, we use the adjust-round protocol to adjust round length at the beginning of each new epoch, so that during next epoch participants have more suitable round time.  

\textbf{Modeling peer-to-peer Communication. }In our model, we no longer guarantee that message will be eventually delivered.
        Message loss could happen randomly but with limitation.
        The adversary can control the messages sent by parties by delaying the messages a random amount of time.
        If the message can be lost, the adversary can set the delay time of this message very large so that the party can never receive this message.
        And once the waiting for block information is more than 2 rounds (current round length), the party gives up listening for this message.
        Because of the different time model we use, the delay is no longer round-based but a relative notion of time.

We model a diffusion network handling messages for every party including the adversary.
        Different from {[}15{]}, we assume that the delivery success rate for messages sent by honest parties in each round should be no less than $\eta $.
        Messages sent by honest parties should be fetched by most of the protocol participants within a round, tolerating some delay or loss.
        And, the length of a round should be reasonable, so that messages could be delivered to most of the participants and parties do not waste time waiting after receiving the messages.
        The details of the corresponding functionality are in \uline{Section A.1}.

\textbf{Genesis Block Distribution.} Similar to {[}15{]}, we need every initial shareholder to start participating after receiving the necessary message.
        We denote the genesis round length as $t_{\textit{start}}$, before which essential message should be delivered.
        Detailed in \uline{Section A.2}.

\textbf{Ledger functionality in UC-model.} The ledger functionality $\boldsymbol{G}_{\textit{\textbf{LEDGER}}}$ used in our model is similar to that in {[}15{]}, but has a few differences.
        Since we no longer globally provide slot number by clock function, the functionality $\boldsymbol{G}_{\textit{\textbf{LEDGER}}}$ acquires current round time stamp from the simulator for the input given by the environment.
        Thus, the simulator keeps track of the execution of participants and maintains corresponding parameters of current slot number and round information.
        The ledger function is detailed in \uline{Section A.4} and the simulator is detailed in \uline{Section C}.

\textbf{Other Hybrids.} Our protocol also needs VRF (verifiable random function) functionality $F_{VRF}$, KES (key-evolving signature) functionality $F_{KES}$, a (global) random oracle functionality $G_{RO}$ as well.
        For the details we refer to {[}14{]}

\chapter{ The New Protocol: Ouroboros AutoSyn}

\section{ Overview and Main Challenges}

The protocol Ouroboros AutoSyn inherits the basic mode of operation from {[}15{]} and some features from {[}16{]}.
        In {[}15{]}, the execution of the protocol depends on a global clock $G_{\text{clock}}$ providing each party with the current global time or slot number allowing an agreement on the current round number at any instant of the execution.
        With a central clock keeping every participant at the same pace, during every round, parties execute fetch information and select chain commands to update local chains.
        And then participants run a local lottery procedure, to decide who should be setting the block of this round according to their stake ratio.
        After that, the slot leader publishes his block by multicasting the new block to $F_{N-MC}$(delay time control by adversary).
        Stake ratio adjusts at the beginning of every new epoch.
        However, in {[}16{]} participants need to adjust their local time by multicasting ``Synchronization Beacon'' stored on the Blockchain and used for the adjustment at the end of the epoch.
        There is sub-procedure running at the end of every epoch, to adjust the difference of the local time between every participant. 

{[}15{]} and {[}16{]} both need all parties to inform the global clock or ``tick'' function before moving to next round.
        The clock function in our protocol only acts like a clock, no longer gathering executing status of each participant and maintaining round information.
        Participants locally execute the protocol in a round structure, for $t_{\textit{round}}$ length of time and calculate the current slot number from the present time $t_{now}$ (obtained from the global clock function $G_{\text{AutoClock}}$).
        In the light of {[}16{]}, every slot leader broadcasts an ``$\textit{Adjust}$'' message to adjust slot length $t_{\textit{round}}$ at the start of every epoch. 

\section{ Party Types}

\begin{table}
\begin{tabularx}{\textwidth}{|p{\dimexpr 0.403\linewidth-2\tabcolsep-2\arrayrulewidth}|p{\dimexpr 0.282\linewidth-2\tabcolsep-\arrayrulewidth}|p{\dimexpr 0.315\linewidth-2\tabcolsep-\arrayrulewidth}|} \hline 
\multirow[t]{2}{*}{\textbf{Resource}} & \multicolumn{2}{p{\dimexpr 0.597\linewidth-2\tabcolsep-\arrayrulewidth}|}{\centering\arraybackslash{}\centering\arraybackslash{}\textbf{Basic types of honest parties}}\\\cline{2-3}
 & \textbf{Resource unavailable} & \textbf{Resource available}\\\hline 
Random oracle $G_{RO}$ & stalled & operational\\\hline 
 Clock $G_{\textit{AutoClock}}$ &  time-unaware &  time-aware\\\hline 
Network $F_{N-MC}$ & Offline & Online\\\hline 
Synchronized state & desynchronized & synchronized\\\hline 
\end{tabularx}
\end{table}

Stalled parties are time-aware, but unable to perform protocol execution, which will also make them desynchronized.
        However, as soon as they are registered to the random oracle function, they will get synchronized by the information from the network as long as they were not offline.
        More details are on {[}15{]}.

\textbf{alert }:${\Leftrightarrow}$ \textbf{operational }$\cap $ \textbf{online }$\cap $ \textbf{synchronized }$\cap $ \textbf{time-aware}

\textbf{active }:${\Leftrightarrow}$ (\textbf{operational }$\cap $ \textbf{online }$\cap $\textbf{time-aware) }$\cup $ \textbf{adversarial }$\cup $ \textbf{time-unaware}

\section{ Technical Overview with Differences to Ouroboros Genesis}

The structure of Ouroboros AutoSyn is similar to {[}15{]}, it has three parts which are registration, inputs of command and interaction with global function.
        And handling interrupts in a UC protocol.
        All operations are given as pseudo-code in the following.
        To underline the changes to Ouroboros Genesis we marked the lines that are new to Ouroboros AutoSyn in \textcolor{color-2E74B5}{blue}\textcolor{color-00B0F0}{.}

\subsection{Time-based round structure}

Ouroboros AutoSyn is a round model based on the counter of global clock $\mathrm{G}_{\textit{AutoClock}}$. $\mathrm{G}_{\textit{AutoClock}}$ is a global function, every registered participant can access it for the current time $\mathrm{t}_{\mathrm{now}}$.
        During every round, participants store the beginning time $\mathrm{t}_{\text{begin}}$ and the end time $\mathrm{t}_{\text{next}}$ of the round.  $\mathrm{t}_{\text{begin}}$ is also the end time of the round before this round, thus $\mathrm{t}_{\text{next}}$ is the start time of the next round.
        The round length is measured in time length and denoted as $\mathrm{t}_{\text{round}}$ which stays the same during each epoch and is adjusted at the beginning of the next epoch.
        Initially, genesis block stores the round length of the first epoch $(\mathrm{t}_{{\text{round}_{1}}})$.
        And naturally, we have 

{\centering $\mathrm{t}_{\text{next}}=\mathrm{t}_{\text{begin}}+\mathrm{t}_{\text{round}}$ (of one round).

\par}Participants get the current slot number by comparing the round length with the current time $\mathrm{t}_{\mathrm{now}}$, to see which slot $\mathrm{t}_{\mathrm{now}}$ is currently at, and denote the current slot number as $sl$. 

Precisely, we have 

\begin{align*}
\mathrm{t}_{\text{begin}}&=\mathrm{t}_{{\text{round}_{1}}}\cdot \mathrm{R}+\ldots +\mathrm{t}_{{\text{round}_{\mathrm{ep}}}}\cdot \left(\mathrm{sl}-\mathrm{R}\cdot (\mathrm{ep}-1)-1\right) \\
\mathrm{t}_{\text{next}}&=\mathrm{t}_{{\text{round}_{1}}}\cdot \mathrm{R}+\ldots +\mathrm{t}_{{\text{round}_{\mathrm{ep}}}}\cdot (\mathrm{sl}-\mathrm{R}\cdot (\mathrm{ep}-1)) 
\end{align*}

And
\begin{equation*}
\mathrm{t}_{\text{begin}}\leq \mathrm{t}_{\mathrm{now}}\leq \mathrm{t}_{\text{next}}
\end{equation*}
$\mathrm{ep}$ denotes the epoch the party is currently at.

\subsection{ Registration and Special Procedures}

A party P needs access to all its resources in order to start operation.
        Once it is registered to all resources it is able to perform basic operations. 

\textbf{Initialization}.
        The first special procedure a party runs through is initialization as in Section \uline{B.3}.
        It is invoked by the participant the first time executing MAINTAIN-LEDGER command.
        If the first round has not begun ($\mathrm{t}_{\mathrm{now}}< 0$), it will inform $\mathrm{F}_{\text{INIT}}$ for initial stake together with Genesis block information containing initial round length $\mathrm{t}_{{\text{round}_{1}}}$.
        We assume that when the protocol starts running, every party which holds an initial share has successfully received their share from $F_{INIT}$, and it's up to them whether to participate or not.
        And if this party joins the execution after the first round begins, it gathers corresponding information and runs the JoinProc procedure to update time ($\mathrm{t}_{\text{next}}$, $\mathrm{t}_{\text{round}}$). 

\textbf{Joining procedure.} The joining procedure (Section \uline{B.4}) is executed by a new party after joining the protocol.
        The newly joined parties will get synchronized by keeping their parameter update with other participants.
        These parties need to wait a while (3$\cdot \mathrm{t}_{{\text{round}_{1}}}$) to gather the current information of the chain and then calculate the right round count information.

\textbf{Get round number from real-time}.
        Since the global clock no longer maintains the current slot information, in order to get slot information, parties need to calculate slot number from $\mathrm{t}_{\mathrm{now}}$ with the parameter they stored or from their local chain.
        Basically, they first calculate the epoch they are currently at by getting the corresponding round length, and then they use the round length of current epoch to determine the slot number.
        The function is detailed in Section \uline{B.14}. (In reality they can store a table for connection between the time period and the slot number for efficiency.
        For simplicity we only use a function to calculate the current round number)

\textbf{Adjustment procedure}.
        At the beginning of each epoch, participants use ``adjust'' message on the chain to adjust the length of next round, which shares the similarity with round number functionality.
        The function is detailed in Section \uline{B.13}.

\subsection{ Mode of Operation for Alert Parties}

        If a party has been keeping up with the pace of the protocol, and keeping $\mathrm{t}_{\text{round}}$ and $\mathrm{t}_{\text{next}}$ updated, then it is synchronized and if it has registered to all the resources then it is an alert party.
        The standard execution of the alert party is as below.
\begin{itemize}
\item[-] Get messages such as transactions $\mathrm{tx}$, chains $\mathrm{N}$, new parties message$\text{ HELLO}$, adjust message$\text{ adjust}$ by invoking FetchInformation (shown in Section \uline{B.9}) and store them in local buffer parameters.
\item[-] Update local chain $C_{loc}$ by invoking SelectChain (Section \uline{B.7}).
\item[-] Invoke UpdateTime function (Section \uline{B.8}) to update time.
        If the party is not stalled, update the slot number by one, check if the current slot is a new slot of an epoch and get the corresponding round length $\mathrm{t}_{\text{round}}$.
        Finally update $\mathrm{t}_{\text{begin}}$, $\mathrm{t}_{\text{next}}$.
        If the party has stalled but local chain updated, then invoke CurrentSlotNumber function to get corresponding round information.
\item[-] Update stake information by invoking UpdateStakeDist (Section \uline{B.12}).
\item[-] Executing main staking procedure StakingProcedure(Section \uline{B.5}), to evaluate leadership of this round.
        If selected as the round leader, issue a new block and send it to $F_{N-MC}^{bc}$ if not delayed, and keep track of last round block delay information for the next procedure.
        At the end of this procedure, whether elected as the leader or delayed, update the KES signing key.
\item[-] If it is selected as the round leader, or during the last two rounds selected as one, it is able to run this AdjustDelay protocol.
        The procedure sends adjust information to the multicast network, which contains the block received within last two rounds, the receiving time, and the proof elected as slot leader. 
\item[-] End this round by invoking FinishRound(Section \uline{B.10}) and waits for next round start time $\mathrm{t}_{\text{next}}$.
        Before running function FinishRound, if ``WELCOME'' message has been received, broadcast its local chain.
\end{itemize}
\textbf{Stake distribution and leader election}

Relationship between relative stake of P ($\upalpha _{\mathrm{p}}^{\mathrm{ep}}$) and probability elected as the slot leader under stake distribution $\mathrm{S}_{\mathrm{ep}}$ is,
\begin{equation*}
\varnothing _{f}\left(\alpha \right)=1-(1-f)^{\alpha }
\end{equation*}

with active slots coefficient $f\in (0,1]$.
        And the threshold is defined as
\begin{equation*}
T_{p}^{ep}=2^{{l_{VRF}}}\phi _{f}(\alpha _{p}^{ep})
\end{equation*}
Independent aggregation is also satisfied. 

Constants TEST used in VRF are to evaluate slot leadership, and NONCE is for the randomness for the next epoch.
        We use $\mathrm{F}_{\mathrm{KES}}$ to produce signature $\upsigma $.

More details can be found in {[}15{]}.

\subsection{ Further Ledger Queries}

\textbf{Submit transactions. }Similar to {[}15{]}, when receiving a transaction input, it will store the transactions information and send the message to multicast network. 

\textbf{Read state. }Similar to {[}15{]}, when receiving a stake-read input, it will export its newest state to the environment.
        The function is detailed in Section B.16.

\subsection{ De-Registration and Re-Joining}

An alert party will lose its status if losing access to some of its resources.
        Similar to {[}15{]}, if a party lost the resource of random oracle (this will be shown as sending deregistration command to $\mathrm{G}_{\mathrm{RO}}$), it can still receive messages and observe the protocol execution.
        And it can still join the protocol.
        If it loses the resource of the network (deregistration of $\mathrm{F}_{\mathrm{N}-\mathrm{MC}}$), it has lost the connection to the network, and remembers that the lost connection has happened.
        And after it has rejoined the network, it invokes JoinProc protocol to wait a few rounds and gather information before getting synchronized.
        The function is detailed in Section B.17.

\section{ The Adjusting Round Procedure of Ouroboros AutoSyn}

\textbf{Adjustment message.} During every round the slot leader issues a message $\textit{Adjust}$ to adjust the slot length.
        Message $\textit{Adjust}$ records last block $B_{last}$ and the receiving time $T_{recv}$, sending party P and the leader proof $(y,\pi )$.
\begin{equation*}
\textit{Adjust}\triangleq (B_{last},T_{recv}, P,y,\pi )
\end{equation*}
\textbf{Record adjust message.} For every participant, if receives an $\text{Adjust}$ message, record its arriving time $T_{adj}$ and stores it as \{($\textit{Adjust}_{1},T_{{adj_{1}}}),${\ldots}$,(\textit{Adjust}_{\mathrm{n}},T_{{adj_{n}}})$\} in buffer.
        If this party is elect as slot leader, it checks and records all of the pair which have not appeared on the block chain.

\textbf{Adjust next round length}.
        At the start of every epoch, every participant needs to adjust the round length by using the adjustment message on the blockchain.
        Considering the common prefix issue, we only use (i) the first half (R/2) rounds in the last epoch and (ii) the second half (R/2) rounds in the epoch before the last one (if the last epoch is the first epoch, we only use (i)).
        Because the round length is different between (i) and (ii), we calculate them separately with different parameters $t_{\textit{round}}^{1}$, $t_{\textit{round}}^{2}$.
        The relation between old round length and the new length $t_{\textit{round}-new}^{1}$ is 
\begin{equation*}
t_{\textit{round}-new}^{1}=\omega _{1}\left(\frac{1}{m}\sum \limits_{1}^{m}t_{i}^{a}-t_{\textit{round}}^{1}\right)+\omega _{2}\left(\frac{1}{m}\sum \limits_{1}^{m}t_{i}^{b}-t_{\textit{round}}^{1}\right)
\end{equation*}
And $t_{i}^{a}$ denotes the latency between block $B_{last}$ sending time and its receiving time, and $t_{i}^{b}$ is the length of the corresponding block adjustment message sending delay. $\omega _{1}$ and $\omega _{2}$ are two parameters and we have $0< \omega _{2}\leq \omega _{1}< 1\& \omega _{2}+\omega _{1}\leq 1$, but usually set $\omega _{1}=0.3, \omega _{2}=0.1$.
        And new round length is the average of them.

\textbf{Current slot number.} For a party who has updated local chain and current time, this party can calculate the current slot number $sl$ by calculating each epoch round length and adding to $t_{\textit{start}}$ until the result is larger than the current time $t_{now}$.
        The function is detailed in Section B.14.

\section{ The Resynchronization Procedure of Ouroboros AutoSyn}

\textbf{New joining party.} If a party runs the protocol for the first time, it will first need to invoke Initialization-AutoSyn function to get the initial stake and the genesis block from $F_{INIT}$.
        And if the party joins the protocol outside the initial round, it needs to invoke JoinProc to update the local information and catch up with the current round.
        The purpose of this function is similar to the one in {[}16{]}.

\textbf{Online stalled party.} If a party is stalled which means this party has been listening to the network and receiving the chains from the network but not executing the protocol, then this party needs to run the CurrentSlotNumber function to update the local parameters in order to keep up with the other participants.
        It needs to update $t_{\textit{round}}$ and $t_{next}$ for the next round execution by invoking SelectChain to update local chain $C_{loc}$ and calculating from it.

\textbf{Offline party.} If a party has lost its connection to the network (the party is aware of this state), it needs to rejoin the protocol.
        Since the party has already initialized it will only need to run JoinProc to update the local information and catch up with the current round.

\chapter{Security Analysis}

\section{Time structure of a round}

\subsection{ Scenario of message delivery}
\textbf{General situation}

Assume $P_{1},P_{2},P_{3},P_{4}$ are honest nodes and are selected as the leaders of $sl_{1},sl_{2},sl_{3},sl_{4}$.
        During $sl_{1}$, $P_{1}$ executes the protocol and broadcasts $B_{1}$ and $P_{2}$ receives this block within $sl_{1}$.
        Thus $P_{2}$ can broadcast its block $B_{2}$ as soon as $sl_{2}$ begins so that this block can be accepted by as many parties as possible.
        Unfortunately, due to network delay, $B_{2}$ was not received during $sl_{2}$ by $P_{3}$.
        After waiting a small amount of time (less than pre-waiting time decided by $P_{3}$), $P_{3}$ sends its block $B_{2}^{\prime}$ linking to $B_{1}$ causing a divergence to $B_{2}$.
        But since $B_{2}$ has already been broadcast a round of time ahead of $B_{2}^{\prime}$ (if $B_{2}$ was not lost to $P_{4}$ in the network), $P_{4}$ has a higher probability of receiving $B_{2}$ than $B_{2}^{\prime}$ and sends $B_{4}$ linking to $B_{2}$.
        Thus we have $B_{1}$-$B_{2}$-$B_{4}$.
        But if $P_{2}$ has a really poor network connection $B_{2}^{\prime}$ could beat $B_{2}$, and we have $B_{1}$-$B_{2}^{\prime}$-$B_{4}$.

\begin{figure}[h]
    \centering
    \begin{tikzpicture}[
        scale=0.9,
        diamond/.style={shape=diamond, fill=black, minimum size=3pt, inner sep=0pt},
        triangle down/.style={shape=regular polygon, regular polygon sides=3, fill=black, minimum size=5pt, inner sep=0pt, rotate=180},
        triangle up/.style={shape=regular polygon, regular polygon sides=3, fill=black, minimum size=5pt, inner sep=0pt}
    ]
        \def\slotwidth{2}
        \def\slotsep{0.5}
        
        \def\yPOne{3}
        \def\yPTwo{1.5}
        \def\yPThree{0}
        \def\yPFour{-1.5}
        
        \def\xstart{0}
        \def\xslOne{2}
        \def\xslTwo{4}
        \def\xslThree{6}
        \def\xslFour{8}
        
        \coordinate (P1start) at (\xstart,\yPOne);
        \coordinate (P1sl1) at (\xslOne,\yPOne);
        \coordinate (P1sl2) at (\xslTwo,\yPOne);
        \coordinate (P1sl3) at (\xslThree,\yPOne);
        \coordinate (P1sl4) at (\xslFour,\yPOne);
        
        \coordinate (P2start) at (\xstart,\yPTwo);
        \coordinate (P2sl1) at (\xslOne,\yPTwo);
        \coordinate (P2sl2) at (\xslTwo,\yPTwo);
        \coordinate (P2sl3) at (\xslThree,\yPTwo);
        \coordinate (P2sl4) at (\xslFour,\yPTwo);
        
        \coordinate (P3start) at (\xstart,\yPThree);
        \coordinate (P3sl1) at (\xslOne,\yPThree);
        \coordinate (P3sl2) at (\xslTwo,\yPThree);
        \coordinate (P3sl3) at (\xslThree,\yPThree);
        \coordinate (P3sl4) at (\xslFour,\yPThree);
    
        \coordinate (P4start) at (\xstart,\yPFour);
        \coordinate (P4sl1) at (\xslOne,\yPFour);
        \coordinate (P4sl2) at (\xslTwo,\yPFour);
        \coordinate (P4sl3) at (\xslThree,\yPFour);
        \coordinate (P4sl4) at (\xslFour,\yPFour);
        
        \draw[very thick] (P1start) -- (P1sl1);
        \draw[dashed, thick] (P1sl1) -- (P1sl2);
        \draw[dashed, thick] (P1sl2) -- (P1sl3);
        \draw[dashed, thick] (P1sl3) -- (P1sl4);
        
        \draw[dashed, thick] (P2start) -- (P2sl1);
        \draw[very thick] (P2sl1) -- (P2sl2);
        \draw[dashed, thick] (P2sl2) -- (P2sl3);
        \draw[dashed, thick] (P2sl3) -- (P2sl4);
        
        \draw[dashed, thick] (P3start) -- (P3sl1);
        \draw[dashed, thick] (P3sl1) -- (P3sl2);
        \draw[very thick] (P3sl2) -- (P3sl3);
        \draw[dashed, thick] (P3sl3) -- (P3sl4);
        
        \draw[dashed, thick] (P4start) -- (P4sl1);
        \draw[dashed, thick] (P4sl1) -- (P4sl2);
        \draw[dashed, thick] (P4sl2) -- (P4sl3);
        \draw[very thick] (P4sl3) -- (P4sl4);
        
        \foreach \y in {\yPOne,\yPTwo,\yPThree,\yPFour} {
            \foreach \x in {\xstart,\xslOne,\xslTwo,\xslThree,\xslFour} {
                \node[diamond] at (\x,\y) {};
            }
        }
        
        \node[above] at (1,\yPOne+0.3) {$sl_1$};
        \node[above] at (3,\yPOne+0.3) {$sl_2$};
        \node[above] at (5,\yPOne+0.3) {$sl_3$};
        \node[above] at (7,\yPOne+0.3) {$sl_4$};
        
        \node[left] at (P1start) {P1};
        \node[left] at (P2start) {P2};
        \node[left] at (P3start) {P3};
        \node[left] at (P4start) {P4};
        
        \coordinate (P1send) at (0.5,\yPOne);
        \node[above] at (0.5,\yPOne+0.3) {$B_{1}$};
        \node[triangle down] at (P1send) {};
        \coordinate (P2recvB1) at (1.5,\yPTwo);
        \coordinate (P3recvB1) at (1.5,\yPThree);
        \coordinate (P4recvB1) at (1.5,\yPFour);
        \node[triangle up] at (P2recvB1) {};
        \node[triangle up] at (P3recvB1) {};
        \node[triangle up] at (P4recvB1) {};
        \draw[dashed] (P1send) -- (P2recvB1);
        \draw[dashed] (P1send) -- (P3recvB1);
        \draw[dashed] (P1send) -- (P4recvB1);
        
        \coordinate (P2send) at (2.5,\yPTwo);
        \node[above] at (2.5,\yPTwo+0.3) {$B_{2}$};
        \node[triangle down] at (P2send) {};
        \coordinate (P1recvB2) at (5,\yPOne);
        \coordinate (P3recvB2) at (5,\yPThree);
        \coordinate (P4recvB2) at (5,\yPFour);
        \node[triangle up] at (P3recvB2) {};
        \node[triangle up] at (P1recvB2) {};
        \node[triangle up] at (P4recvB2) {};
        \draw[dashed] (P2send) -- (P3recvB2);
        \draw[dashed] (P2send) -- (P1recvB2);
        \draw[dashed] (P2send) -- (P4recvB2);
        
        \coordinate (P3send) at (4.5,\yPThree);
        \node[above] at (4.5,\yPThree+0.3) {$B_{2}^{\prime}$};

        \node[triangle down] at (P3send) {};
        \coordinate (P1recvB3) at (5.5,\yPOne);
        \coordinate (P2recvB3) at (5.5,\yPTwo);
        \coordinate (P4recvB3) at (5.5,\yPFour);
        \node[triangle up] at (P4recvB3) {};
        \node[triangle up] at (P1recvB3) {};
        \node[triangle up] at (P2recvB3) {};
        \draw[dashed] (P3send) -- (P4recvB3);
        \draw[dashed] (P3send) -- (P1recvB3);
        \draw[dashed] (P3send) -- (P2recvB3);
        
        \coordinate (P4send) at (6.5,\yPFour);
        \node[above] at (6.5,\yPFour+0.3) {$B_{4}$};
        \node[triangle down] at (P4send) {};
        \coordinate (P1recvB4) at (7.5,\yPOne);
        \coordinate (P2recvB4) at (7.5,\yPTwo);
        \coordinate (P3recvB4) at (7.5,\yPThree);
        \node[triangle up] at (P1recvB4) {};
        \node[triangle up] at (P2recvB4) {};
        \node[triangle up] at (P3recvB4) {};
        \draw[dashed] (P4send) -- (P1recvB4);
        \draw[dashed] (P4send) -- (P2recvB4);
        \draw[dashed] (P4send) -- (P3recvB4);

        \coordinate (legendpos) at (10,3);
        \node[triangle down] at (legendpos) {};
        \node[right] at ([xshift=8pt]legendpos) {Send time};
        \coordinate (legendpos2) at (10,1.5);
        \node[triangle up] at (legendpos2) {};
        \node[right] at ([xshift=8pt]legendpos2) {Receive time};
        
    \end{tikzpicture}
    \caption{Timing diagram showing message delivery between four processes across four time slots, generating $B_{1}$-$B_{2}$-$B_{4}$}
\end{figure}
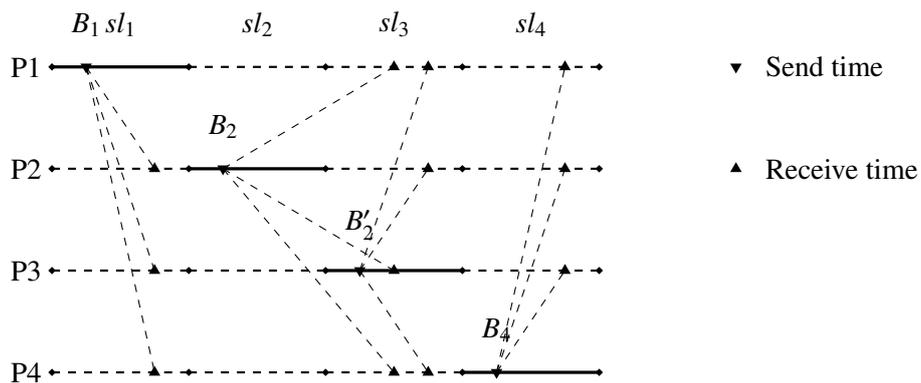

\begin{figure}[H]
    \centering
    \begin{tikzpicture}[
        scale=0.9,
        diamond/.style={shape=diamond, fill=black, minimum size=3pt, inner sep=0pt},
        triangle down/.style={shape=regular polygon, regular polygon sides=3, fill=black, minimum size=5pt, inner sep=0pt, rotate=180},
        triangle up/.style={shape=regular polygon, regular polygon sides=3, fill=black, minimum size=5pt, inner sep=0pt}
    ]
        \def\slotwidth{2}
        \def\slotsep{0.5}
        
        \def\yPOne{3}
        \def\yPTwo{1.5}
        \def\yPThree{0}
        \def\yPFour{-1.5}
        
        \def\xstart{0}
        \def\xslOne{2}
        \def\xslTwo{4}
        \def\xslThree{6}
        \def\xslFour{8}
        
        \coordinate (P1start) at (\xstart,\yPOne);
        \coordinate (P1sl1) at (\xslOne,\yPOne);
        \coordinate (P1sl2) at (\xslTwo,\yPOne);
        \coordinate (P1sl3) at (\xslThree,\yPOne);
        \coordinate (P1sl4) at (\xslFour,\yPOne);
        
        \coordinate (P2start) at (\xstart,\yPTwo);
        \coordinate (P2sl1) at (\xslOne,\yPTwo);
        \coordinate (P2sl2) at (\xslTwo,\yPTwo);
        \coordinate (P2sl3) at (\xslThree,\yPTwo);
        \coordinate (P2sl4) at (\xslFour,\yPTwo);
        
        \coordinate (P3start) at (\xstart,\yPThree);
        \coordinate (P3sl1) at (\xslOne,\yPThree);
        \coordinate (P3sl2) at (\xslTwo,\yPThree);
        \coordinate (P3sl3) at (\xslThree,\yPThree);
        \coordinate (P3sl4) at (\xslFour,\yPThree);
    
        \coordinate (P4start) at (\xstart,\yPFour);
        \coordinate (P4sl1) at (\xslOne,\yPFour);
        \coordinate (P4sl2) at (\xslTwo,\yPFour);
        \coordinate (P4sl3) at (\xslThree,\yPFour);
        \coordinate (P4sl4) at (\xslFour,\yPFour);
        
        \draw[very thick] (P1start) -- (P1sl1);
        \draw[dashed, thick] (P1sl1) -- (P1sl2);
        \draw[dashed, thick] (P1sl2) -- (P1sl3);
        \draw[dashed, thick] (P1sl3) -- (P1sl4);
        
        \draw[dashed, thick] (P2start) -- (P2sl1);
        \draw[very thick] (P2sl1) -- (P2sl2);
        \draw[dashed, thick] (P2sl2) -- (P2sl3);
        \draw[dashed, thick] (P2sl3) -- (P2sl4);
        
        \draw[dashed, thick] (P3start) -- (P3sl1);
        \draw[dashed, thick] (P3sl1) -- (P3sl2);
        \draw[very thick] (P3sl2) -- (P3sl3);
        \draw[dashed, thick] (P3sl3) -- (P3sl4);
        
        \draw[dashed, thick] (P4start) -- (P4sl1);
        \draw[dashed, thick] (P4sl1) -- (P4sl2);
        \draw[dashed, thick] (P4sl2) -- (P4sl3);
        \draw[very thick] (P4sl3) -- (P4sl4);
        
        \foreach \y in {\yPOne,\yPTwo,\yPThree,\yPFour} {
            \foreach \x in {\xstart,\xslOne,\xslTwo,\xslThree,\xslFour} {
                \node[diamond] at (\x,\y) {};
            }
        }
        
        \node[above] at (1,\yPOne+0.3) {$sl_1$};
        \node[above] at (3,\yPOne+0.3) {$sl_2$};
        \node[above] at (5,\yPOne+0.3) {$sl_3$};
        \node[above] at (7,\yPOne+0.3) {$sl_4$};
        
        \node[left] at (P1start) {P1};
        \node[left] at (P2start) {P2};
        \node[left] at (P3start) {P3};
        \node[left] at (P4start) {P4};
        
        \coordinate (P1send) at (0.5,\yPOne);
        \node[above] at (0.5,\yPOne+0.3) {$B_{1}$};
        \node[triangle down] at (P1send) {};
        \coordinate (P2recvB1) at (1.5,\yPTwo);
        \coordinate (P3recvB1) at (1.5,\yPThree);
        \coordinate (P4recvB1) at (1.5,\yPFour);
        \node[triangle up] at (P2recvB1) {};
        \node[triangle up] at (P3recvB1) {};
        \node[triangle up] at (P4recvB1) {};
        \draw[dashed] (P1send) -- (P2recvB1);
        \draw[dashed] (P1send) -- (P3recvB1);
        \draw[dashed] (P1send) -- (P4recvB1);
        
        \coordinate (P2send) at (2.5,\yPTwo);
        \node[above] at (2.5,\yPTwo+0.3) {$B_{2}$};
        \node[triangle down] at (P2send) {};
        \coordinate (P1recvB2) at (6,\yPOne);
        \coordinate (P3recvB2) at (6,\yPThree);
        \coordinate (P4recvB2) at (6,\yPFour);
        \node[triangle up] at (P3recvB2) {};
        \node[triangle up] at (P1recvB2) {};
        \node[triangle up] at (P4recvB2) {};
        \draw[dashed] (P2send) -- (P3recvB2);
        \draw[dashed] (P2send) -- (P1recvB2);
        \draw[dashed] (P2send) -- (P4recvB2);
        
        \coordinate (P3send) at (4.5,\yPThree);
        \node[above] at (4.5,\yPThree+0.3) {$B_{2}^{\prime}$};

        \node[triangle down] at (P3send) {};
        \coordinate (P1recvB3) at (5.5,\yPOne);
        \coordinate (P2recvB3) at (5.5,\yPTwo);
        \coordinate (P4recvB3) at (5.5,\yPFour);
        \node[triangle up] at (P4recvB3) {};
        \node[triangle up] at (P1recvB3) {};
        \node[triangle up] at (P2recvB3) {};
        \draw[dashed] (P3send) -- (P4recvB3);
        \draw[dashed] (P3send) -- (P1recvB3);
        \draw[dashed] (P3send) -- (P2recvB3);
        
        \coordinate (P4send) at (6.5,\yPFour);
        \node[above] at (6.5,\yPFour+0.3) {$B_{4}$};
        \node[triangle down] at (P4send) {};
        \coordinate (P1recvB4) at (7.5,\yPOne);
        \coordinate (P2recvB4) at (7.5,\yPTwo);
        \coordinate (P3recvB4) at (7.5,\yPThree);
        \node[triangle up] at (P1recvB4) {};
        \node[triangle up] at (P2recvB4) {};
        \node[triangle up] at (P3recvB4) {};
        \draw[dashed] (P4send) -- (P1recvB4);
        \draw[dashed] (P4send) -- (P2recvB4);
        \draw[dashed] (P4send) -- (P3recvB4);

        \coordinate (legendpos) at (10,3);
        \node[triangle down] at (legendpos) {};
        \node[right] at ([xshift=8pt]legendpos) {Send time};
        \coordinate (legendpos2) at (10,1.5);
        \node[triangle up] at (legendpos2) {};
        \node[right] at ([xshift=8pt]legendpos2) {Receive time};
        
    \end{tikzpicture}
    \caption{Timing diagram showing message delivery between four processes across four time slots, generating $B_{1}$-$B_{2}^{\prime}$-$B_{4}$}
\end{figure}

\textbf{Empty slot and delay slot.}

\begin{figure}[h]
    \centering
    \begin{tikzpicture}[
        scale=1.2,
        diamond/.style={shape=diamond, fill=black, minimum size=3pt, inner sep=0pt},
        triangle down/.style={shape=regular polygon, regular polygon sides=3, fill=black, minimum size=4pt, inner sep=0pt, rotate=180},
        triangle up/.style={shape=regular polygon, regular polygon sides=3, fill=black, minimum size=4pt, inner sep=0pt}
    ]
        \coordinate (start) at (0,0);
        \coordinate (mid) at (3,0);
        \coordinate (end) at (6,0);
        
        \coordinate (P1start) at (0,1);
        \coordinate (P1mid) at (3,1);
        \coordinate (P1end) at (6,1);
        
        \coordinate (P2start) at (0,-1);
        \coordinate (P2mid) at (3,-1);
        \coordinate (P2end) at (6,-1);
        
        \draw[dashed, thick] (P1start) -- (P1mid);
        \draw[dashed, thick] (P1mid) -- (P1end);
        
        \draw[dashed, thick] (P2start) -- (P2mid);
        \draw[very thick] (P2mid) -- (P2end);
        
        \node[diamond] at (P1start) {};
        \node[diamond] at (P1mid) {};
        \node[diamond] at (P1end) {};
        \node[diamond] at (P2start) {};
        \node[diamond] at (P2mid) {};
        \node[diamond] at (P2end) {};
        
        \node[above] at (1.5,1) {$sl_1$};
        \node[above] at (4.5,1) {$sl_2$};

        \node[left] at (P1start) {P1};
        \node[left] at (P2start) {P2};
        
        \node[triangle down] at (3.8,-1) {};
        \node[above] at (3.8, -0.3) {$B_{1}$};
        \node[triangle up] at (5.2,1) {};
        
        \draw[dashed] (3.8,-1) -- (5.2,1);

        \coordinate (legendpos) at (8,1.5);
        \node[triangle down] at (legendpos) {};
        \node[right] at ([xshift=8pt]legendpos) {Send time};
        \coordinate (legendpos2) at (8,0.5);
        \node[triangle up] at (legendpos2) {};
        \node[right] at ([xshift=8pt]legendpos2) {Receive time};
    \end{tikzpicture}
    \caption{Example timeline diagram showing two processes P1 and P2 with slot segments}
\end{figure}
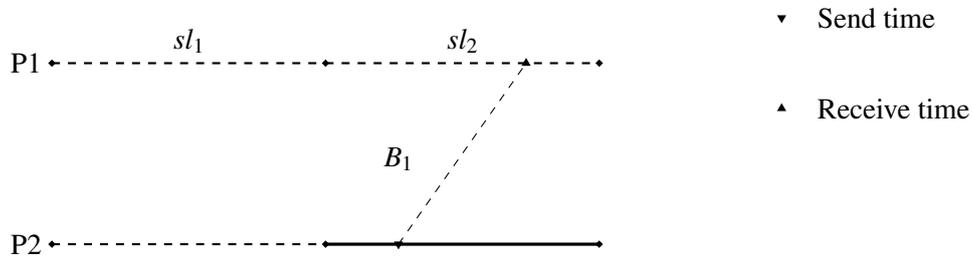

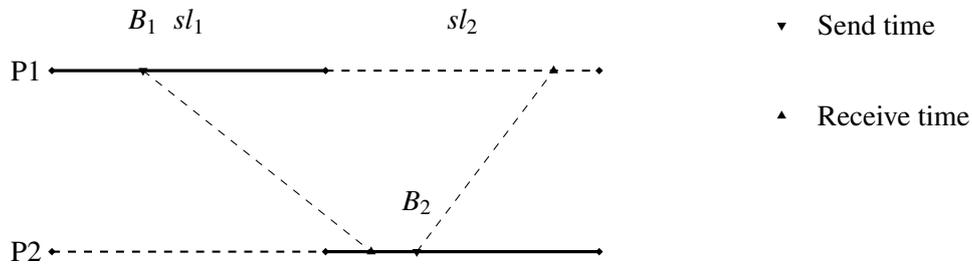
\begin{figure}[h]
\centering
\begin{tikzpicture}[
    scale=1.2,
    diamond/.style={shape=diamond, fill=black, minimum size=3pt, inner sep=0pt},
    triangle down/.style={shape=regular polygon, regular polygon sides=3, fill=black, minimum size=4pt, inner sep=0pt, rotate=180},
    triangle up/.style={shape=regular polygon, regular polygon sides=3, fill=black, minimum size=4pt, inner sep=0pt}
]
    \def\yPOne{1}
    \def\yPTwo{-1}
    
    \def\xstart{0}
    \def\xslOne{3}
    \def\xslTwo{6}
    \def\xend{6}
    
    \coordinate (P1start) at (\xstart,\yPOne);
    \coordinate (P1sl1) at (\xslOne,\yPOne);
    \coordinate (P1sl2) at (\xslTwo,\yPOne);
    \coordinate (P1end) at (\xend,\yPOne);
    
    \coordinate (P2start) at (\xstart,\yPTwo);
    \coordinate (P2sl1) at (\xslOne,\yPTwo);
    \coordinate (P2sl2) at (\xslTwo,\yPTwo);
    \coordinate (P2end) at (\xend,\yPTwo);
    
    \draw[very thick] (P1start) -- (P1sl1);
    \draw[dashed, thick] (P1sl1) -- (P1sl2);
    
    \draw[dashed, thick] (P2start) -- (P2sl1);
    \draw[very thick] (P2sl1) -- (P2sl2);
    
    \node[diamond] at (P1start) {};
    \node[diamond] at (P1sl1) {};
    \node[diamond] at (P1sl2) {};
    \node[diamond] at (P2start) {};
    \node[diamond] at (P2sl1) {};
    \node[diamond] at (P2sl2) {};
    
    \node[above] at (1.5,\yPOne+0.3) {$sl_1$};
    \node[above] at (4.5,\yPOne+0.3) {$sl_2$};
    
    \node[left] at (P1start) {P1};
    \node[left] at (P2start) {P2};
    
    \coordinate (P1send) at (1,\yPOne);
    \node[above] at (1, \yPOne+0.3) {$B_{1}$};
    \node[triangle down] at (P1send) {};

    \coordinate (P1recv) at (3.5,\yPTwo);
    \node[triangle up] at (P1recv) {};
    
    \coordinate (P2send) at (4,\yPTwo);
    \node[above] at (4, \yPTwo+0.3) {$B_{2}$};
    \node[triangle down] at (P2send) {};
    
    \coordinate (P2recv) at (5.5,\yPOne);
    \node[triangle up] at (P2recv) {};
    
    \draw[dashed] (P1send) -- (P1recv);
    
    \draw[dashed] (P2send) -- (P2recv);

    \coordinate (legendpos) at (8,1.5);
    \node[triangle down] at (legendpos) {};
    \node[right] at ([xshift=8pt]legendpos) {Send time};
    \coordinate (legendpos2) at (8,0.5);
    \node[triangle up] at (legendpos2) {};
    \node[right] at ([xshift=8pt]legendpos2) {Receive time};
    
\end{tikzpicture}
\caption{Timeline diagram showing interaction between P1 and P2 across two time slots}
\end{figure}

Considering the first and second graph, $P_{2}$ has not received anything sent from the last slot. $P_{2}$ cannot tell the difference whether the last round is empty or that the message has been delayed. $P_{2}$ can wait until pre-waiting ends and run the protocol or just send the message without waiting.
        If the last round is empty and he waits too long, this could cause his message not to be delivered to the next slot leader, thus the next slot leader competing with him.
        But if he doesn't wait, this could lead to him competing with the last round.
        So by default, we set the party waiting until pre-waiting time ends, but a party can set this waiting time shorter according to its network. 

\textbf{Delay-attack.}

\begin{figure}[h]
    \centering
    \begin{tikzpicture}[
        scale=1.2,
        diamond/.style={shape=diamond, fill=black, minimum size=3pt, inner sep=0pt},
        triangle down/.style={shape=regular polygon, regular polygon sides=3, fill=black, minimum size=4pt, inner sep=0pt, rotate=180},
        triangle up/.style={shape=regular polygon, regular polygon sides=3, fill=black, minimum size=4pt, inner sep=0pt}
    ]
        \def\yPOne{1}
        \def\yPTwo{-1}
        
        \def\xstart{0}
        \def\xslOne{3}
        \def\xslTwo{6}
        \def\xend{6}
        
        \coordinate (P1start) at (\xstart,\yPOne);
        \coordinate (P1sl1) at (\xslOne,\yPOne);
        \coordinate (P1sl2) at (\xslTwo,\yPOne);
        \coordinate (P1end) at (\xend,\yPOne);
        
        \coordinate (P2start) at (\xstart,\yPTwo);
        \coordinate (P2sl1) at (\xslOne,\yPTwo);
        \coordinate (P2sl2) at (\xslTwo,\yPTwo);
        \coordinate (P2end) at (\xend,\yPTwo);
        
        \draw[very thick] (P1start) -- (P1sl1);
        \draw[dashed, thick] (P1sl1) -- (P1sl2);
        
        \draw[dashed, thick] (P2start) -- (P2sl1);
        \draw[very thick] (P2sl1) -- (P2sl2);
        
        \node[diamond] at (P1start) {};
        \node[diamond] at (P1sl1) {};
        \node[diamond] at (P1sl2) {};
        \node[diamond] at (P2start) {};
        \node[diamond] at (P2sl1) {};
        \node[diamond] at (P2sl2) {};
        
        \node[above] at (1.5,\yPOne+0.3) {$sl_1$};
        \node[above] at (4.5,\yPOne+0.3) {$sl_2$};
        
        \node[left] at (P1start) {P1};
        \node[left] at (P2start) {P2};
        
        \coordinate (P1send) at (1,\yPOne);
        \node[above] at (1, \yPOne+0.3) {$B_{1}$};
        \node[triangle down] at (P1send) {};
    
        \coordinate (P1recv) at (2.5,\yPTwo);
        \node[triangle up] at (P1recv) {};
        
        \coordinate (P2send) at (4,\yPTwo);
        \node[above] at (4, \yPTwo+0.3) {$B_{1}^{\prime}$};
        \node[triangle down] at (P2send) {};
        
        \coordinate (P2recv) at (5.5,\yPOne);
        \node[triangle up] at (P2recv) {};
        
        \draw[dashed] (P1send) -- (P1recv);
        
        \draw[dashed] (P2send) -- (P2recv);
    
        \coordinate (legendpos) at (8,1.5);
        \node[triangle down] at (legendpos) {};
        \node[right] at ([xshift=8pt]legendpos) {Send time};
        \coordinate (legendpos2) at (8,0.5);
        \node[triangle up] at (legendpos2) {};
        \node[right] at ([xshift=8pt]legendpos2) {Receive time};
        
    \end{tikzpicture}
    \caption{Timeline diagram showing interaction between P1 and P2 across two time slots}
\end{figure}
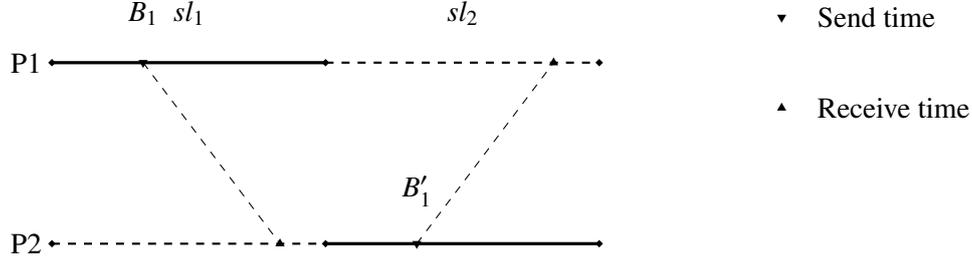

Since a party can make a block ignoring the last slot, the adversary can initiate delay-attack, by pretending not to have received the block sent from the previous round.
        If the adversary did so, it would be competing with the last round leader by network speed.
        But the last round leader has the advantage because it has sent its message a round ahead of the adversary.
        Usually the adversary would fail unless it has a really fast network.

\textbf{Maximum protocol executing time}.

Participants need to deliver their message within $t_{run}$, so that the message has enough time to be delivered to other participants.
        In order that the adversary could affect the next slot leader by deliberately delaying the message sending time, the signing key of the KES function is updated during every round after the party has been executing for $t_{run}$ length of time.

\subsection{ New definitions.}

Now we can analyze all the situations if delay happens.
        We use the notion of characteristic string {[}15{]}, and in the light of discussion above.
        We introduce the Real-Reduction mapping.

\textbf{Definition 1 (Real-Reduction mapping)}\textbf{. }We define the function $\rho _{r}$\textit{:} $\{0,1,\bot \}^{\mathrm{*}}\rightarrow \{0,\bot ,1\}^{\mathrm{*}}$ inductively as follows:

\begin{align*}
\rho _{r}\left(\epsilon \right)&=\epsilon , \\
\rho _{r}\left(1| | w'\right)&=1| | \rho _{r}\left(w'\right), \\
\rho _{r}\left(0| | w'\right)&=\begin{cases} 0| | \rho _{r}\left(w'\right) \textit{if last round has been received}, \\
\bot | | \rho _{r}\left(w'\right) \textit{otherwise}
\end{cases}  
\end{align*}

We argue that the protocol execution of \textbf{\textit{Ouroboros-AutoSyn}} satisfies \textbf{Definition 1}, for more details see \uline{Section E}.

We need to set a bound on the percentage of the message delay (or lost) for the alert party. 

\textbf{Definition 2 (Delivery ratio) }\textit{The chance that during an active round the message sent from the slot leader will be successfully sent to }$\mathrm{P}(j)$ ($\mathrm{P}(j)\neq \mathrm{\varnothing }$) \textit{according to {[}Genesis, Definition 8{]} is more than }$\eta $\textit{. }

The possibility that a message has not been successfully delivered (delay could happen, or message just simply lost, never sent to the next slot leader) is less than $(1-\eta )$.
        As for message sent by adversary, we suppose those will always be delivered on time.
        Here, we set a bound on the delivery ratio $\eta \geq 2/3$.

In the light of discussion above, we can conclude that the probability that $\rho _{r}\left(0| | w'\right)=0| | \rho _{r}\left(w'\right)$ is more than $\eta $ in Reduction mapping $\rho _{r}$.

Proof (sketch).
        When $\rho _{r}\left(0| | w'\right)=0| | \rho _{r}\left(w'\right)$ happens, there are a few situations, such as that the message has been successfully delivered, or there are a few $\bot $ behind $0$.
        Or, the honest leader behind $0$ received the message during the executing period.
        Thus, we have the possibility is larger than $\eta $.

(Remark of \textbf{2-rounds listening time})\textbf{. }In {[}15{]}, $\Updelta $ is the maximum slot delay number of the message.
        However, in our model, we have already taken message delay into consideration, so that messages sent by parties can be received by most of the other online parties within one round.
        But, in case that delay happens, for safety reason, participants wait one more round for the message to spread in the network, so that they can have their local chain stabilized.
        We consider the chain stabilized after 2 rounds of time’s waiting, and 2-rounds listening time is also used in the analysis.

\section{ Blockchain Security Properties}

We now define the standard security properties of Blockchain protocols: common prefix, chain growth and chain quality.
        Similar to {[}15{]}.
        We stress that for every onset of the slots means at the time $t_{\textit{begin}}$ of the round of the party.

 \textbf{Common Prefix (CP); with parameter} $k\in N$.
         The chains $\mathrm{C}_{1},\mathrm{C}_{2}$ possessed by two alert parties at the onset of the slots $\mathrm{sl}_{1}\leq \mathrm{sl}_{2}$, are such that $\mathrm{C}_{1}^{\lceil \mathrm{k}}\leq \mathrm{C}_{2}$, where $\mathrm{C}_{1}^{\lceil \mathrm{k}}$ denotes the chain obtained by removing the last $k$ blocks from $\mathrm{C}_{1}$, and $\leq $ denotes the prefix relation.

 \textbf{Chain Growth (CG); with parameters} $\uptau \in (0,1]$ \textbf{and} $s\in N$.
         Consider a chain C possessed by an alert party at the onset of slot sl.
         Let $\mathrm{sl}_{1}$ and $\mathrm{sl}_{2}$ be two previous slots for which $\mathrm{sl}_{1}+s\leq \mathrm{sl}_{2}\leq sl$, so $\mathrm{sl}_{1}$ is at least s slots prior to $\mathrm{sl}_{2}$.
         Then $\left| \mathrm{C}\left[\mathrm{sl}_{1}\colon \mathrm{sl}_{2}\right]\right| \geq \uptau \cdot \mathrm{s}$.
         We call $\uptau $ the speed coefficient.

 \textbf{Chain Quality} \textbf{(CQ); with parameters} $\upmu \in (0,1]$ \textbf{and} $k\in N$.
         Consider any portion of length at least $k$ of the chain possessed by an alert party at the onset of a slot; the ratio of blocks originating from alert parties in this portion is at least $\upmu $.
         We call $\upmu $ the chain quality coefficient.

Note that previous work identified and studied a stronger version of chain growth( denoted below as CG2), which controls the relative growth of chains held by potentially distinct honest parties.

\textbf{(Strong) Chain Growth (CG2); with parameters} $\uptau \in (0,1]$ \textbf{and }$s\in N$.
        Consider the chains $\mathrm{C}_{1},\mathrm{C}_{2}$ possessed by two alert parties at the onset of the slots $\mathrm{sl}_{1}, \mathrm{sl}_{2}$ with $\mathrm{sl}_{1}$ at least s slots prior to $\mathrm{sl}_{2}$.
        Then it holds that $\mathrm{len}(\mathrm{C}_{2})-\mathrm{len}(\mathrm{C}_{1})\geq \uptau \cdot \mathrm{s}$.
        We call $\uptau $ the speed coefficient.

\textbf{Existential Chain Quality (}$\exists $\textbf{CQ); with parameter} $s\in N$.
        Consider a chain C possessed by an alert party at the onset of slot $sl$.
        Let $\mathrm{sl}_{1}$ and $\mathrm{sl}_{2}$ be two previous slots for which $\mathrm{sl}_{1}+s\leq \mathrm{sl}_{2}\leq sl$.
        Then $\mathrm{C}[\mathrm{sl}_{1}\colon \mathrm{sl}_{2}]$ contains at least one alertly generated block

\section{ Security of Ouroboros AutoSyn with maxvalid-mc}

We adapt the notion of alert ratio and participating ratio from {[}15{]}, and for detailed analysis see Section E.

\textbf{Theorem 1.} \textit{Consider the execution of Ouroboros-AutoSyn with adversary A and environment Z in the setting with static }$F_{N-MC}$\textit{ registration.
        Let f be the active-slot coefficient, let }$\eta $\textit{ be the lower bound on message deliver ratio.
        Let} $\alpha , \beta \in [0,1]$\textit{ denote a lower bound on the alert and participating stake ratios throughout the whole execution, respectively.
        Let R and L denote the epoch length and the total lifetime of the system (in slots), and let Q be the total number of queries issued to }$G_{RO}$\textit{.
        If for some }$\varepsilon \in (0,1)$\textit{ we have}
\begin{equation*}
\upalpha (1-f)^{2}\eta > (1+\epsilon )/2
\end{equation*}
And $R\geq 144\Delta /\epsilon \beta f\eta $ then Ouroboros-AutoSyn achieves the following guarantees:
\begin{itemize}
\item[-] \textbf{Common prefix.} The probability that \textbf{Ouroboros-AutoSyn} violates the common prefix property with parameter k is no more than 
\end{itemize}

\begin{equation*}
\epsilon _{CP}(k)\triangleq \frac{19L}{\varepsilon ^{4}}\exp \left(2-\frac{\varepsilon ^{4}k}{18}\right)+\upvarepsilon _{\text{lift}}
\end{equation*}

\begin{itemize}
\item[-] \textbf{Chain growth.} The probability that \textbf{Ouroboros-AutoSyn} violates the chain growth property with parameter $\mathrm{s}\geq 96/(\upepsilon \beta f\eta )$ and $\tau _{CG}=\beta f\eta /16$ is no more than
\end{itemize}

\begin{equation*}
\epsilon _{CG}\left(\tau _{CG},s\right)\triangleq \frac{sL^{2}}{2}\exp \left(-\frac{\left(\epsilon f\beta \eta \right)^{2}s}{256}\right)+\upvarepsilon _{\text{lift}}
\end{equation*}

\begin{itemize}
\item[-] \textbf{Existential chain quality}.
        The probability that \textbf{Ouroboros-AutoSyn} violates the existential chain quality property with parameter $\mathrm{s}\geq 24/(\upepsilon \beta f\eta )$ is no more than
\end{itemize}

\begin{equation*}
\epsilon _{\exists CG}(s)\triangleq (\mathrm{s}+1)L^{2}\exp (-\frac{\left(\epsilon f\beta \eta \right)^{2}s}{64})+\upvarepsilon _{\text{lift}}
\end{equation*}

\begin{itemize}
\item[-] \textbf{Chain quality.} The probability that \textbf{Ouroboros-AutoSyn} violates the chain quality property with parameter $\mathrm{k}=96/(\upepsilon \beta f\eta )$\textit{ and }$\mu =\epsilon \beta f\eta /16$\textit{ is no more than}
\end{itemize}

\begin{equation*}
\epsilon _{CQ}(\mu ,k)\triangleq \frac{kL^{2}}{2}\exp (-\frac{\left(\epsilon f\beta \eta \right)^{2}k}{256})+\upvarepsilon _{\text{lift}}
\end{equation*}
Where $\upvarepsilon _{\text{lift}}$ is a shorthand for the quantity
\begin{equation*}
\upvarepsilon _{\text{lift}}\triangleq QL\cdot \left[R^{3}\cdot \exp \left(-\frac{\left(\epsilon f\beta \eta \right)^{2}R}{768}\right)+\frac{38R}{\varepsilon ^{4}}\cdot \exp \left(2-\frac{\varepsilon ^{4}f\beta \eta R}{864}\right) \right]
\end{equation*}
\section{ Adopting the maxvalid-bg Rule}

\textbf{Theorem 2.} \textit{Consider the protocol Ouroboros-AutoSyn using maxvalid-bg as described in Section 3, executed in the setting with static  }$F_{N-MC}$\textit{ registration, under the same assumptions as in Theorem 1.
        If the maxvalid-bg parameters, k and s, satisfy }

{\centering $k> 384/\upepsilon \beta \eta $     and     $R/6\geq s=k/(4f)\geq 96/\epsilon \beta f\eta $

\par}\textit{then the guarantees given in Theorem 1 for common prefix, chain growth, chain quality, and existential chain quality are still valid except for an additional error probability}
\begin{equation*}
\exp \left(\ln L-\Omega \left(k\right)\right)+\epsilon _{CG}\left(\beta f\eta /16,k/\left(4f\right)\right)+\epsilon _{\exists CG}(k/(4f))+\epsilon _{CP}(k\beta \eta /64)
\end{equation*}
Proof is similar to {[}15{]}.

\section{ Newly Joining Parties}

\textbf{\textit{Lemma 2.}}\textit{ Consider the same setting as Lemma 1 and let }$t_{\textit{round}}$ \textit{be the round length (contains the network delay), }$\eta $\textit{ be the message deliver ratio.
        Consider an honest party }$U_{p}$\textit{ in slot }$sl$\textit{, which newly joined the protocol execution (and hence being registered to the network) at some slots }$sl_{join}\leq sl$\textit{.
        If party }$U_{p}$ \textit{is considered synchronized in slot }$sl$\textit{, according to Definition 6 in {[}15{]} with parameter }$t_{sync}\geq 2\cdot t_{\textit{round}}/\eta $\textit{, then it has also received its synchronizing chain.\newline
}

\textit{Proof.
        This follows from lemma 2 in {[}15{]}, with average delay value as }$t_{\textit{round}}/\eta $\textit{ (based on real time).}

\textbf{\textit{Remark 2(Self-synchronization).  }}\textit{In procedure JoinProc, }$t_{join}(3\cdot t_{{\textit{round}_{1}}})$\textit{ is the waiting time before the party being operational.
        Before }$t_{join}$\textit{ ends the newly joining party will receive its synchronizing chain except with error probability }$\epsilon _{CG2}$\textit{ of the event that } $\overline{U}$\textit{ does not adopt a new chain during a period of }$t_{join}$\textit{.
        And we have the relation }$t_{join}=3\cdot t_{{\textit{round}_{1}}}$\textit{ and }$\eta \geq 2/3$\textit{ so we got }$t_{join}\geq t_{sync}$\textit{.}

\textit{Rejoining party is considered adversely even it does not behave in adversarial way.}

\textbf{\textit{Corollary 1. }}\textit{Consider the protocol }\textbf{Ouroboros-AutoSyn} \textit{as described in Section 3, executed in an environment with dynamic }$F_{N-MC}$\textit{-registrations and deregistrations.
        Then, under the assumptions of Theorem 2, the guarantees it gives for common prefix, chain growth, and chain quality are valid also in this general setting.}

\section{ Composable Guarantees}

Message maximum delay measured in time is $2\cdot t_{\textit{round}}/\eta $ on average.
        And same as 2-rounds listening time settings, parameter $\text{Delay}$ is set as 2 rounds. 

\textbf{Theorem 3.}\textit{ Let }k \textit{be the common-prefix parameter and let }R \textit{be the epoch-length parameter (restricted as in Theorem 2), let }$t_{\textit{round}}$ \textit{be the round length (contains the network delay), }$\eta $\textit{ be the message successful deliver ratio, let }$\tau _{CG}$ \textit{and }$\mu $\textit{ be the speed and chain-quality coefficients, respectively (both defined as in Theorem 1), and let }$\upalpha $\textit{ and }$\upbeta $ \textit{refer to the respective bounds on the participation ratios (as in Theorem 1).
        Let }$G_{\textit{Ledger}}$ \textit{be the ledger functionality defined in Section A.4 and instantiate its parameters by:}
\begin{equation*}
\text{windowSize}=k \mathrm{and} \text{ Delay}=2\text{ (rounds)}
\end{equation*}
\begin{equation*}
\textit{maxTime}_{\textit{window}}\geq \frac{\textit{windowSize}}{\tau _{CG}}\text{ and }\textit{advBlcks}_{\textit{window}}\geq (1-\mu )\textit{windowSize}
\end{equation*}
The protocol Ouroboros-AutoSyn (with access to its specified hybrids) securely UC-realizes $G_{\textit{LEDGER}}$ under the assumptions required by Theorem 1 (which are formally enforceable by a real-world wrapper functionality as given in Section D).
        In addition, the corresponding simulation is perfect except with negligible probability in the parameter k when setting $\mathrm{R}\geq \mathrm{w}(\log k)$.

Proof.
        The difference from theorem 3 in {[}15{]} is that the ledger functionality no longer reads time from clock functionality, but from the simulator.
        Simulator maintains a central clock functionality by reading time from the simulated parties.
        As long as theorem 5 holds, the simulated alert parties will stay synchronized (local round number advances at the same pace).

New parties will get synchronized within $t_{join}$ after they joined the execution, for $t_{join}\geq 2\cdot t_{\textit{round}}/\eta $.
        Proof of other parts can be found in {[}15{]}.

\textbf{Reference}
\begin{enumerate}[{[1]}]
\item Christoph Lenzen, Thomas Locher, and Roger Wattenhofer.
        Clock synchronization with bounded global and local skew.
        In 49th FOCS, pages 509\textendash{}518.
        IEEE Computer Society Press, October 2008.
\item Barbara B.
        Simons, Jennifer Lundelius Welch, and Nancy A.
        Lynch.
        An overview of clock synchronization.
        In Barbara B.
        Simons and Alfred Z.
        Spector, editors, Fault-Tolerant Distributed Computing, volume 448 of LNCS.
        Springer, Heidelberg, 1990.
\item Andrew Miller, Yu Xia, Kyle Croman, Elaine Shi, and Dawn Song.
        The honey badger of BFT protocols.
        In Edgar R.
        Weippl, Stefan Katzenbeisser, Christopher Kruegel, Andrew C.
        Myers, and Shai Halevi, editors, Proceedings of the 2016 ACM SIGSAC Conference on Computer and Communications Security, Vienna, Austria, October 24-28, 2016, pages 31\textendash{}42.
        ACM, 2016
\item Miguel Castro and Barbara Liskov.
        Practical byzantine fault tolerance.
        In Proceedings of the Third USENIX Symposium on Operating Systems Design and Implementation (OSDI), New Orleans, Louisiana, USA, February 22-25, 1999, pages 173\textendash{}186, 1999.
\item Christian Cachin, Klaus Kursawe, Frank Petzold, and Victor Shoup.
        Secure and efficient asynchronous broadcast protocols.
        In Joe Kilian, editor, Advances in Cryptology - CRYPTO 2001, 21st Annual International Cryptology Conference, Santa Barbara, California, USA, August 19-23, 2001, Proceedings, volume 2139 of Lecture Notes in Computer Science, pages 524\textendash{}541.
        Springer, 2001.
\item Sandro Coretti, Juan A.
        Garay, Martin Hirt, and Vassilis Zikas.
        Constant-round asynchronous multi-party computation based on one-way functions.
        In Jung Hee Cheon and Tsuyoshi Takagi, editors, Advances in Cryptology - ASIACRYPT 2016 - 22nd International Conference on the Theory and Application of Cryptology and Information Security, Hanoi, Vietnam, December 4-8, 2016, Proceedings, Part II, volume 10032 of Lecture Notes in Computer Science, pages 998\textendash{}1021, 2016.
\item Satoshi Nakamoto.
        Bitcoin: A peer-to-peer electronic cash system, 2008. \url{http://bitcoin.org/bitcoin.pdf}.
\item Vitalik Buterin.
        A next-generation smart contract and decentralized application platform, 2013. https:// github.com/ethereum/wiki/wiki/White-Paper.
\item Sunoo Park, Krzysztof Pietrzak, Albert Kwon, Jo¨el Alwen, Georg Fuchsbauer, and Peter Gazi.
        Spacemint: A cryptocurrency based on proofs of space.
        IACR Cryptology ePrint Archive, 2015:528, 2015.
\item Tal Moran and Ilan Orlov.
        Proofs of space-time and rational proofs of storage.
        IACR Cryptology ePrint Archive, 2016:35, 2016.
\item Castro M, Liskov B.
        Practical byzantine fault tolerance and proactive recovery{[}J{]}.
        ACM Transactions on Computer Systems, 2002, 20(4): 398-461.
\item \textcolor{color-404040}{Garay J A, Kiayias A, Leonardos N, et al.
        The Bitcoin Backbone Protocol: Analysis and Applications{[}C{]}. theory and application of cryptographic techniques, 2015: 281-310.}
\item \textcolor{color-404040}{Kiayias A, Russell A, David B, et al.
        Ouroboros: A Provably Secure Proof-of-Stake Blockchain Protocol{[}C{]}. international cryptology conference, 2017: 357-388.}
\item \textcolor{color-404040}{David B, Ga{\v{z}}i P, Kiayias A, et al.
        Ouroboros Praos: An Adaptively-Secure, Semi-synchronous Proof-of-Stake Blockchain{[}C{]}. theory and application of cryptographic techniques, 2018: 66-98.}
\item \textcolor{color-404040}{Badertscher C, Gazi P, Kiayias A, et al.
        Ouroboros Genesis: Composable Proof-of-Stake Blockchains with Dynamic Availability{[}C{]}. computer and communications security, 2018: 913-930.}
\item \textcolor{color-404040}{Badertscher C, Gazi P, Kiayias A, et al.
        Ouroboros Chronos: Permissionless Clock Synchronization via Proof-of-Stake.{[}J{]}.
        IACR Cryptology ePrint Archive, 2019.}
\item \textcolor{color-404040}{Kerber T, Kiayias A, Kohlweiss M, et al.
        Ouroboros Crypsinous: Privacy-Preserving Proof-of-Stake{[}C{]}. ieee symposium on security and privacy, 2019: 157-174.}
\item \textcolor{color-404040}{Sasson E B, Chiesa A, Garman C, et al.
        Zerocash: Decentralized Anonymous Payments from Bitcoin{[}C{]}. ieee symposium on security and privacy, 2014: 459-474.}
\item Ran Canetti.
        Universally composable security: A new paradigm for cryptographic protocols.
        In 42nd FOCS, pages 136\textendash{}145.
        IEEE Computer Society Press, October 2001.
\item \textcolor{color-404040}{Canetti R, Hogan K, Malhotra A, et al.
        A Universally Composable Treatment of Network Time{[}J{]}. ieee computer security foundations symposium, 2017: 360-375.}
\item King, S., Nadal, S.: Ppcoin: Peer-to-peer crypto-currency with proof-of-stake (2012). \href{https://peercoin.net/assets/paper/peercoin-paper.pdf.\%20Accessed\%20June\%202017}{https://peercoin.net/assets/paper/peercoin-paper.pdf.
        Accessed June 2017}
\item \textcolor{color-404040}{Bentov I, Pass R, Shi E, et al.
        Snow White: Provably Secure Proofs of Stake.{[}J{]}.
        IACR Cryptology ePrint Archive, 2016.}
\item Gilad, Y., Hemo, R., Micali, S., Vlachos, G., Zeldovich, N.: Algorand: scaling byzantine agreements for cryptocurrencies. https://people.csail.mit.edu/nickolai/ papers/gilad-algorand-eprint.pdf
\end{enumerate}

\begin{appendices}
\chapter{ Completing the Setup Functionality Description}

\section{ The Communication Network}

\textbf{Function }$\boldsymbol{F}_{\boldsymbol{N}-\boldsymbol{MC}}$ 

The functionality is parameterized with a set possible senders and receivers P.
        Any newly registered (resp. deregistered) party is added to (resp. deleted from) P.

We use $F_{N-MC}^{\mathrm{tx}}$ to denote the network for sending transactions, $F_{N-MC}^{\mathrm{adj}}$ to denote the network for sending adjustment messages and $F_{N-MC}^{\mathrm{bc}}$ to denote the network for sending blockchains.
        Since 3 kinds of networks have similar functions, we use $F_{N-MC}$ for simplicity.
        Also, we need to model delivery ratio so we introduce a new notion $rd$.
        If $rd\coloneqq 0$ or $1$, it means message delay has not been set by the adversary.
        After adversary set the delay time, the function will change $rd$ according to $2or3$.

\begin{table}
\begin{tabularx}{\textwidth}{|p{\dimexpr 0.361\linewidth-2\tabcolsep-2\arrayrulewidth}|p{\dimexpr 0.222\linewidth-2\tabcolsep-\arrayrulewidth}|p{\dimexpr 0.417\linewidth-2\tabcolsep-\arrayrulewidth}|} \hline 
\centering\arraybackslash{} & \centering\arraybackslash{}\textbf{Set} & \centering\arraybackslash{}\textbf{Probability}\\\hline 
\centering\arraybackslash{}$rd\coloneqq 0$ & \centering\arraybackslash{}No & \centering\arraybackslash{}$\eta $\\\hline 
\centering\arraybackslash{}$rd\coloneqq 1$ & \centering\arraybackslash{}No & \centering\arraybackslash{}$1-\eta $\\\hline 
\centering\arraybackslash{}$rd\coloneqq 2$ & \centering\arraybackslash{}Yes & \centering\arraybackslash{}$\eta $\\\hline 
\centering\arraybackslash{}$rd\coloneqq 3$ & \centering\arraybackslash{}Yes & \centering\arraybackslash{}$1-\eta $\\\hline 
\end{tabularx}
\end{table}
\textbf{Honest sender multicast.} Upon receiving (multicast, sid, m, $\textcolor{color-2E74B5}{t}_{\textcolor{color-2E74B5}{n}\textcolor{color-2E74B5}{e}\textcolor{color-2E74B5}{x}\textcolor{color-2E74B5}{t}}$) from some $U_{p}$ ${\in}$P, where P =\{$U_{1}$,..., $U_{n}$\} denotes the current party set, send (clock-read, $sid_{C}$) to $G_{\textit{AutoClock}}$ get current time $t_{now}$, choose n new unique message-IDs $mid_{1}$,..., $mid_{n}$, initialize 2n new delay variables $D_{{mid_{1}}}$ := ... := $D_{{mid_{n}}}\coloneqq t_{now}$, $D_{mid_{1}}^{MAX}\colon =${\ldots} $\coloneqq D_{mid_{n}}^{MAX}\coloneqq \textcolor{color-2E74B5}{t}_{\textcolor{color-2E74B5}{n}\textcolor{color-2E74B5}{e}\textcolor{color-2E74B5}{x}\textcolor{color-2E74B5}{t}}$, and \textcolor{color-2E74B5}{n new state valuables} $\textcolor{color-2E74B5}{r}\textcolor{color-2E74B5}{d}_{\textcolor{color-2E74B5}{1}}\textcolor{color-2E74B5}{,}\textcolor{color-2E74B5}{\ldots }\textcolor{color-2E74B5}{,}\textcolor{color-2E74B5}{r}\textcolor{color-2E74B5}{d}_{\textcolor{color-2E74B5}{n}}$\textcolor{color-2E74B5}{.
        For each }$\textcolor{color-2E74B5}{r}\textcolor{color-2E74B5}{d}_{\textcolor{color-2E74B5}{i}}$\textcolor{color-2E74B5}{ in \{}$\textcolor{color-2E74B5}{r}\textcolor{color-2E74B5}{d}_{\textcolor{color-2E74B5}{1}}\textcolor{color-2E74B5}{,}\textcolor{color-2E74B5}{\ldots }\textcolor{color-2E74B5}{,}\textcolor{color-2E74B5}{r}\textcolor{color-2E74B5}{d}_{\textcolor{color-2E74B5}{n}}$\textcolor{color-2E74B5}{\} do a random draw }$\textcolor{color-2E74B5}{j}$\textcolor{color-2E74B5}{ in \{0, 1\} with probability }$\textcolor{color-2E74B5}{\eta }$\textcolor{color-2E74B5}{ that }$\textcolor{color-2E74B5}{j}\textcolor{color-2E74B5}{=}\textcolor{color-2E74B5}{0}$\textcolor{color-2E74B5}{, and probability }$\textcolor{color-2E74B5}{(}\textcolor{color-2E74B5}{1}\textcolor{color-2E74B5}{-}\textcolor{color-2E74B5}{\eta }\textcolor{color-2E74B5}{)}$\textcolor{color-2E74B5}{ that }$\textcolor{color-2E74B5}{j}\textcolor{color-2E74B5}{=}\textcolor{color-2E74B5}{1}$\textcolor{color-2E74B5}{ and set }$\textcolor{color-2E74B5}{r}\textcolor{color-2E74B5}{d}_{\textcolor{color-2E74B5}{i}}\textcolor{color-2E74B5}{=}\textcolor{color-2E74B5}{j}$ .Set $\vec{M}$:= $\vec{M}${\textbar}{\textbar}(m, $mid_{1}$, $D_{{mid_{1}}}$, $U_{1}$, $\textcolor{color-2E74B5}{r}\textcolor{color-2E74B5}{d}_{\textcolor{color-2E74B5}{1}}$){\textbar}{\textbar}...{\textbar}{\textbar}(m, $mid_{n}$, $D_{{mid_{n}}}$, $U_{n}$, $\textcolor{color-2E74B5}{r}\textcolor{color-2E74B5}{d}_{\textcolor{color-2E74B5}{n}}$),and send ( multicast, sid, m,  $U_{P}$, $\textcolor{color-2E74B5}{t}_{\textcolor{color-2E74B5}{n}\textcolor{color-2E74B5}{e}\textcolor{color-2E74B5}{x}\textcolor{color-2E74B5}{t}}$, ( $U_{1}$, $mid_{1},rd_{1}$), ...,( $U_{n}$, $mid_{n},rd_{n}$)) to the adversary. 

\textbf{Adversarial sender (partial) multicast}.
        Upon receiving (multicast, sid, ( $m_{{i_{1}}}$, $U_{{i_{1}}}$,  $\textcolor{color-4472C4}{r}\textcolor{color-4472C4}{d}_{{\textcolor{color-4472C4}{i}_{\textcolor{color-4472C4}{1}}}}$\textcolor{color-4472C4}{,} $\textcolor{color-2E74B5}{t}_{{\textcolor{color-2E74B5}{n}\textcolor{color-2E74B5}{e}\textcolor{color-2E74B5}{x}\textcolor{color-2E74B5}{t}_{{\textcolor{color-4472C4}{i}_{\textcolor{color-4472C4}{1}}}}}}$), ...,( $m_{{i_{l}}}$, $U_{{i_{l}}}$, $\textcolor{color-4472C4}{r}\textcolor{color-4472C4}{d}_{{\textcolor{color-4472C4}{i}_{\textcolor{color-4472C4}{l}}}}$\textcolor{color-4472C4}{,} $\textcolor{color-2E74B5}{t}_{{\textcolor{color-2E74B5}{n}\textcolor{color-2E74B5}{e}\textcolor{color-2E74B5}{x}\textcolor{color-2E74B5}{t}_{{\textcolor{color-4472C4}{i}_{\textcolor{color-4472C4}{l}}}}}}$) from the adversary with \{$U_{{i_{1}}}$,..., $U_{{i_{l}}}$\}${\subseteq}$P, choose $l$ new unique message-IDs $mid_{{i_{1}}}$,..., $mid_{{i_{l}}}$, initialize $2l$ new variables $D_{{mid_{{i_{1}}}}}\coloneqq ${\ldots} $\coloneqq D_{{mid_{{i_{l}}}}}\textcolor{color-2E74B5}{\coloneqq }$ $t_{now}$, and $D_{mid_{{i_{1}}}}^{MAX}$:= $\textcolor{color-2E74B5}{t}_{{\textcolor{color-2E74B5}{n}\textcolor{color-2E74B5}{e}\textcolor{color-2E74B5}{x}\textcolor{color-2E74B5}{t}_{{\textcolor{color-4472C4}{i}_{\textcolor{color-4472C4}{1}}}}}}$\textcolor{color-2E74B5}{,{\ldots},} $D_{mid_{{i_{l}}}}^{MAX}$ :\textbf{\textcolor{color-4472C4}{= }}$\textcolor{color-2E74B5}{t}_{{\textcolor{color-2E74B5}{n}\textcolor{color-2E74B5}{e}\textcolor{color-2E74B5}{x}\textcolor{color-2E74B5}{t}_{{\textcolor{color-4472C4}{i}_{\textcolor{color-4472C4}{l}}}}}}$)\textcolor{color-4472C4}{.} Then set $\vec{M}$:= $\vec{M}${\textbar}{\textbar}($m_{{i_{1}}}$, $mid_{{i_{1}}}$, $D_{{mid_{{i_{1}}}}}$, $U_{{i_{1}}}$,  $\textcolor{color-4472C4}{r}\textcolor{color-4472C4}{d}_{{\textcolor{color-4472C4}{i}_{\textcolor{color-4472C4}{1}}}}$) {\textbar}{\textbar}...{\textbar}{\textbar} ( $m_{{i_{l}}}$, $mid_{{i_{l}}}$, $D_{{mid_{{i_{l}}}}}$, $U_{{i_{l}}}, \textcolor{color-4472C4}{r}\textcolor{color-4472C4}{d}_{{\textcolor{color-4472C4}{i}_{\textcolor{color-4472C4}{l}}}}$), and send ( multicast, sid, ( $m_{{i_{1}}}$,$U_{{i_{1}}}, mid_{{i_{1}}}, \textcolor{color-4472C4}{r}\textcolor{color-4472C4}{d}_{{\textcolor{color-4472C4}{i}_{\textcolor{color-4472C4}{1}}}}$) ,..., ( $m_{{i_{l}}}$,$U_{{i_{l}}}, mid_{{i_{l}}}, \textcolor{color-4472C4}{r}\textcolor{color-4472C4}{d}_{{\textcolor{color-4472C4}{i}_{\textcolor{color-4472C4}{l}}}}$)) to the adversary.

\textbf{\textcolor{color-2E74B5}{Adversarial message redelivery.}} Upon receiving (mix, sid, mid, $mid'$) from the adversary, if corresponding $rd,rd'\in \{0,1\}$, also mid and $mid'$ are message-IDs registered in the current $\vec{M}$, then swap the tuple of corresponding $rd$ of each mid in $\vec{M}$ as $\left(m,mid,D_{mid},U,rd'\right)$ and $(m',mid',D_{mid'},,U',rd)$, also return(swap, sid) to the adversary.
        Otherwise, ignore this message.

\textbf{Adding adversarial delays}.
        Upon receiving (delays, sid,( $T_{{mid_{{i_{1}}}}}$, $mid_{{i_{1}}}$),...,( $T_{{mid_{{i_{l}}}}}$ , $mid_{{i_{l}}}$)) from the adversary do the following for each pair ($T_{{mid_{{i_{j}}}}}$, $mid_{{i_{j}}}$):  

If $rd_{{i_{j}}}=1$ and mid is a message-ID registered in the current $\vec{M}$, then set $D_{{mid_{{i_{j}}}}}$ := $D_{{mid_{{i_{j}}}}}$+$T_{{mid_{{i_{j}}}}}$ and $rd_{{i_{j}}}\coloneqq 3$.

If $rd_{{i_{j}}}=0$ and $D_{{mid_{{i_{j}}}}}$+$T_{{mid_{{i_{l}}}}}$ ${\leq}$$D_{mid_{{i_{j}}}}^{MAX}$ and mid is a message-ID registered in the current $\vec{M}$, then set $D_{{mid_{{i_{j}}}}}$ := $D_{{mid_{{i_{j}}}}}$+$T_{{mid_{{i_{j}}}}}$ , $rd_{{i_{j}}}=2$.
        Otherwise, ignore this pair.

(Whenever honest player fetching from the network, read from the $t_{now}$ the clock and adjust the count-down time and send those count-down is 0) 

\textbf{Honest party fetching}.
        Upon receiving (fetch, sid) from $U_{p}$ ${\in}$P (or from A on behalf of $U_{p}$  if $U_{p}$  is corrupted): Send (clock-read, $sid_{C}$) to $G_{\textit{AutoClock}}$ get current time $t_{now}$.

Let $\vec{M}_{0}^{U_{p}}$ denote the subvector $\vec{M}$(initially $\vec{M}_{0}^{U_{p}}=\bot $ ), including all tuples of the form $(m, mid,D_{mid}, U_{p}, rd)$ with $D_{mid}\leq t_{now}$ and $\textcolor{color-2E74B5}{r}\textcolor{color-2E74B5}{d}\textcolor{color-2E74B5}{=}\textcolor{color-2E74B5}{2}$\textcolor{color-2E74B5}{ (or }$\textcolor{color-2E74B5}{r}\textcolor{color-2E74B5}{d}\textcolor{color-2E74B5}{=}\textcolor{color-2E74B5}{3}$\textcolor{color-2E74B5}{ )} (in the same order as they appear in $\vec{M}$).
        Delete all entries in $\vec{M}_{0}^{U_{p}}$ from $\vec{M}$, and send $\vec{M}_{0}^{U_{p}}$ to $U_{p}$. 

\textbf{Adversarial reordering messages.} Upon receiving (swap, sid, mid, $mid'$) from the adversary, if mid and $mid'$ are message-IDs registered in the current $\vec{M}$, then swap the tuple (m, mid, $T_{mid},\cdot ,rd$) and( m, $mid'$, $T_{mid'}\cdot ,rd'$) in $\vec{M}$.
        Return(swap, sid) to the adversary.

\section{ Modeling Synchrony}

The basic functionality is to capture clock-functionality.
        Our model uses time-based synchrony.

\textbf{Functionality }$\boldsymbol{G}_{\textit{\textbf{AutoClock}}}$:

The global functionality $\boldsymbol{G}_{\textit{\textbf{AutoClock}}}$ provides a universal reference clock.
        When required, it provides an abstract notion of time similar to the reference clock in {[}20{]}.
        It is monotonic and participants cannot forge the time.
        We denote current time as $\mathrm{NOW}$. 

$\boldsymbol{G}_{\textit{\textbf{AutoClock}}}$ maintains integer $\mathrm{NOW}$ and $\text{NEXT}$.
        It also manages the set P of registered identities, i.e., parties $U_{p}=(\mathrm{pid},\mathrm{sid})$, the set $F$ of functionalities (together with their session identifier).
        Initially, $P\colon =\mathrm{\varnothing }$ , $F\colon =\mathrm{\varnothing }$, $\mathrm{NOW}$= $-t_{\textit{start}}$, $\text{NEXT}=0$ which stands for genesis round.

For each session $\mathrm{sid}$ the clock maintains a variable $\uptau _{sid}$.
        For each identity $U_{p}\colon =(\mathrm{pid},\mathrm{sid})\in \mathrm{P}$ it manages variable $\mathrm{d}_{{U_{p}}}$ \textcolor{color-2E74B5}{and }$\textcolor{color-2E74B5}{\mathrm{t}}_{{\text{next}_{{\textcolor{color-2E74B5}{U}_{\textcolor{color-2E74B5}{p}}}}}}$.
        For each pair $(\mathrm{F},\mathrm{sid})\in \mathrm{F}$ it manages variable $\mathrm{d}_{(\mathrm{F},\mathrm{sid})}$ (all integer variables are initially 0).
\begin{itemize}
\item[-] \textcolor{color-2E74B5}{IncrementTime: Maintains an integer }$\mathrm{NOW}$\textcolor{color-2E74B5}{, while }$\textcolor{color-2E74B5}{\mathrm{NOW}}\textcolor{color-2E74B5}{< }\textcolor{color-2E74B5}{\text{NEXT}}$\textcolor{color-2E74B5}{ keep update }$\mathrm{NOW}$\textcolor{color-2E74B5}{ ${\leftarrow}$ }$\mathrm{NOW}$\textcolor{color-2E74B5}{ + 1 with a steady pace.}
\item[-] Upon receiving (clock-update, $\mathrm{sid}_{C}$, $\mathrm{t}_{next}$) from some party $U_{p}\in \mathrm{P}$ set $\mathrm{t}_{{\text{next}_{{U_{p}}}}}=\mathrm{t}_{next}$(if $\mathrm{t}_{next}=\bot $, set $\mathrm{t}_{{\text{next}_{{U_{p}}}}}=\bot $) set $\mathrm{d}_{{U_{p}}}\colon =1$; execute Round-Update and forward (clock-update, $\mathrm{sid}_{C}$, $\mathrm{t}_{next}$), $U_{p}$) to A 
\item[-] Upon receiving (clock-update, $\mathrm{sid}_{C}$) from some functionality $\mathrm{F}$ in a session $\mathrm{sid}$ such that $(\mathrm{F},\mathrm{sid})\in \mathrm{F}$ set $\mathrm{d}_{(\mathrm{F},\mathrm{sid})}\colon =1$, execute Round-Update and return (clock-update, $\mathrm{sid}_{C}$,F) to this instance of F 
\item[-] Upon receiving (clock-read, $sid_{C}$) from any participant (including the environment on behalf of a party, the adversary, or any ideal\textemdash{}shared or local\textemdash{}functionality) return (clock-read, $sid$,  $\mathrm{NOW}$) to the requestor (where  $sid$ is the $sid$ of the calling instance).
\end{itemize}
Procedure Round-Update: For each session $\mathrm{sid}$ do: If $\mathrm{d}_{(\mathrm{F},\mathrm{sid})}\colon =1$ for all $F\in \mathrm{F}$ and $\mathrm{d}_{{U_{p}}}=1$ for all honest parties $U_{p}=(\cdot ,\mathrm{sid})\in \mathrm{P}$. \textcolor{color-2E74B5}{Select a value }$\textcolor{color-2E74B5}{\mathrm{t}}_{{\text{next}_{{\textcolor{color-2E74B5}{U}_{\textcolor{color-2E74B5}{p}}}}}}$\textcolor{color-2E74B5}{ among }$\textcolor{color-2E74B5}{U}_{\textcolor{color-2E74B5}{p}}\textcolor{color-2E74B5}{\in }\textcolor{color-2E74B5}{\mathrm{P}}$\textcolor{color-2E74B5}{ that most parties have the same value.
        Then set }$\text{NEXT}$ $\textcolor{color-2E74B5}{\colon }\textcolor{color-2E74B5}{=}\textcolor{color-2E74B5}{\mathrm{t}}_{\textcolor{color-2E74B5}{r}\textcolor{color-2E74B5}{u}\textcolor{color-2E74B5}{n}}\textcolor{color-2E74B5}{+}\textcolor{color-2E74B5}{\mathrm{t}}_{{\text{next}_{{\textcolor{color-2E74B5}{U}_{\textcolor{color-2E74B5}{p}}}}}} $  and reset $\mathrm{d}_{(\mathrm{F},\mathrm{sid})}\colon =0$  and $\mathrm{d}_{{U_{p}}}\colon =0$  for all parties $U_{p}=(\cdot ,\mathrm{sid})\in \mathrm{P}$.

\section{ The Genesis Block Distribution}

$\boldsymbol{F}_{\boldsymbol{INIT}}$ 

The functionality $F_{INIT}$ is parameterized by the set $U_{1}$\textit{, . . . , }$U_{n}$ of initial stakeholders \textit{n }and their respective stakes $s_{1}$\textit{, . . . , }$s_{N}$.\textbf{\textcolor{color-2E74B5}{ It also stores }}$\textcolor{color-2E74B5}{\boldsymbol{t}}_{{\textcolor{color-2E74B5}{\textit{\textbf{round}}}_{\textcolor{color-2E74B5}{\mathbf{1}}}}}$\textbf{\textcolor{color-2E74B5}{ for the first round to start with.}} It maintains the set of registered parties \textit{P.}
\begin{itemize}
\item[-] Upon receiving any message from a party, the functionality first sends (\textit{clock-read}, $sid_{C}$) to the clock to receive \textbf{\textcolor{color-2E74B5}{the current time}}.
        Subsequently:
\begin{itemize}
\item[${\CIRCLE}$] \textbf{\textcolor{color-2E74B5}{If the first (genesis) round has not begun}} and the message is request from some initial stakeholder $U_{i}$ of the form (ver\_keys, sid, $U_{i}$,$\mathrm{v}_{\mathrm{i}}^{\mathrm{vrf}}$, $\mathrm{v}_{\mathrm{i}}^{\mathrm{kes}}$) then $F_{INIT}$ stores the verification keys tuple ($U_{i}$\textit{, }$\mathrm{v}_{\mathrm{i}}^{\mathrm{vrf}}$\textit{,} $\mathrm{v}_{\mathrm{i}}^{\mathrm{kes}}$) and acknowledges its receipt.
        If some of the registered public keys are equal, it outputs an error and halts.
        Otherwise, it samples and stores a random value.  $\eta _{1}\xleftarrow{\$ }\{0,1\}^{\uplambda }$and constructs a genesis block ($S_{1}$\textit{, }$\eta _{1},\textcolor{color-2E74B5}{\boldsymbol{t}}_{\textcolor{color-2E74B5}{\textit{\textbf{start}}}}\textcolor{color-2E74B5}{,} \textcolor{color-2E74B5}{\boldsymbol{t}}_{{\textcolor{color-2E74B5}{\textit{\textbf{round}}}_{\textcolor{color-2E74B5}{\mathbf{1}}}}}$), where 

$S_{1}=\left(U_{1},\mathrm{v}_{1}^{\mathrm{vrf}},\mathrm{v}_{1}^{\mathrm{kes}},s_{1}\right),..., \left(U_{n},\mathrm{v}_{\mathrm{n}}^{\mathrm{vrf}},\mathrm{v}_{\mathrm{n}}^{\mathrm{kes}},s_{n}\right).$ 
\item[${\CIRCLE}$] \textbf{\textcolor{color-2E74B5}{If the first (genesis) round has already begun}} then do the following
\begin{description}
\item[*]If any of the \textit{n }initial stakeholders has not sent a request of the above form, i.e., a (ver\_keys, sid, $U_{i}$,$\mathrm{v}_{\mathrm{i}}^{\mathrm{vrf}}$, $\mathrm{v}_{\mathrm{i}}^{\mathrm{kes}}$)-message, to $F_{INIT}$ in the genesis round then $F_{INIT}$ outputs an error and halts
\item[*]Otherwise, if the currently received input is a request of the form (\textit{genblock\_req}, sid,$U_{i}$) from any (initial or not) stakeholder U, $F_{INIT}$ sends (\textit{genblock}, sid, ($S_{1}$\textit{, }$\eta _{1},\textcolor{color-2E74B5}{\boldsymbol{t}}_{\textcolor{color-2E74B5}{\textit{\textbf{start}}}}\textcolor{color-2E74B5}{,} \textcolor{color-2E74B5}{\boldsymbol{t}}_{{\textcolor{color-2E74B5}{\textit{\textbf{round}}}_{\textcolor{color-2E74B5}{\mathbf{1}}}}}$) to the requester.
\end{description}

\end{itemize}

\end{itemize}
\section{ The Ouroboros AutoSyn Ledger}

The ledger functionality is similar to the one in {[}15{]}, except for a few details. 

\subsection{Functionality $\boldsymbol{G}_{\textit{\textbf{LEDGER}}}$}

\textbf{General:} The functionality is parameterized by four algorithms, Validate, ExtendPolicy, Blockify, and predict-time, along with three parameters: windowSize, Delay ${\in}$ N, and $\mathrm{S}_{\text{initStake}}$ := \{($\mathrm{U}_{1}$, $\mathrm{s}_{1}$), . . . , ($\mathrm{U}_{\mathrm{n}}$, $\mathrm{s}_{\mathrm{n}}$)\}.
        The functionality manages variables state, NxtBC, buffer, $\uptau _{L}$, and $\vec{\uptau }_{\textit{state}}$, as described above.
        The variables are initialized as follows: state := $\vec{\uptau }_{\textit{state}}$ := NxtBC := ${\upvarepsilon}$, buffer := ${\varnothing}$, $\uptau _{L}$ = 0.
        For each party $\mathrm{U}_{\mathrm{P}}$ ${\in}$ P the functionality maintains a pointer $\mathrm{pt}_{i}$ (initially set to 1) and a current-state view $\text{state}_{p}$ := ${\upvarepsilon}$ (initially set to empty).
        The functionality also keeps track of the timed honest-input sequence in a vector $\vec{\mathrm{I}}_{\mathrm{H}}^{\mathrm{T}}$ (initially $\vec{\mathrm{I}}_{\mathrm{H}}^{\mathrm{T}}$ := ${\upvarepsilon}$).

\textbf{Party Management}: The functionality maintains the set of registered parties P, the (sub-)set of honest parties H ${\subseteq}$ P, and the (sub-set) of de-synchronized honest parties $P_{DS}$ ${\subset}$ H (as discussed below).
        The sets P, H, $P_{DS}$ are all initially set to ${\varnothing}$.
        When a (currently unregistered) honest party is registered at the ledger, if it is registered with the clock and the global \textit{RO} already, then it is added to the party sets H and P and the current time of registration is also recorded; if the current time is $\uptau _{L}$ {\textgreater} 0, it is also added to $P_{DS}$.
        Similarly, when a party is deregistered, it is removed from both P (and therefore also from $P_{DS}$ or H).
        The ledger maintains the invariant that it is registered (as a functionality) to the clock whenever H$\neq $ ${\varnothing}$.

\textbf{Handling initial stakeholders}: If during round ${\uptau}$ = 0 ($\textcolor{color-2E74B5}{t}_{\textcolor{color-2E74B5}{n}\textcolor{color-2E74B5}{o}\textcolor{color-2E74B5}{w}}\textcolor{color-2E74B5}{< }\textcolor{color-2E74B5}{0}$), the ledger did not receive a registration from each initial stakeholder, i.e., $U_{P}$ ${\in}$ $S_{\textit{initStake}}$, the functionality outputs an error and halts.

-------------------------------------------------------------------------------

\textbf{Upon receiving any input} $I$ from any party $P$ or the adversary, \textcolor{color-2E74B5}{send (time-read, }$\textcolor{color-2E74B5}{s}\textcolor{color-2E74B5}{i}\textcolor{color-2E74B5}{d}$\textcolor{color-2E74B5}{) to }$\textcolor{color-2E74B5}{\textit{simulator}}$\textcolor{color-2E74B5}{ and upon receiving response (time-read,} $\textcolor{color-2E74B5}{s}\textcolor{color-2E74B5}{i}\textcolor{color-2E74B5}{d}$\textcolor{color-2E74B5}{,}$\textcolor{color-2E74B5}{\uptau }$\textcolor{color-2E74B5}{)} set $\uptau _{L}$ := ${\uptau}$ and do the following if ${\uptau}$ {\textgreater} 0 (otherwise, ignore input): 
\begin{description}
\item[1.]Updating synchronized/desynchronized party set:
\item[(a)]Let $\hat{P}$ ${\subseteq}$ $P_{DS}$ denote the set of desynchronized honest parties that have been registered (continuously) to the ledger, the network, and the $G_{RO}$ since $\uptau \mathrm{'}< \uptau _{L}-\text{Delay}$.
        Set $P_{DS}$ := $P_{DS}\backslash \hat{P}$
\item[(b)]For any synchronized party $U_{P}$ ${\in}$H\textbackslash $P_{DS}$, if $U_{P}$ is not registered to the network, then consider it desynchronized, i.e., set $P_{DS}$ ${\cup}$\{$U_{P}$\}.
\end{description}

\begin{enumerate}
\setcounter{enumi}{1}
\item If  $I$ was received from an honest party $U_{P}$ ${\in}$P:
\end{enumerate}

\begin{enumerate}[{(a)}]
\item Set $\vec{I}_{H}^{T}\coloneqq \vec{I}_{H}^{T}| | (I,U_{P},\uptau _{L})$ 
\item Compute $\vec{N}=\left(\vec{N}_{1},\ldots ,\vec{N}_{t}\right)\coloneqq \textit{ExtendPolicy}(\vec{I}_{H}^{T}, \textit{state},\textit{NxtBC},\textit{buffer},\vec{\uptau }_{\textit{state}})$ and if $\vec{N}\neq \varepsilon $ set $\text{state}\colon =\text{state}| | \text{Blockify}(\vec{N}_{1})| | \ldots | | \text{Blockify}(\vec{N}_{t})$ and $\vec{\uptau }_{\textit{state}}\coloneqq \vec{\uptau }_{\textit{state}}| | \vec{\uptau }_{L}^{t}$, where $\vec{\uptau }_{L}^{t}$ = $\uptau _{L}| | \ldots | | \uptau _{L}$.
\item For each BTX ${\in}$buffer: if Validate (BTX, state, buffer) = 0 then delete BTX from buffer.
        Also reset $\textit{NxtBC}$:=$\varepsilon $.
\item If there exists $U_{j}\in H\backslash P_{DS}$ such that {\textbar}state{\textbar}-$pt_{j}> \textit{windowSize}$ or $pt_{j}< | \textit{state}_{j}| $, then set $pt_{k}\coloneqq | \textit{state}| $ for all $U_{k}\in H\backslash P_{DS}$//
\end{enumerate}
3.
        If the calling party $\mathrm{U}_{P}$ is stalled or time-unaware, then no further actions are taken.
        Otherwise, depending on the above input $I$ and its sender’s ID, $G_{\textit{LEDGER}}$ executes the corresponding code from the following list:
\begin{itemize}
\item[${\CIRCLE}$] \textit{Submitting a transaction: }
\end{itemize}
  If $I$ = (submit ,sid ,tx) and is received from a party $U_{P}\in \mathrm{P}$ or from A (on behalf of a corrupted party $U_{P}$) do the following

  (a) Choose a unique transaction ID txid and set \textbf{BTX} := (tx, txid, $\uptau _{L}$, $U_{P}$) 

(b) If Validate(\textbf{BTX}, state, buffer) = 1, then buffer := buffer${\cup}$\{\textbf{ BTX} \}. 

\textcolor{color-A6A6A6}{//BTX does not exist in buffer and state}

(c) Send (submit, \textbf{BTX} ) to A.
\begin{itemize}
\item[${\CIRCLE}$] \textit{Reading the state: }

If $I$ = (read, sid) is received from a party $U_{P}$ ${\in}$P then set $\textit{state}_{P}\colon =\text{state}| _{\min \{{\mathrm{pt}_{P}},| \text{state}| \}}$  and return (read, sid, $\textit{state}_{P}$) to the requester.
        If the requester is A then send (state, buffer, $\vec{I}_{H}^{T}$) to A. 
\item[${\CIRCLE}$] \textit{Maintaining the ledger state: }

\textcolor{color-2E74B5}{If} $\textcolor{color-2E74B5}{I}$\textcolor{color-2E74B5}{ = (maintain-ledger, sid ,minerID) is received by an honest party }$\textcolor{color-2E74B5}{U}_{\textcolor{color-2E74B5}{P}}$\textcolor{color-2E74B5}{ ${\in}$P and (after updating }$\vec{\textcolor{color-2E74B5}{I}}_{\textcolor{color-2E74B5}{H}}^{\textcolor{color-2E74B5}{T}}$\textcolor{color-2E74B5}{ as above) predict-time(}$\vec{\textcolor{color-2E74B5}{I}}_{\textcolor{color-2E74B5}{H}}^{\textcolor{color-2E74B5}{T}}$\textcolor{color-2E74B5}{) =}$\hat{\textcolor{color-2E74B5}{\uptau }}\textcolor{color-2E74B5}{\leq }\textcolor{color-2E74B5}{\uptau }_{\textcolor{color-2E74B5}{L}}$\textcolor{color-2E74B5}{, send }$\textcolor{color-2E74B5}{I}$\textcolor{color-2E74B5}{ to A.
        Else do nothing(we no longer require to update the global clock function) }
\item[${\CIRCLE}$] The adversary proposing the next block:

If I = (NEXT-BLOCK, $\text{hFlag}$, ( $txid_{1}$,..., $txid_{l}$)) is sent from the adversary, update  $\text{NxtBC }$as follows: 

(a) Set $\text{listOfTxid }$${\leftarrow}$ $\varepsilon $ 

(b) For $i=1,\ldots ,l$ do: if there exists BTX := (x, txid, minerID, $\uptau _{L}$, $\mathrm{U}_{j}$)${\in}$buffer with ID txid = $\text{txid}_{i}$ then set $\text{listOfTxid }\colon =\text{listOfTxid}| | \text{txid}_{i}$. 

(c) Finally, set $\text{NxtBC }\colon =\text{ NxtBC}| | \left(\text{hFlag},\text{listOfTxid}\right)$and output (next-block, ok) to A 
\item[${\CIRCLE}$] The adversary setting state-slackness: 

If I = (set-slack,($\mathrm{U}_{i1}$,$\widehat{Pt}_{i1}$),...,($\mathrm{U}_{il}$,$\widehat{Pt}_{il}$)), with \{$\mathrm{U}_{i1}$,..., $\mathrm{U}_{il}$\}${\subseteq}$ $H\backslash P_{DS}$ is received from the adversary A do the following:

(a) If for all $\mathrm{j}\in [l]$ : {\textbar}state{\textbar}${-}$$\widehat{Pt}_{ij}$${\leq}$$\text{windowSize}$ and $\widehat{Pt}_{ij}$ ${\geq}${\textbar}$\text{state}_{\mathrm{ij}}| $, set $pt_{i\mathrm{j}}$ := $\widehat{Pt}_{ij}$ for every $\mathrm{j}\in [l]$ and return (set-slack, ok) to A. 

(b) Otherwise set $pt_{i1}$ :={\textbar}state{\textbar} for all $\mathrm{j}\in [l]$.
\item[${\CIRCLE}$] The adversary setting the state for desynchronized parties: 

If I = (desync-state,( $\mathrm{U}_{i1}$ , $\mathrm{state}\mathrm{'}_{\mathrm{i}1}$),...,( $\mathrm{U}_{il} $ , $\mathrm{state}\mathrm{'}_{\mathrm{il}}$)), with \{$\mathrm{U}_{i1}$,..., $\mathrm{U}_{il}$\}${\subseteq}$$P_{DS} $ is received from the adversary A, set $\text{state}_{\mathrm{ij}}$:= $\mathrm{state}\mathrm{'}_{\mathrm{ij}} $ for each $\mathrm{j}\in [l]$ and return (desync-state, ok) to A.
\end{itemize}
\chapter{ Completing the AutoSyn Protocol Description}

\section{ Registration Procedure}

Protocol \textbf{Registration-AutoSyn} ($\mathrm{P},\mathrm{sid},\mathrm{Reg},G$)
\begin{enumerate}[{1:}]
\item If $G\in \{G_{RO},\textcolor{color-2E74B5}{G}_{\textcolor{color-2E74B5}{\textit{AutoClock}}}\}$ then send (REGISTER, sid) to $G$, set registration status to registered with $G$, and output the value received by $G$.
\item End if
\item If $G=G_{\textit{LEDGER}}$ then
\item     If the party is not registered with $G_{RO}$ then or already registered with all setups ignore this input
\item     Else

    For each $\mathrm{F}\in \{F_{INIT},F_{VRF},F_{KES}\}$ do
\item Send(REGISTER, sid) to $\mathrm{F}$, set its registration status as registered with $\mathrm{F}$,                        but do not output the received values.

End for
\item Send (REGISTER, sid) to $F_{N-MC}$
\item If this is the first registration invocation for this ITI, then set isInit $\leftarrow $ false
\item Output (REGISTER, sid,P) once completing the registration with all above resources $\mathrm{F}$
\end{enumerate}
         End if 

End if

\section{ The Main Protocol Instance}

\textbf{Global Variable:}
\begin{itemize}
\item[-] Read-only: R, k, f, s, $t_{\textit{start}}$(static): genesis creating time; 
\item[-] Read-write: $t_{\textit{round}}$: length of one slot, readjusted at the end of the epoch; $sl$ current slot number, start from 0; $t_{\textit{begin}}$: the start time of this round; $t_{next}$: the end time of this slot; $t_{now}$: present time, get from $\boldsymbol{G}_{\textit{\textbf{AutoClock}}} $ Buffer has 2 kinds as $\textit{TxBuffer},$  $\textit{AdjBuffer}$,$v_{p}^{vrf}$, $v_{p}^{kes}$, $\tau $,\textbf{ ep}, sl\textbf{, }$C_{loc}, T_{p}^{ep}$, $t_{on}$ $\textcolor{color-2E74B5}{t}_{\textcolor{color-2E74B5}{r}\textcolor{color-2E74B5}{e}\textcolor{color-2E74B5}{c}}$\textcolor{color-2E74B5}{ : current round block receive time}
\end{itemize}
\textbf{Registration/Deregistration:}
\begin{itemize}
\item[-] Upon receiving input (REGISTER, R), where R${\in}$\{$G_{\textit{LEDGER}}$, $G_{RO},G_{\textit{AutoClock}}$\} execute protocol Registration-AutoSyn(P, sid, Reg, R). 
\item[-] Upon receiving input (DE-REGISTER, R), where R${\in}$\{$G_{\textit{LEDGER}}$, $G_{RO},G_{\textit{AutoClock}}$\} execute protocol Deregistration- AutoSyn (P, sid, Reg, R).
\item[-] Upon receiving input (IS-REGISTER, sid) return (REGISTER, sid,1) if the local registry Reg indicates that this party has successfully completed a registration with R=$G_{\textit{LEDGER}}$ (and did not DE-REGISTER since then).
        Otherwise, return (REGISTER, sid,0).
\end{itemize}
\textbf{Interacting with the Ledger:}

Upon receiving a ledger-specific input I ${\in}$\{(submit,...),(read,...),(maintain-ledger,...)\} verify first that all resources are available.
        If not all resources are available, then ignore the input; else (i.e., the party is operational and time-aware) execute one of the following steps depending on the input I:
\begin{itemize}
\item[-] If $I=(\text{SUBMIT},\mathrm{sid},\mathrm{tx})$ then set $\textit{TxBuffer}$ $\leftarrow $ $\textit{TxBuffer}| | \mathrm{tx}$, and send $\left(\text{MULTICAST},\mathrm{sid},\mathrm{tx}\right)to F_{N-MC}^{tx}$
\item[-] If $I=(\text{MAINTAIN}-\text{LEDGER},\mathrm{sid},\text{minerID})$ then invoke protocol\textbf{ LedgerMaintance }($C_{loc},U_{P},\text{ sid}, k, s, R,f$); if halts then halt the protocol \textbf{LedgerMaintance}  execution(all future input is ignored)
\item[-] If $I=(\text{READ},\mathrm{sid})$ then invoke protocol \textbf{ReadState}($\mathrm{k}, C_{loc},U_{P},\mathrm{sid},R,f$)
\end{itemize}
\textbf{Handling calls to the shared setup:}
\begin{itemize}
\item[-] Upon receiving (clock-read\textit{, }$\mathrm{sid}_{\mathrm{C}}$) forward it to $G_{\textit{AutoClock}}$ and output $G_{\textit{AutoClock}}$’s response.
\item[-] Upon receiving (clock-update, $\mathrm{sid}_{\mathrm{C}}$), record that a clock-update was received in the current round. \textcolor{color-2E74B5}{If this instance is currently time-aware but otherwise stalled or offline, then evolve the KES signing key by sending (USign,sid,Up,0,} $\textcolor{color-2E74B5}{s}\textcolor{color-2E74B5}{l}$\textcolor{color-2E74B5}{) to }$\textcolor{color-2E74B5}{\mathrm{F}}_{\textcolor{color-2E74B5}{K}\textcolor{color-2E74B5}{E}\textcolor{color-2E74B5}{S}}$\textcolor{color-2E74B5}{, where }$\textcolor{color-2E74B5}{s}\textcolor{color-2E74B5}{l}$\textcolor{color-2E74B5}{ is the current local round, and forward (clock-update,} $\mathrm{sid}_{\textcolor{color-2E74B5}{\mathrm{C}}}$\textcolor{color-2E74B5}{,} $\textcolor{color-2E74B5}{\bot }$\textcolor{color-2E74B5}{) to }$\textcolor{color-2E74B5}{G}_{\textcolor{color-2E74B5}{\textit{AutoClock}}}$\textcolor{color-2E74B5}{.} Furthermore, consider any active interruptible execution as completed.
\item[-] Upon receiving (eval\textit{, }$\mathrm{sid}_{\mathrm{RO}}, x$) forward the query to $\mathrm{G}_{\mathrm{RO}}$ and output $\mathrm{G}_{\mathrm{RO}}$’s response.
\end{itemize}
\section{ The Initialization Function}

$S_{adj}\leftarrow \left(sl_{i},y_{i},\pi _{i}\right)| \left| \left(sl_{i},y_{i},\pi _{i}\right)\right| | \ldots $  Semi-Adjust info

$\textcolor{color-2E74B5}{a}\textcolor{color-2E74B5}{d}\textcolor{color-2E74B5}{j}\leftarrow \textit{Adjust}| | \textit{Adjust}| | \ldots $ 

\textbf{Initialization-AutoSyn} ($U_{P}$, sid, R)
\begin{enumerate}[{1:}]
\item Send (KeyGen, sid, $U_{P}$) to $F_{VRF}$ and $F_{KES}$; receiving (VerificationKey, sid, $v_{p}^{kes}$), respectively
\item \textcolor{color-2E74B5}{Set }$\textcolor{color-2E74B5}{t}_{\textcolor{color-2E74B5}{\textit{begin}}}\textcolor{color-2E74B5}{\leftarrow }\textcolor{color-2E74B5}{0}$\textcolor{color-2E74B5}{.}
\item Send (clock-read, $sid_{C}$) to $G_{\textit{AutoClock}}$ get current time $t_{now}$.
\item If $\textcolor{color-2E74B5}{\boldsymbol{t}}_{\textcolor{color-2E74B5}{\boldsymbol{now}}}\textcolor{color-2E74B5}{< }\textcolor{color-2E74B5}{\mathbf{0}}$ then
\item    Send (ver\_kes, sid, $U_{P}$, $v_{p}^{vrf}$, $v_{p}^{kes}$) to $F_{INIT}$ to claim stake and ($\textcolor{color-2E74B5}{t}_{{\textcolor{color-2E74B5}{\textit{round}}_{\textcolor{color-2E74B5}{1}}}}$) from the genesis block.
        And set $\textcolor{color-2E74B5}{t}_{\textcolor{color-2E74B5}{\textit{round}}}\textcolor{color-2E74B5}{\leftarrow }\textcolor{color-2E74B5}{t}_{{\textcolor{color-2E74B5}{\textit{round}}_{\textcolor{color-2E74B5}{1}}}}$\textcolor{color-2E74B5}{. }
\item    Set $\textcolor{color-2E74B5}{t}_{\textcolor{color-2E74B5}{n}\textcolor{color-2E74B5}{e}\textcolor{color-2E74B5}{x}\textcolor{color-2E74B5}{t}}\textcolor{color-2E74B5}{\leftarrow }$ $\textcolor{color-2E74B5}{0}$ and invoke \textbf{FinishRound} ($U_{P},\textcolor{color-2E74B5}{t}_{\textcolor{color-2E74B5}{n}\textcolor{color-2E74B5}{e}\textcolor{color-2E74B5}{x}\textcolor{color-2E74B5}{t}}$).// \textcolor{gray}{Wait till }$\textcolor{gray}{t}_{\textcolor{gray}{n}\textcolor{gray}{e}\textcolor{gray}{x}\textcolor{gray}{t}}$
\item End if
\item If $\textcolor{color-2E74B5}{\boldsymbol{t}}_{\textcolor{color-2E74B5}{\boldsymbol{now}}}\textcolor{color-2E74B5}{\geq }\textcolor{color-2E74B5}{\mathbf{0}}$ then 
\item    If $F_{INIT}$ signals an error then
\item       Halt the execution.
\item    End if
\item    Send(genblock\_req, sid, $U_{P}$) to $F_{INIT}$
\item    Receive from $F_{INIT}$ the response (genblock, sid, G = ($S_{1}$, $\eta _{1}\textcolor{color-2E74B5}{,}\textcolor{color-2E74B5}{\boldsymbol{t}}_{{\textcolor{color-2E74B5}{\textit{\textbf{round}}}_{\textcolor{color-2E74B5}{\mathbf{1}}}}}$), where
\item                     $S_{1}=(\left(S_{1}, v_{1}^{vrf}, v_{1}^{kes},s_{1}\right),\ldots ,\left(S_{1}, v_{n}^{vrf}, v_{n}^{kes},s_{n}\right))$
\item    Set $C_{loc}\leftarrow $  (G).
\item    Set $T_{p}^{ep}\leftarrow $ $2^{{l_{VRF}}}\varnothing _{f}(\alpha _{p}^{ep})$as the threshold for stakeholder $U_{P}$ for epoch\textbf{ ep}, where           $\alpha _{p}^{ep} $  is the relative stake of stakeholder $U_{P}$ in $S_{ep}$ and $l_{VRF}$ denotes the output      length of $F_{VRF}$.
\item    \textcolor{color-2E74B5}{Invoke JoinProc(}$\textcolor{color-2E74B5}{C}_{\textcolor{color-2E74B5}{l}\textcolor{color-2E74B5}{o}\textcolor{color-2E74B5}{c}}\textcolor{color-2E74B5}{,}\textcolor{color-2E74B5}{s}\textcolor{color-2E74B5}{i}\textcolor{color-2E74B5}{d}\textcolor{color-2E74B5}{,}\textcolor{color-2E74B5}{\mathrm{R}}\textcolor{color-2E74B5}{,}\textcolor{color-2E74B5}{\boldsymbol{t}}_{{\textcolor{color-2E74B5}{\textit{\textbf{round}}}_{\textcolor{color-2E74B5}{\mathbf{1}}}}}$\textcolor{color-2E74B5}{) to catch up with current round time and get current slot number }$sl$\textcolor{color-2E74B5}{, round length }$\textcolor{color-2E74B5}{t}_{\textcolor{color-2E74B5}{\textit{round}}}$\textcolor{color-2E74B5}{ and next round start time }$\textcolor{color-2E74B5}{t}_{\textcolor{color-2E74B5}{n}\textcolor{color-2E74B5}{e}\textcolor{color-2E74B5}{x}\textcolor{color-2E74B5}{t}}$\textcolor{color-2E74B5}{. }
\item End if
\item Set isInit $\leftarrow $ true, $t_{on}\leftarrow sl$ and $t_{work}\leftarrow 0$
\end{enumerate}
GLOBAL VARIABLES: The protocol stores $v_{p}^{vrf}$, $v_{p}^{kes}$, $\tau $,\textbf{ ep}, sl\textbf{, }$C_{loc}$, $T_{p}^{ep}$, isInit, $t_{on}$, $\textcolor{color-2E74B5}{\boldsymbol{t}}_{\textcolor{color-2E74B5}{\boldsymbol{now}}}$\textbf{\textcolor{color-2E74B5}{, }}$\textcolor{color-2E74B5}{\boldsymbol{t}}_{\textcolor{color-2E74B5}{\boldsymbol{next}}}$\textbf{\textcolor{color-2E74B5}{, }}$\textcolor{color-2E74B5}{\boldsymbol{t}}_{\textcolor{color-2E74B5}{\textit{\textbf{begin}}}}$\textbf{\textcolor{color-2E74B5}{, }}$\textcolor{color-2E74B5}{\boldsymbol{t}}_{\textcolor{color-2E74B5}{\textit{\textbf{round}}}}$\textbf{\textcolor{color-2E74B5}{,}} $\textcolor{color-2E74B5}{t}_{\textcolor{color-2E74B5}{r}\textcolor{color-2E74B5}{e}\textcolor{color-2E74B5}{c}}$\textcolor{color-2E74B5}{,} $\textcolor{color-2E74B5}{S}_{\textcolor{color-2E74B5}{a}\textcolor{color-2E74B5}{d}\textcolor{color-2E74B5}{j}}$ to make each of them accessible by all protocol parts.

\section{ New Party Joining procedure}

\textbf{\textcolor{color-2E74B5}{JoinProc(}}$\textcolor{color-2E74B5}{C}_{\textcolor{color-2E74B5}{l}\textcolor{color-2E74B5}{o}\textcolor{color-2E74B5}{c}}\textcolor{color-2E74B5}{,}\textcolor{color-2E74B5}{s}\textcolor{color-2E74B5}{i}\textcolor{color-2E74B5}{d}\textcolor{color-2E74B5}{,}\textcolor{color-2E74B5}{\mathrm{R}}\textcolor{color-2E74B5}{,}\textcolor{color-2E74B5}{\boldsymbol{t}}_{{\textcolor{color-2E74B5}{\textit{\textbf{round}}}_{\textcolor{color-2E74B5}{\mathbf{1}}}}}$\textbf{\textcolor{color-2E74B5}{) }}
\begin{enumerate}[{1:}]
\item Send (HELLO, sid, $U_{P}$, $v_{p}^{vrf}$, $v_{p}^{kes}$) to $F_{N-MC}^{new}$.
\item Send (clock-read, $sid_{C}$) to $G_{\textit{AutoClock}}$ update current time and set $t_{next}\coloneqq t_{now}+3*\textcolor{color-2E74B5}{\boldsymbol{t}}_{{\textcolor{color-2E74B5}{\textit{\textbf{round}}}_{\textcolor{color-2E74B5}{\mathbf{1}}}}}$ 
\item Invoke \textbf{FinishRound}($U_{P},t_{next}$) 
\item Invoke \textbf{FetchInformation}($U_{P}$, sid) to receive the newest messages for this round; denote the output by $tx\leftarrow (tx_{1},\ldots ,tx_{k})$, $N\leftarrow \left\{\left(C_{1}, D_{{mid_{1}}}\right),\ldots ,\left(C_{m},D_{{mid_{m}}}\right)\right\}$, $S_{adj}\leftarrow $ $\{(\textit{Adjust}_{1}, D_{{mid_{1}}}),\ldots ,(\textit{Adjust}_{\mathrm{n}},D_{{mid_{n}}})\}$.
\item Set $\textit{TxBuffer}\leftarrow \textit{TxBuffer}| | tx$, $\textit{AdjustBuffer}\leftarrow \textit{AdjustBuffer}| | S_{adj}$.
\item Invoke \textbf{SelectChain}($U_{P}, $ sid, $C_{loc}$, N, k ,s, R, f) to update $C_{loc}$
\item Invoke \textbf{CurrentSlotNumber}($t_{now}$, $C_{loc},R$) to get $sl, t_{next},t_{\textit{round}}$.
\item Invoke \textbf{FinishRound} ($U_{P},\textcolor{color-2E74B5}{t}_{\textcolor{color-2E74B5}{n}\textcolor{color-2E74B5}{e}\textcolor{color-2E74B5}{x}\textcolor{color-2E74B5}{t}}$) 
\end{enumerate}
Outputs: The protocol outputs  $\boldsymbol{sl}$\textbf{,} $\boldsymbol{t}_{{\textit{\textbf{round}}_{\mathbf{ep}}}},\boldsymbol{t}_{\mathbf{next}}$ to its caller (but not to Z).

\section{ Staking Procedure}

\textbf{StakingProcedure} ($U_{P}, sid. k, ep, sl, \textit{buffer},\textcolor{color-2E74B5}{\textit{AdjustBuffer}}, C_{loc}\textcolor{color-2E74B5}{,} \textcolor{color-2E74B5}{a}\textcolor{color-2E74B5}{d}\textcolor{color-2E74B5}{j}\textcolor{color-2E74B5}{,}\textcolor{color-2E74B5}{S}_{\textcolor{color-2E74B5}{a}\textcolor{color-2E74B5}{d}\textcolor{color-2E74B5}{j}}$)
\begin{enumerate}[{1:}]
\item Send(EvalProve, sid, $\eta _{j}${\textbar}{\textbar} sl {\textbar}{\textbar} NONCE) to $F_{VRF}$, denote the response from $F_{VRF}$ by (Evaluated,sid, $y_{\rho }$, $\pi _{\rho }$).
\item Send(EvalProve, sid, $\eta _{j}${\textbar}{\textbar} sl {\textbar}{\textbar} TEST) to $F_{VRF}$, denote the response from $F_{VRF}$ by (Evaluated,sid, y, $\pi $).
\item If y {\textless} $T_{p}^{ep}$ then
\item \textcolor{color-A6A6A6}{//pre-waiting}
\item     Set $\textcolor{color-2E74B5}{\textit{TxBuffe}}\textcolor{color-2E74B5}{r}\textcolor{color-2E74B5}{'}\leftarrow \textcolor{color-2E74B5}{\textit{TxBuffer}}\textcolor{color-2E74B5}{,}\textcolor{color-2E74B5}{ }\vec{N}$ $\leftarrow $ $tx_{U_{P}}^{base-tx}$, and st $\leftarrow $ $\text{blockify}_{OG}$($\vec{N}$)
\item     Repeat
\item         Parse $\textcolor{color-2E74B5}{\textit{TxBuffe}}\textcolor{color-2E74B5}{r}\textcolor{color-2E74B5}{'}$ as sequence ($tx_{1}$,{\ldots}, $tx_{n}$)
\item         For i=1 to n do
\item            If $\text{ValidTx}_{OG}$($tx_{i}$, $\overrightarrow{st}| | st$) = 1 then
\item                 $\vec{N}\leftarrow \vec{N}${\textbar}{\textbar} $tx_{i}$
\item                 Remove tx from $\textcolor{color-2E74B5}{\textit{TxBuffe}}\textcolor{color-2E74B5}{r}\textcolor{color-2E74B5}{'}$
\item                 Set st $\leftarrow $ $\text{blockify}_{OG}$($\vec{N}$)

           End if

         End for
\item      Until $\vec{N}$ does not increase anymore
\item      \textcolor{color-2E74B5}{Set} $\textcolor{color-2E74B5}{\mathrm{a}}\textcolor{color-2E74B5}{\leftarrow }$ $\textcolor{color-2E74B5}{\varnothing }$\textcolor{color-2E74B5}{ and parse }$\textcolor{color-2E74B5}{\textit{AdjustBuffer}}$\textcolor{color-2E74B5}{ as \{(}$\textcolor{color-2E74B5}{\textit{Adjust}}_{\textcolor{color-2E74B5}{1}}\textcolor{color-2E74B5}{,}\textcolor{color-2E74B5}{T}_{{\textcolor{color-2E74B5}{a}\textcolor{color-2E74B5}{d}\textcolor{color-2E74B5}{j}_{\textcolor{color-2E74B5}{1}}}}\textcolor{color-2E74B5}{)}\textcolor{color-2E74B5}{,}$\textcolor{color-2E74B5}{{\ldots}}$\textcolor{color-2E74B5}{,}\textcolor{color-2E74B5}{(}\textcolor{color-2E74B5}{\textit{Adjust}}_{\textcolor{color-2E74B5}{\mathrm{n}}}\textcolor{color-2E74B5}{,}\textcolor{color-2E74B5}{T}_{{\textcolor{color-2E74B5}{a}\textcolor{color-2E74B5}{d}\textcolor{color-2E74B5}{j}_{\textcolor{color-2E74B5}{n}}}}\textcolor{color-2E74B5}{)}$\textcolor{color-2E74B5}{\}}
\item      \textcolor{color-2E74B5}{For} $\textcolor{color-2E74B5}{(}\textcolor{color-2E74B5}{\textit{Adjust}}_{\textcolor{color-2E74B5}{\mathrm{i}}}\textcolor{color-2E74B5}{,}\textcolor{color-2E74B5}{T}_{{\textcolor{color-2E74B5}{a}\textcolor{color-2E74B5}{d}\textcolor{color-2E74B5}{j}_{\textcolor{color-2E74B5}{i}}}}\textcolor{color-2E74B5}{)}$\textcolor{color-2E74B5}{ in \{(}$\textcolor{color-2E74B5}{\textit{Adjust}}_{\textcolor{color-2E74B5}{1}}\textcolor{color-2E74B5}{,}\textcolor{color-2E74B5}{T}_{{\textcolor{color-2E74B5}{a}\textcolor{color-2E74B5}{d}\textcolor{color-2E74B5}{j}_{\textcolor{color-2E74B5}{1}}}}\textcolor{color-2E74B5}{)}\textcolor{color-2E74B5}{,}$\textcolor{color-2E74B5}{{\ldots}}$\textcolor{color-2E74B5}{,}\textcolor{color-2E74B5}{(}\textcolor{color-2E74B5}{\textit{Adjust}}_{\textcolor{color-2E74B5}{\mathrm{n}}}\textcolor{color-2E74B5}{,}\textcolor{color-2E74B5}{T}_{{\textcolor{color-2E74B5}{a}\textcolor{color-2E74B5}{d}\textcolor{color-2E74B5}{j}_{\textcolor{color-2E74B5}{n}}}}\textcolor{color-2E74B5}{)}$\textcolor{color-2E74B5}{\} do }
\item \textcolor{color-2E74B5}{          If }$\textcolor{color-2E74B5}{\textit{Adjust}}_{\textcolor{color-2E74B5}{\mathrm{i}}}$\textcolor{color-2E74B5}{ has not appeared on }$\textcolor{color-2E74B5}{C}_{\textcolor{color-2E74B5}{l}\textcolor{color-2E74B5}{o}\textcolor{color-2E74B5}{c}}$\textcolor{color-2E74B5}{ then}
\item \textcolor{color-2E74B5}{             Set }$\textcolor{color-2E74B5}{\mathrm{a}}\textcolor{color-2E74B5}{\leftarrow }\textcolor{color-2E74B5}{\mathrm{a}}$\textcolor{color-2E74B5}{{\textbar}{\textbar} }$\textcolor{color-2E74B5}{(}\textcolor{color-2E74B5}{\textit{Adjust}}_{\textcolor{color-2E74B5}{\mathrm{i}}}\textcolor{color-2E74B5}{,}\textcolor{color-2E74B5}{T}_{{\textcolor{color-2E74B5}{a}\textcolor{color-2E74B5}{d}\textcolor{color-2E74B5}{j}_{\textcolor{color-2E74B5}{i}}}}\textcolor{color-2E74B5}{)}$\textcolor{color-2E74B5}{ and remove }$\textcolor{color-2E74B5}{(}\textcolor{color-2E74B5}{\textit{Adjust}}_{\textcolor{color-2E74B5}{\mathrm{i}}}\textcolor{color-2E74B5}{,}\textcolor{color-2E74B5}{T}_{{\textcolor{color-2E74B5}{a}\textcolor{color-2E74B5}{d}\textcolor{color-2E74B5}{j}_{\textcolor{color-2E74B5}{i}}}}\textcolor{color-2E74B5}{)}$\textcolor{color-2E74B5}{ from} $\textcolor{color-2E74B5}{\textit{AdjustBuffer}}$

\textcolor{color-2E74B5}{          End if}

\textcolor{color-2E74B5}{     End for}

    \textcolor{color-A6A6A6}{ // every Block’s }$\textcolor{color-A6A6A6}{\textit{\textbf{Adjust}}}_{\textcolor{color-A6A6A6}{\mathbf{i}}}$\textcolor{color-A6A6A6}{ only appears once on the Chain with its first receive time.}
\item    \textcolor{color-2E74B5}{If }$\textcolor{color-2E74B5}{t}_{\textcolor{color-2E74B5}{n}\textcolor{color-2E74B5}{o}\textcolor{color-2E74B5}{w}}\textcolor{color-2E74B5}{< }\textcolor{color-2E74B5}{t}_{\textcolor{color-2E74B5}{\textit{begin}}}\textcolor{color-2E74B5}{+}\textcolor{color-2E74B5}{t}_{\textcolor{color-2E74B5}{r}\textcolor{color-2E74B5}{u}\textcolor{color-2E74B5}{n}}$\textcolor{color-2E74B5}{:}\textcolor{color-A6A6A6}{ // Sending time must before }$\textcolor{color-A6A6A6}{t}_{\textcolor{color-A6A6A6}{r}\textcolor{color-A6A6A6}{u}\textcolor{color-A6A6A6}{n}}$
\item        Set crt = ($U_{P}$, y, $\pi $), $\rho =(y_{\rho }, \pi _{\rho })$ and h $\leftarrow $ H(head($C_{loc}$))
\item        Send (Usign, sid , $U_{P}, \left(h, st, sl\textcolor{color-2E74B5}{,}\textcolor{color-2E74B5}{t}_{\textcolor{color-2E74B5}{n}\textcolor{color-2E74B5}{o}\textcolor{color-2E74B5}{w}}, crt, \rho \right), sl,\textcolor{color-2E74B5}{\mathbf{a}}$) to $F_{KES}$, denote the response from $F_{KES}$ by (Signature, sid, $\left(h, st,sl, \textcolor{color-2E74B5}{t}_{\textcolor{color-2E74B5}{n}\textcolor{color-2E74B5}{o}\textcolor{color-2E74B5}{w}}, crt, \rho \right),sl,\textcolor{color-2E74B5}{\mathbf{a}},\sigma $)
\item        Set B $\leftarrow \left(h, st, sl,\textcolor{color-2E74B5}{t}_{\textcolor{color-2E74B5}{n}\textcolor{color-2E74B5}{o}\textcolor{color-2E74B5}{w}}, crt, \rho ,\textcolor{color-2E74B5}{\mathbf{a}}\textcolor{color-2E74B5}{,}\sigma \right)$ and update $C_{loc}$ $\leftarrow $ $C_{loc}| | B$
\item        Send(MULTICAST, sid, $C_{loc}$) to $F_{N-MC}^{bc}$ and proceed from upon next activation of this procedure.
\item    \textcolor{color-2E74B5}{Else }
\item        Evolve the KES signing key by sending (USign, sid , $U_{P}$, 0, sl) to $F_{KES}$. and set the anchor at the end of procedure to resume on next maintenance activation.
\item    \textcolor{color-2E74B5}{If }$\textcolor{color-2E74B5}{t}_{\textcolor{color-2E74B5}{r}\textcolor{color-2E74B5}{e}\textcolor{color-2E74B5}{c}}\textcolor{color-2E74B5}{\neq }\textcolor{color-2E74B5}{\bot }$ \textcolor{color-2E74B5}{ then }
\item          \textcolor{color-2E74B5}{Set }$\textcolor{color-2E74B5}{\mathrm{adj}}\textcolor{color-2E74B5}{\leftarrow }\textcolor{color-2E74B5}{\mathrm{adj}}\textcolor{color-2E74B5}{| }\textcolor{color-2E74B5}{| }\textcolor{color-2E74B5}{(}\textcolor{color-2E74B5}{B}_{\textcolor{color-2E74B5}{l}\textcolor{color-2E74B5}{a}\textcolor{color-2E74B5}{s}\textcolor{color-2E74B5}{t}}\textcolor{color-2E74B5}{,}\textcolor{color-2E74B5}{t}_{\textcolor{color-2E74B5}{r}\textcolor{color-2E74B5}{e}\textcolor{color-2E74B5}{c}}\textcolor{color-2E74B5}{,} \textcolor{color-2E74B5}{P}\textcolor{color-2E74B5}{,}\textcolor{color-2E74B5}{y}$\textcolor{color-2E74B5}{, }$\textcolor{color-2E74B5}{\pi }\textcolor{color-2E74B5}{)}$\textcolor{color-2E74B5}{.}\textcolor{gray}{// later to broadcast it}
\item     \textcolor{color-2E74B5}{Else}
\item \textcolor{color-2E74B5}{         Set }$\textcolor{color-2E74B5}{S}_{\textcolor{color-2E74B5}{a}\textcolor{color-2E74B5}{d}\textcolor{color-2E74B5}{j}}\leftarrow \textcolor{color-2E74B5}{S}_{\textcolor{color-2E74B5}{a}\textcolor{color-2E74B5}{d}\textcolor{color-2E74B5}{j}}\textcolor{color-2E74B5}{| }\textcolor{color-2E74B5}{| }\left(\textcolor{color-2E74B5}{s}\textcolor{color-2E74B5}{l}\textcolor{color-2E74B5}{,}\textcolor{color-2E74B5}{y}\textcolor{color-2E74B5}{,}\textcolor{color-4472C4}{\pi }\right)$\textcolor{gray}{//tell the next round to capture if there is a block received from last slots}

\item Else
\item     Evolve the KES signing key by sending (USign, sid , $U_{P}$, 0, sl) to $F_{KES}$ and set the anchor at the end of procedure to resume on next maintenance activation.

End if
\end{enumerate}
\section{ Ledger Maintain Procedure}

The following steps are executed in an$(\text{MAINTAIN}-\text{LEDGER},\mathrm{sid},\text{minerID})$- interruptible manner 

\textbf{LedgerMaintaince}($C_{loc}$, $U_{P},\text{ sid},\mathrm{k},\mathrm{s},\mathrm{R},f$)
\begin{enumerate}[{1:}]
\item If isInit is false then invoke Initialization-AutoSyn($U_{P}$, sid, R); if Initialization-AutoSyn halts then halt(this will abort the execution)

End if
\item Guarantee $v_{p}^{vrf}$, $v_{p}^{kes}$, $\tau $, ep, sl, $C_{loc}$, $T_{p}^{ep}$, isInit, $t_{on}$, $t_{now}$, $t_{next}$, $t_{\textit{begin}}$, $t_{\textit{round}}$, $t_{rec}$, $S_{adj}$ is readable.
\item If been offline before (been deregistered to $F_{N-mc}$)
\item      Invoke JoinProc($C_{loc},sid,\mathrm{R},\boldsymbol{t}_{{\textit{\textbf{round}}_{\mathbf{1}}}}$) to get $sl$, $t_{\textit{round}}$ and $t_{next}$.
\item End if 
\item Invoke \textbf{FetchInformation}($U_{P}$, sid) to receive the newest messages for this round; denote the output by $tx\leftarrow (tx_{1},\ldots ,tx_{k})$, $N\leftarrow \left\{\left(C_{1}, D_{{mid_{1}}}\right),\ldots ,\left(C_{m},D_{{mid_{m}}}\right)\right\}$,$\textit{AdjBuffer}_{sl}\leftarrow $ $\{(\textit{Adjust}_{1}, D_{{mid_{1}}}),\ldots ,(\textit{Adjust}_{\mathrm{n}},D_{{mid_{n}}})\}$ and record the flag WELCOME if not yet recorded for this round.
\item Set $\textit{TxBuffer}\leftarrow \textit{TxBuffer}| | tx$, $\textit{AdjBuffer}\leftarrow \textit{AdjBuffer}| | \textit{AdjBuffer}_{sl}$.
\item Invoke \textbf{SelectChain}$(U_{P},\mathrm{sid} ,C_{loc}, N,k ,s, R, f,\textcolor{color-2E74B5}{t}_{\textcolor{color-2E74B5}{r}\textcolor{color-2E74B5}{e}\textcolor{color-2E74B5}{c}})$ to replace $C_{loc}$.
\item Invoke \textbf{UpdateTime}($U_{P},R,C_{loc},\textcolor{color-2E74B5}{t}_{\textcolor{color-2E74B5}{r}\textcolor{color-2E74B5}{u}\textcolor{color-2E74B5}{n}})$ to update $\textcolor{color-2E74B5}{t}_{\textcolor{color-2E74B5}{n}\textcolor{color-2E74B5}{o}\textcolor{color-2E74B5}{w}}\textcolor{color-2E74B5}{,}\textcolor{color-2E74B5}{t}_{\textcolor{color-2E74B5}{n}\textcolor{color-2E74B5}{e}\textcolor{color-2E74B5}{x}\textcolor{color-2E74B5}{t}}\textcolor{color-2E74B5}{,}\textcolor{color-2E74B5}{t}_{\textcolor{color-2E74B5}{\textit{begin}}}\textcolor{color-2E74B5}{,}\textcolor{color-2E74B5}{t}_{\textcolor{color-2E74B5}{\textit{round}}},\mathrm{sl},\text{and ep}$.
\item Set $\mathrm{t}_{\mathrm{on}}\leftarrow sl, \textcolor{color-2E74B5}{t}_{\textcolor{color-2E74B5}{r}\textcolor{color-2E74B5}{e}\textcolor{color-2E74B5}{c}}\textcolor{color-2E74B5}{\leftarrow }\textcolor{color-2E74B5}{\bot }$\textcolor{color-2E74B5}{, }$\textcolor{color-2E74B5}{a}\textcolor{color-2E74B5}{d}\textcolor{color-2E74B5}{j}\textcolor{color-2E74B5}{\leftarrow }\textcolor{color-2E74B5}{\bot }$\textcolor{color-2E74B5}{, }$\textcolor{color-2E74B5}{S}_{\textcolor{color-2E74B5}{a}\textcolor{color-2E74B5}{d}\textcolor{color-2E74B5}{j}}\textcolor{color-2E74B5}{\leftarrow }\textcolor{color-2E74B5}{\bot }$\textcolor{color-2E74B5}{.}
\item Invoke \textbf{UpdateStakeDist}(k, $U_{P}$, R, f) to update $S_{ep}$, $\alpha _{p}^{ep}$, $T_{p}^{ep}$, and  $\eta _{ep }$
\item Call \textbf{StakingProcedure}($U_{P}$, sid, k , ep, sl , buffer, $\textcolor{color-2E74B5}{\textit{AdjustBuffer}}$, $\textcolor{color-2E74B5}{a}\textcolor{color-2E74B5}{d}\textcolor{color-2E74B5}{j}$, $C_{loc}$, $S_{adj}$)
\item If WELCOME = 1 in this round then send (MULTICAST, sid,$C_{loc}$) to $F_{N-MC}^{bc}$ and (MULTICAST, sid, $\textit{TxBuffer}$) to $F_{N-MC}^{tx}$. 

End if.
\item Invoke \textbf{\textcolor{color-2E74B5}{AdjustDelay }}\textbf{(}$U_{P},\text{ sid}, S_{adj}, N,sl,\textcolor{color-2E74B5}{\boldsymbol{adj}}$\textbf{) }to deliver adjustment information.
\item Invoke \textbf{FinishRound} ($U_{P},t_{next}$)
\end{enumerate}

\section{ Selection Chain Procedure }

\textbf{SelectChain}($U_{P},sid, C_{loc},\textcolor{color-2E74B5}{N}\textcolor{color-2E74B5}{=}\left\{\left(\textcolor{color-2E74B5}{C}_{\textcolor{color-2E74B5}{1}}\textcolor{color-2E74B5}{,} \textcolor{color-2E74B5}{D}_{{\textcolor{color-2E74B5}{m}\textcolor{color-2E74B5}{i}\textcolor{color-2E74B5}{d}_{\textcolor{color-2E74B5}{1}}}}\right)\textcolor{color-2E74B5}{,}\textcolor{color-2E74B5}{\ldots }\textcolor{color-2E74B5}{,}\left(\textcolor{color-2E74B5}{C}_{\textcolor{color-2E74B5}{m}}\textcolor{color-2E74B5}{,}\textcolor{color-2E74B5}{D}_{{\textcolor{color-2E74B5}{m}\textcolor{color-2E74B5}{i}\textcolor{color-2E74B5}{d}_{\textcolor{color-2E74B5}{m}}}}\right)\right\},k,s,R,f$, $\textcolor{color-2E74B5}{t}_{\textcolor{color-2E74B5}{r}\textcolor{color-2E74B5}{e}\textcolor{color-2E74B5}{c}}$)
\begin{enumerate}[{1:}]
\item Initialize  $N_{\textit{valid}}\leftarrow 0$
\item For i = 1,{\ldots}, M do
\item     Invoke Protocol IsValidChain($U_{P},sid, \mathrm{k},C_{i},R,f$); 
\item     if returns true then update$N_{\textit{valid}}\leftarrow N_{\textit{valid}}\cup C_{i}$

End for 
\item Execute Algorithm maxvalid-bg($C_{loc}$, $N_{\textit{valid}},\mathrm{k},\mathrm{s},\mathrm{f}$) and receive its output $C_{max}$
\item \textcolor{color-2E74B5}{If }$\textcolor{color-2E74B5}{C}_{\textcolor{color-2E74B5}{m}\textcolor{color-2E74B5}{a}\textcolor{color-2E74B5}{x}}\textcolor{color-2E74B5}{\neq }\textcolor{color-2E74B5}{C}_{\textcolor{color-2E74B5}{l}\textcolor{color-2E74B5}{o}\textcolor{color-2E74B5}{c}}$\textcolor{color-2E74B5}{ then, get corresponding } $\textcolor{color-2E74B5}{D}_{\textcolor{color-2E74B5}{m}\textcolor{color-2E74B5}{i}\textcolor{color-2E74B5}{d}}$\textcolor{color-2E74B5}{ from }$\textcolor{color-2E74B5}{N}$\textcolor{color-2E74B5}{, and set }$\textcolor{color-2E74B5}{t}_{\textcolor{color-2E74B5}{r}\textcolor{color-2E74B5}{e}\textcolor{color-2E74B5}{c}}\textcolor{color-2E74B5}{\leftarrow }\textcolor{color-2E74B5}{D}_{\textcolor{color-2E74B5}{m}\textcolor{color-2E74B5}{i}\textcolor{color-2E74B5}{d}}$
\item Replace $C_{loc}$ by $C_{max}$
\end{enumerate}
Outputs: The protocol outputs $C_{max}$ to its caller (but not to Z).

\section{ Update Time Procedure}

\textbf{UpdateTime}($U_{P},R,C_{loc},\textcolor{color-2E74B5}{t}_{\textcolor{color-2E74B5}{r}\textcolor{color-2E74B5}{u}\textcolor{color-2E74B5}{n}})$
\begin{enumerate}[{1:}]
\item Send (clock-read, $sid_{C}$) to $G_{\textit{AutoClock}}$ update current time $t_{now}$
\item If $\textcolor{color-2E74B5}{t}_{\textcolor{color-2E74B5}{r}\textcolor{color-2E74B5}{u}\textcolor{color-2E74B5}{n}}+t_{next}> t_{now}$\textcolor{color-2E74B5}{ then}
\item     Set $sl\leftarrow sl+1$.
\item     If $\textcolor{color-2E74B5}{s}\textcolor{color-2E74B5}{l} \textcolor{color-2E74B5}{m}\textcolor{color-2E74B5}{o}\textcolor{color-2E74B5}{d} \textcolor{color-2E74B5}{R}\textcolor{color-2E74B5}{=}\textcolor{color-2E74B5}{1}$ and $\textcolor{color-2E74B5}{e}\textcolor{color-2E74B5}{p}\textcolor{color-2E74B5}{> }\textcolor{color-2E74B5}{1}$ then (this is a new epoch)
\item \textcolor{color-2E74B5}{        Invoke }\textbf{\textcolor{color-2E74B5}{AdjustingNextRoundLength}}\textcolor{color-2E74B5}{(}$\textcolor{color-2E74B5}{C}_{\textcolor{color-2E74B5}{l}\textcolor{color-2E74B5}{o}\textcolor{color-2E74B5}{c}}$\textcolor{color-2E74B5}{) to get new }$\textcolor{color-2E74B5}{t}_{\textcolor{color-2E74B5}{\textit{round}}}$

End if
\item     Set $t_{\textit{begin}}\leftarrow t_{next} and t_{next}\leftarrow t_{next}+t_{\textit{round}}$
\item Else \textcolor{color-A6A6A6}{// Party has been stalled.}
\item     Invoke CurrentSlotNumber($t_{now}$, $C_{loc}, R$) to get $sl, t_{next},t_{\textit{round}}$.

End if
\item Set $ep\leftarrow \lceil sl/R\rceil $.
\end{enumerate}
Output : The protocol output $sl,\mathrm{ep},t_{\textit{round}},t_{next}$ to the caller (but not to Z)

\section{ Fetch Information Procedure}

\textbf{FetchInformation}($U_{P}$, sid)
\begin{enumerate}[{1:}]
\item Send (fetch, sid) to $F_{N-MC}^{tx}$; denote the response from $F_{N-MC}^{tx}$ by (fetch, sid, $b_{1}$)
\item Extract received transactions $tx\leftarrow (tx_{1},\ldots ,tx_{k})$ from $b_{1}$.
        If $b_{1}=\bot $, $tx=\bot $.
\item Send (fetch, sid) to $F_{N-MC}^{bc}$; denote the response from $F_{N-MC}^{bc}$ by (fetch, sid, $b_{2}$))
\item Extract received chains $(C_{1},\ldots ,C_{m})$ and corresponding arriving time $D_{mid}$ from $b_{2}$.
        Denote as $N\leftarrow \left\{\left(C_{1}, D_{{mid_{1}}}\right),\ldots ,\left(C_{M},D_{{mid_{m}}}\right)\right\}$.
        If $b_{2}=\bot $, $N=\bot $.
\item Send (fetch, sid) to $F_{N-MC}^{adj}$; denote the response from $F_{N-MC}^{adj}$ by (fetch, sid, $b_{3}$) $D_{mid}^{adj}$
\item Extract received adjustment message $\textit{Adjust}$ and corresponding arriving time $D_{mid}$ from $b_{3}$.
        Denote as $\textit{AdjustBuffer}\leftarrow \{(\textit{Adjust}_{1}, D_{{mid_{1}}}),\ldots ,(\textit{Adjust}_{\mathrm{n}},D_{{mid_{n}}})\}$.
        If $b_{3}=\bot $, $adj=\bot $.
\item IF a message (HELLO, sid, ·) was received then

    set welcome = 1 
\item ELSE

    set welcome = 0 

END IF
\end{enumerate}
Output: The protocol outputs $tx$, $N$,WELCOME, $\textit{AdjustBuffer}$ to its caller (but not to Z).

\section{ Waiting Procedure}

\textbf{FinishRound} ($U_{P},t_{next}$)
\begin{enumerate}[{1:}]
\item Send (clock-read, $sid_{C}$) to $G_{\textit{AutoClock}}$ update current time $t_{now}$.
\item While $t_{now}< t_{next}$ do
\item    Send (clock-read, $sid_{C}$) to $G_{\textit{AutoClock}}$ update current time $t_{now}$.
\item    Give up activation(set the anchor here)

End while
\end{enumerate}
\section{ Delay Adjusting Procedure}

\textbf{AdjustDelay }($U_{P}$, sid,$\text{ adj}$, $\textcolor{color-2E74B5}{S}_{\textcolor{color-2E74B5}{a}\textcolor{color-2E74B5}{d}\textcolor{color-2E74B5}{j}}$\textcolor{color-2E74B5}{,} $N$, $sl,\textcolor{color-2E74B5}{a}\textcolor{color-2E74B5}{d}\textcolor{color-2E74B5}{j}$)
\begin{enumerate}[{1:}]
\item For each ($sl_{i},y_{i},\pi _{i}$) in  $S_{adj}$
\item    $\text{Adjust}\leftarrow (\bot ,\bot , P,y$, $\pi )$
\item    For each $\left(C, D_{mid}\right)$ in $N$ do 
\item       If the create round of last block of C (denote as $B_{last}$) is $sl_{i}$ then
\item          Set $\text{Adjust}\leftarrow (B_{last},D_{mid}, P,y,\pi )$. 
\item          Set $adj\leftarrow adj| | \textit{Adjust}$
\item          Delete ($sl_{i},y_{i},\pi _{i}$) in $S_{adj}$
\item       End if
\item    End for
\item    If $sl-sl_{i}> 2$ then delete $sl_{i},y_{i},\pi _{i}$) in $S_{adj}$.
\item End for 
\item Send (MULTICAST, sid, $\mathrm{adj}$) to $F_{N-MC}^{adj}$, if $\mathrm{adj}\neq \bot $ and set anchor at end of procedure to resume on next maintenance activation.
\end{enumerate}
\section{ Update Stake Procedure}

\textbf{UpdateStakeDist}($\mathrm{k}, \mathrm{U}_{\mathrm{p}},R,f$)
\begin{enumerate}[{1:}]
\item Set $\mathrm{S}_{\mathrm{ep}}$ to be the stakeholder distribution at the end of epoch $\mathrm{ep}-2$ in $\mathrm{C}_{\mathrm{loc}}$ in case $\mathrm{ep}> 2$ (and keep the initial stake distribution in case $\mathrm{ep}< 2$).
\item Set $\alpha _{p}^{ep}$ to be the relative stake of $\mathrm{U}_{\mathrm{p}}$ in $\mathrm{S}_{\mathrm{ep}}$ and $T_{p}^{ep}\leftarrow 2^{{l_{VRF}}}\varnothing _{f}(\alpha _{p}^{ep})$
\item Set $\eta _{ep}\leftarrow H(\eta _{ep-1}| \left| ep\right| | v)$ where $v$ is the concatenation of VRF outputs $y_{\rho }$ from all blocks in $\mathrm{C}_{\mathrm{loc}}$ from the first 2R/3 slots of epoch $\mathrm{ep}-1$
\end{enumerate}
Output : The protocol output $\mathrm{sl},\text{ ep}, \boldsymbol{t}_{\boldsymbol{next}}$, $\mathrm{S}_{\mathrm{ep}},\alpha _{p}^{ep},T_{p}^{ep},\eta _{ep }$ to the caller (but not to Z)

\section{ Adjusting Next Round Length Procedure}

\textbf{AdjustingNextRoundLength}($C_{loc}$)
\begin{enumerate}[{1:}]
\item Set $adj_{1},adj_{2}\leftarrow \varnothing $
\item Set $C_{loc}^{1}$ as first R/2 rounds of block in last epoch, and $t_{\textit{round}}^{1}$ is its according round length
\item Set $C_{loc}^{2}$ as last R/2 rounds of block in the epoch before last epoch, and $t_{\textit{round}}^{2}$ is its according round length
\item For j =1, 2
\item    For each $(\textit{\textbf{Adjust}}_{\mathbf{i}},\boldsymbol{T}_{{\boldsymbol{adj}_{\boldsymbol{i}}}})$ in each block in $\boldsymbol{B}$\textbf{ of }$C_{loc}^{j}$
\item       If $\textit{\textbf{Adjust}}_{\mathbf{i}}$ has not appeared in $\textit{adjust}$
\item          Read $t_{B}$($t_{now}$ of $\boldsymbol{B}$) from $\boldsymbol{B}$ 
\item          Parse $\textit{\textbf{Adjust}}_{\mathbf{i}}$ as $(\boldsymbol{B}_{\boldsymbol{last}},\boldsymbol{T}_{\boldsymbol{recv}}, \boldsymbol{P},y$, $\pi )$
\item          Read $t_{{\boldsymbol{B}_{\textit{\textbf{conflict}}}}-send}$($t_{now}$ of $\boldsymbol{B}_{\boldsymbol{last}}$) from $\boldsymbol{B}_{\boldsymbol{last}}$
\item          If $\boldsymbol{T}_{\boldsymbol{recv}}-t_{{\boldsymbol{B}_{\textit{\textbf{conflict}}}}}\leq 2\cdot t_{\textit{round}}^{j}$ 
\item             $adj_{\mathrm{j}}\leftarrow adj_{\mathrm{j}}| | (\textit{\textbf{Adjust}}_{\mathbf{i}} ,\boldsymbol{T}_{\boldsymbol{recv}}-t_{{\boldsymbol{B}_{\boldsymbol{last}}}},\boldsymbol{T}_{\boldsymbol{recv}}-\boldsymbol{T}_{{\boldsymbol{adj}_{\boldsymbol{i}}}})$,
\item             // $\textcolor{color-AEAAAA}{\boldsymbol{B}}_{\textcolor{color-AEAAAA}{\boldsymbol{last}}} \textcolor{color-AEAAAA}{\boldsymbol{and}} \textcolor{color-AEAAAA}{\textit{\textbf{Adjust}}}_{\textcolor{color-AEAAAA}{\mathbf{i}}}$\textbf{\textcolor{color-AEAAAA}{ send duration time}}
\item          End if
\item       End if
\item    End for
\item End for
\item Set $t_{\textit{round}-new}^{1}=t_{\textit{round}}^{1}$ and $t_{\textit{round}-new}^{2}=t_{\textit{round}}^{2}$
\item For each $(\textit{\textbf{Adjust}}_{\mathbf{i}},t_{i}^{a},t_{i}^{b})$ in $adj_{1}=\{\left(\textit{\textbf{Adjust}}_{\mathbf{1}},t_{1}^{a},t_{1}^{b}\right),\ldots ,(\textit{\textbf{Adjust}}_{\mathbf{m}},t_{m}^{a},t_{m}^{b})\}$
\item     Set $t_{\textit{round}-new}^{1}=\omega _{1}\left(\frac{1}{m}\sum \limits_{1}^{m}t_{i}^{a}-t_{\textit{round}}^{1}\right)+\omega _{2}\left(\frac{1}{m}\sum \limits_{1}^{m}t_{i}^{b}-t_{\textit{round}}^{1}\right)$
\item End for
\item For each $(\textit{\textbf{Adjust}}_{\mathbf{i}},t_{i}^{a},t_{i}^{b})$ in $adj_{2}=\{\left(\textit{\textbf{Adjust}}_{\mathbf{1}},t_{1}^{a},t_{1}^{b}\right),\ldots ,(\textit{\textbf{Adjust}}_{\mathbf{m}},t_{m}^{a},t_{m}^{b})\}$
\item     Set $t_{\textit{round}-new}^{2}=\omega _{1}\left(\frac{1}{m}\sum \limits_{1}^{m}t_{i}^{a}-t_{\textit{round}}^{2}\right)+\omega _{2}\left(\frac{1}{m}\sum \limits_{1}^{m}t_{i}^{b}-t_{\textit{round}}^{2}\right)$
\item End for
\item Set $t_{\textit{round}-new}=(t_{\textit{round}-new}^{1}+t_{\textit{round}-new}^{2})/2$
\end{enumerate}
Output: The protocol output $t_{new-\textit{round}}$ to the caller (but not to Z)

\section{ Current Slot Number Procedure}

\textbf{CurrentSlotNumber}($t_{now}$, $C_{loc},R$)
\begin{enumerate}[{1:}]
\item If $t_{now}\leq 0$
\item    Return 0 \textcolor{color-A6A6A6}{// Genesis round.}
\item End if
\item Get $t_{{\textit{round}_{1}}}$from Genesis Block
\item Set $t_{0}\coloneqq 0$, $sl\coloneqq 0, ep\coloneqq 1$ 
\item If $t_{0}+t_{{\textit{round}_{1}}}*R\geq t_{now}$ :
\item    $sl=sl+$  $\lceil (t_{now}-t_{0})/t_{{\textit{round}_{1}}}\rceil $
\item    Return $sl$ 
\item Else:
\item    $t_{0}\coloneqq t_{0}+t_{{\textit{round}_{1}}}*R$ 
\item    $sl=sl+\mathrm{R}$ 
\item End if
\item While $t_{0}\leq t_{now}$: do
\item    Invoke \textbf{AdjustingNextRoundLength}($C_{loc}^{\prime}$) to get $t_{{\textit{round}_{ep}}}$;$C_{loc}^{\prime}$ is $C_{loc}$ part before ep.
\item    If $t_{0}+t_{{\textit{round}_{ep}}}*R\leq t_{now}$:
\item       $t_{0}=t_{0}+t_{{\textit{round}_{ep}}}*R$ 
\item       $sl=sl+\mathrm{R}$ 
\item       $ep=\mathrm{ep}+1$ 
\item    Else:
\item       $sl=sl+$  $\lceil (t_{now}-t_{0})/t_{{\textit{round}_{ep}}}\rceil $
\item       Set $t_{next}\leftarrow t_{0}+(sl-ep*R)*t_{{\textit{round}_{ep}}}$
\item       Return $sl, t_{next}, t_{{\textit{round}_{ep}}}$
\item    End if
\item End while
\end{enumerate}
Output: The protocol output $sl, t_{next}, t_{{\textit{round}_{ep}}}$ to the caller (but not to Z)

\section{ Chain Validation Procedure}

\textbf{IsValidChain}($U_{P},sid, \mathrm{k},C_{i},R,f$); 
\begin{enumerate}[{1:}]
\item If C contains future blocks, empty epochs, starts with a block other than G, or encodes an invalid state with isvalidstate($\overrightarrow{st}$) =0 then
\item Return false

End if
\item For each epoch ep do
\item    Set $S_{ep}^{C}$ to be takeholder distribution at the end of epoch ep -2 in C
\item    Set $\alpha _{p'}^{ep,C} $  to be the relative stake of any party $U_{p'}$ in $S_{ep}^{C}$ and $T_{p'}^{ep,C}$${\leftarrow}$ $2^{{l_{VRF}}}\varnothing _{P}$($\alpha _{p'}^{ep,C}$).
\item    $\eta _{ep}^{C}$${\leftarrow}$ H($\eta _{ep-1}^{C}${\textbar}{\textbar}ep{\textbar}{\textbar}v) where v is the concatenation of the VRF outputs $y_{\rho }$ from all     blocks in C from the first two-thirds slots of epoch ep-1, and $\eta _{1}^{C}$ $=$ $\eta _{1}$ from G
\item    For each block B in C from epoch ep do
\item        Parse B as  $\left(h, st, sl, crt, \rho ,\textcolor{color-2E74B5}{\mathbf{adj}}\textcolor{color-2E74B5}{,}\sigma \right)$
\item        Set badhash ${\leftarrow}$ (h $\neq $ H($B^{-1}$)), where $B^{-1}$ is the last block in C before B ( possibly the Genesis block).
\item        Parse crt as ($U_{P}$, y, $\pi $) for some $p'$.
\item        Send (Verify, sid, $\eta _{ep}\left| \left| sl\right| \right| \text{TEST}, y,\pi ,v_{p'}^{vrf}$) to $F_{VRF}$,

            Denote its response by (Verified, sid,$\eta _{ep}\left| \left| sl\right| \right| TEST, y, \pi $, $b_{1}$).
\item        Send (Verify, sid, $\eta _{ep}\left| \left| sl\right| \right| \text{NONCE}, y_{p},\pi _{p},v_{p'}^{vrf}$) to $F_{VRF}$,

Denote its response by (Verified, sid,$\eta _{ep}\left| \left| sl\right| \right| \text{NONCE}, y_{p},\pi _{p}$, $b_{2}$)
\item        Set badvrf ${\leftarrow}$ ($b_{1}=0\cup y\geq T_{p'}^{ep,C}$)
\item        Send (Verify, sid, (h, st,sl ,crt, $\rho $), sl, $\textcolor{color-2E74B5}{\mathbf{adj}}$, $\sigma $, $v_{p'}^{vrf}$) to $F_{KES}$,

            Denote its response by (Verified, sid  (h, st,sl ,crt, $\rho $), sl , $\textcolor{color-2E74B5}{\mathbf{adj}}$ $b_{3}$)
\item        Set badsig ${\leftarrow}$ ($b_{3}=0$)
\item        \textbf{\textcolor{color-2E74B5}{For each }}$\textcolor{color-2E74B5}{(}\textcolor{color-2E74B5}{\textit{\textbf{Adjust}}}_{\textcolor{color-2E74B5}{\mathbf{i}}}\textcolor{color-2E74B5}{,}\textcolor{color-2E74B5}{\boldsymbol{T}}_{{\textcolor{color-2E74B5}{\boldsymbol{adj}}_{\textcolor{color-2E74B5}{\boldsymbol{i}}}}}\textcolor{color-2E74B5}{)}$\textbf{\textcolor{color-2E74B5}{ in }}$\textcolor{color-2E74B5}{\mathbf{adj}}$\textbf{\textcolor{color-2E74B5}{:}}
\item           \textcolor{color-A6A6A6}{//Check if }$\textcolor{color-A6A6A6}{\textit{\textbf{Adjust}}}_{\textcolor{color-A6A6A6}{\mathbf{i}}}$\textcolor{color-A6A6A6}{ is right}
\item \textbf{\textcolor{color-2E74B5}{          Parse }}$\textcolor{color-2E74B5}{\textit{\textbf{Adjust}}}_{\textcolor{color-2E74B5}{\mathbf{i}}}$\textbf{\textcolor{color-2E74B5}{ as }}$\textcolor{color-2E74B5}{(}\textcolor{color-2E74B5}{\boldsymbol{B}}_{{\textcolor{color-2E74B5}{\textit{\textbf{conflict}}}_{\textcolor{color-2E74B5}{\mathbf{i}}}}}\textcolor{color-2E74B5}{,}\textcolor{color-2E74B5}{\boldsymbol{T}}_{{\textcolor{color-2E74B5}{\boldsymbol{recv}}_{\textcolor{color-2E74B5}{\mathbf{i}}}}}\textcolor{color-2E74B5}{,} \textcolor{color-2E74B5}{\boldsymbol{P}}_{\textcolor{color-2E74B5}{\mathbf{i}}}\textcolor{color-2E74B5}{,} \textcolor{color-2E74B5}{\boldsymbol{y}}_{{\rho _{\textcolor{color-2E74B5}{\mathbf{i}}}}}\textcolor{color-2E74B5}{,}\textcolor{color-2E74B5}{\boldsymbol{\pi }}_{{\rho _{\textcolor{color-2E74B5}{\mathbf{i}}}}}\textcolor{color-2E74B5}{)}$
\item           \textbf{\textcolor{color-2E74B5}{If }}$\textcolor{color-2E74B5}{\textit{\textbf{Adjust}}}_{\textcolor{color-2E74B5}{\mathbf{i}}}$\textbf{\textcolor{color-2E74B5}{ appeared on }}$\textcolor{color-2E74B5}{\boldsymbol{C}}_{\textcolor{color-2E74B5}{\boldsymbol{i}}}$\textbf{\textcolor{color-2E74B5}{ before:}}
\item \textbf{\textcolor{color-2E74B5}{               Set }}$\textcolor{color-2E74B5}{\textbf{badadj}}\textcolor{color-2E74B5}{\leftarrow }\textcolor{color-2E74B5}{\mathbf{true}}$

\textbf{\textcolor{color-2E74B5}{          Else }}
\item \textbf{\textcolor{color-2E74B5}{               Set }}$\textcolor{color-2E74B5}{\textbf{badadj}}\textcolor{color-2E74B5}{\leftarrow }\textcolor{color-2E74B5}{\textbf{false}}$

\textbf{\textcolor{color-2E74B5}{          End if}}
\item \textbf{\textcolor{color-2E74B5}{          Send (Verify, sid, }}$\textcolor{color-2E74B5}{\boldsymbol{\eta }}_{\textcolor{color-2E74B5}{\boldsymbol{ep}}}$\textbf{\textcolor{color-2E74B5}{{\textbar}{\textbar}}}$\textcolor{color-2E74B5}{\boldsymbol{sl}}$\textbf{\textcolor{color-2E74B5}{{\textbar}{\textbar}}}$\textcolor{color-2E74B5}{\textbf{NONCE}}\textcolor{color-2E74B5}{,} \textcolor{color-2E74B5}{\boldsymbol{y}}_{{\rho _{\textcolor{color-2E74B5}{\mathbf{i}}}}}\textcolor{color-2E74B5}{,}\textcolor{color-2E74B5}{\boldsymbol{\pi }}_{{\rho _{\textcolor{color-2E74B5}{\mathbf{i}}}}}\textcolor{color-2E74B5}{,}\textcolor{color-2E74B5}{\boldsymbol{v}}_{\textcolor{color-2E74B5}{\boldsymbol{p}}\textcolor{color-2E74B5}{\mathbf{'}}}^{\textcolor{color-2E74B5}{\boldsymbol{vrf}}}$\textbf{\textcolor{color-2E74B5}{)}}
\item \textbf{\textcolor{color-2E74B5}{               Denote its response by (Verified, sid,}}$\textcolor{color-2E74B5}{\boldsymbol{\eta }}_{\textcolor{color-2E74B5}{\boldsymbol{ep}}}\left| \left| \textcolor{color-2E74B5}{\boldsymbol{sl}}\right| \right| \textcolor{color-2E74B5}{\textbf{NONCE}}\textcolor{color-2E74B5}{,}\textcolor{color-2E74B5}{\boldsymbol{y}}_{{\rho _{\textcolor{color-2E74B5}{\mathbf{i}}}}}\textcolor{color-2E74B5}{,}\textcolor{color-2E74B5}{\boldsymbol{\pi }}_{{\rho _{\textcolor{color-2E74B5}{\mathbf{i}}}}}$\textbf{\textcolor{color-2E74B5}{,}} $\textcolor{color-2E74B5}{\boldsymbol{b}}_{{\textcolor{color-2E74B5}{\mathbf{4}}_{\textcolor{color-2E74B5}{\boldsymbol{i}}}}}$\textbf{\textcolor{color-2E74B5}{)}}
\item \textbf{\textcolor{color-2E74B5}{          Set }}$\textcolor{color-2E74B5}{\textbf{badadj}}\textcolor{color-2E74B5}{\leftarrow }\textcolor{color-2E74B5}{\textbf{badadj}}\cup \textcolor{color-2E74B5}{(}\textcolor{color-2E74B5}{\boldsymbol{b}}_{{\textcolor{color-2E74B5}{\mathbf{4}}_{\textcolor{color-2E74B5}{\boldsymbol{i}}}}}\textcolor{color-2E74B5}{=}\textcolor{color-2E74B5}{\mathbf{0}}\textcolor{color-2E74B5}{)}$

\textbf{\textcolor{color-2E74B5}{       End for}}
\item        If  ($\text{badhash}\cup \text{badvrf}\cup \text{badsig}\cup \textcolor{color-2E74B5}{\textbf{badadj}}$) then
\item           Return false

       End if

   End for

End for
\item Return true
\end{enumerate}
\section{ Read State Procedure}

\textbf{ReadState}($\mathrm{k}, C_{loc},U_{P},\mathrm{sid},R,f$)
\begin{enumerate}[{1:}]
\item If isInit is false output the empty state (READ, sid, $\upvarepsilon $) to Z.
        Otherwise, do the following:
\item Invoke \textbf{FetchInformation}($U_{P}$, sid) to receive the newest messages for this round; denote the output by $tx\leftarrow (tx_{1},\ldots ,tx_{k})$, $N\leftarrow \left\{\left(C_{1}, D_{{mid_{1}}}\right),\ldots ,\left(C_{m},D_{{mid_{m}}}\right)\right\}$, $adj\leftarrow $ $\{(\textit{Adjust}_{1}, D_{{mid_{1}}}),\ldots ,(\textit{Adjust}_{\mathrm{n}},D_{{mid_{n}}})\}$; Record if WELCOME=1 (for late use).
\item Invoke \textbf{UpdateTime}($U_{P})$ to update $t_{now}$ sl, ep
\item Set $t_{on}\leftarrow sl$
\item Invoke \textbf{SelectChain}$(U_{P},\mathrm{sid} ,C_{loc}, N,k ,s, R, f)$
\item Extract the state $\overrightarrow{st}$ from the current local chain $C_{loc}$
\item Output (Read, sid, $\overrightarrow{st}^{\left\lceil k\right\rceil}$) to Z
\end{enumerate}
\chapter{ The Simulator}

{\centering \textbf{Simulator }$\boldsymbol{S}_{\boldsymbol{ledg}}$\textbf{ (Part 1 \textendash{} Main Structure)}

\par}The simulator is slightly different from {[}15{]}.
        Adversary running at the same time with the simulated parties during every round and also in a black box way.

\textbf{Overview:}
\begin{itemize}
\item[-] The simulator internally emulates all local UC functionalities by running the code (and keeping the state) of $\mathrm{F}_{\mathrm{VRF}}$, $\mathrm{F}_{\mathrm{KES}}$, $\mathrm{F}_{\text{INIT}}$, $F_{N-MC}^{bc}$ , $F_{N-MC}^{tx}$, and $F_{N-MC}^{adj}$ . 
\item[-] The simulator mimics the execution of Ouroboros-Genesis for each honest party $\mathrm{U}_{\mathrm{P}}$ (including their state and the interaction with the hybrids).
\item[-] The simulator emulates a view towards the adversary A in a black-box way, i.e., by internally running adversary A and simulating his interaction with the protocol (and hybrids) as detailed below for each hybrid.
        To simplify the description, we assume A does not violate the requirements by the wrapper $\mathrm{W}_{\mathrm{OG}}^{\mathrm{PoS}}(\cdot )$ as this would imply no interaction between $\mathrm{S}_{\text{ledger}}$ (i.e., the emulated hybrids) and A.
\item[-] For global functionalities, the simulator simply relays the messages sent from A to the global functionalities (and returns the generated replies).
        Recall that the ideal world consists of the dummy parties, the ledger functionality, the clock, and the global random oracle.
\item[-] The running time of simulate functions (e.g. $F_{N-MC}$, SimulateMaintence), and commands(quarries to the clock $G_{Auto-\textit{Clock}}$, the ledger $G_{\textit{LEDGER}}$) should be the same as the running time of those in the real world.
        Environment can’t tell the difference between real world and the simulator just by the time difference.
\end{itemize}
\textbf{Party sets:}
\begin{itemize}
\item[-] As defined in the main body of this paper, honest parties are categorized.
        We denote $S_{\textit{alert}}$ the alert parties (synchronized and executing the protocol) and use $S_{\text{syncStalled}}$ shorthand for parties that are synchronized (and hence time aware and online) but stalled.
        Finally, we denote by $P_{DS}$ all honest but de-synchronized parties (both operational or stalled). 
\item[-] For each registered honest party, the simulator maintains the local state containing in particular the local chain $\mathrm{C}_{loc}^{(U_{p})}$, the time $t_{on}$ it remembers when last being online.
        For every participant, it maintains variable $\mathrm{t}_{\textit{begin}}^{(U_{p})}$, $\mathrm{t}_{next}^{(U_{p})}$ as the begin and the end time of this round and $\mathrm{t}_{\textit{round}}^{(U_{p})}$ as the length of this round ($t_{next}=t_{\textit{round}}+t_{\textit{begin}}$ and initially each $t_{next}=t_{\textit{round}}=t_{\textit{begin}}=0$.
        For each party $U_{p}$, the simulator stores flags $\text{updateState}_{P,{\uptau _{L}},}$, $\text{updateTime}_{P,{\uptau _{L}},}$, and $\text{updateInitTime}_{P,{\uptau _{L}},}$  (initially \textit{false}) to remember whether this party has completed its core maintenance tasks in (objective) round $sl=\uptau _{L}$ to update the state and its time (where the initial time for each party is a separate case), respectively.
        Note that a registered party is registered with all its local hybrids. 
\item[-] Upon any activation, the simulator will query the current party set from the ledger, the clock, and the random oracle to evaluate in which category an honest party belongs to.
        If a new honest party is registered to the ledger, it internally runs the initialization procedure of \textit{Ouroboros-AutoSyn} and update 
\item[-] We assume that the simulator queries upon any activation for the sequence $\vec{I}_{H}^{T}$, and maintains a clock functionality($sl$).
        We note that the simulator is capable of determining predict-time(·) of $G_{\textit{LEDGER}}$
\item[-] Simulator maintains a clock functionality (current logical slot number $sl$) and stores $T_{\textit{round}}$ current logical round length, $T_{\textit{begin}}$ the begin time of current logical round, $T_{next}$ the end time of current logical round,.
        Initially, $T_{\textit{round}}=t_{{\textit{round}_{1}}}$, $T_{\textit{begin}}=0$, $T_{next}=t_{{\textit{round}_{1}}},sl=0$.
        And $\textit{\textbf{updatecompleted}}_{sl}$ is used to monitor the round simulation of clock (initially false) and will be set true once any alert party has updated its local round number.
        These parameters above stores clock information and monitors the execution of local clock function. 
\end{itemize}
\textbf{Message from the Ledger}
\begin{itemize}
\item[-] Upon receiving (\textit{submit}, BTX) from $\mathrm{G}_{\textit{LEDGER}}$ where BTX := (tx, txid, ${\uptau}$, $\mathrm{U}_{\mathrm{P}}$) forward (\textit{multicast}, sid, tx) to the simulated network $\mathrm{F}_{N-MC}$ in the name of $\mathrm{U}_{\mathrm{P}}$.
        Output the answer of $\mathrm{F}_{N-MC}$ to the adversary
\item[-] Upon receiving (\textit{maintain-ledger}, sid, minerID) from $G_{\textit{LEDGER}}$ extract from $\vec{I}_{H}^{T}$ the party $U_{P}$ that issued this query.
        If $U_{P}$ has already completed its round-task, then ignore this request.
        Otherwise, execute SimulateMaintence ($U_{P},$ $t_{now}$).
\item[-] Upon receiving(time-read, $sid_{C}$) from $G_{\textit{LEDGER}}$, send (clock-read, $sid_{C}$) to $G_{\textit{AutoClock}}$ get current time $t_{now}$ {\ldots} and send (time-read, $sid_{C}$) to $G_{\textit{LEDGER}}$

{\centering \textbf{Simulator }$\boldsymbol{S}_{\boldsymbol{ledg}}$\textbf{ (Part 2 \textendash{}Black-Box Interaction)}\par}
\end{itemize}
Simulation of Functionality $F_{INIT}$ towards A
\begin{itemize}
\item[-] The simulator relays back and forth the communication between the (internally emulated) $\mathrm{F}_{\text{INIT}}$ functionality and the adversary \textit{A }acting on behalf of a corrupted party
\item[-] If at time $sl=0$ ($t_{now}\leq 0$), a corrupted party $U_{p}\in S_{\textit{initStake}}$ registers via (ver\_keys, sid, $U_{p}$,$\mathrm{v}_{U_{p}}^{\mathrm{vrf}}$, $\mathrm{v}_{U_{p}}^{\mathrm{kes}}$ ) to $F_{INIT}$, then input( register,sid) to $G_{\textit{LEDGER}}$ on behalf of $U_{p}$.
\end{itemize}
Simulation of Functionalities $F_{KES},F_{VRF}$ towards A
\begin{itemize}
\item[-] The simulator relays back and forth the communication between the (internally emulated) hybrids and the adversary A (either direct communication, communication to A caused by emulating the actions of honest parties, or communication of A on behalf of a corrupted party). 
\end{itemize}
Simulation of Network $F_{N-MC}^{bc}$ (over which chains are sent) towards A:
\begin{itemize}
\item[-] Upon receiving (multicast, sid,($C_{{i_{1}}}$, $U_{{i_{1}}}$),...,($C_{{i_{l}}}$, $U_{{i_{l}}}$))with a list of chains with corresponding parties from A (or on behalf some corrupted $P\in P_{net}$), then 
\begin{enumerate}
\item Relay this input to the simulate Network functionality and record its response to A
\end{enumerate}

\begin{enumerate}
\setcounter{enumii}{1}
\item Provide A with the recorded output of the simulated network
\end{enumerate}

\item[-] Upon receiving  (multicast, sid, C) from A on behalf of some corrupted party P, then 
\begin{enumerate}
\item Relay this input to the simulate Network functionality and record its response to A
\item Provide A with the recorded output of the simulated network
\end{enumerate}

\item[-] Upon receiving (fetch, sid) from A on behalf of some corrupted party $P\in P_{net}$, forward the request to the simulated $F_{N-MC}^{bc}$ and return whatever is returned to A.
\item[-] Upon receiving (delays, sid, (${T_{mid}}_{{i_{1}}}$, $mid_{{i_{1}}}$),...,(${T_{mid}}_{{i_{l}}}$, $mid_{{i_{l}}}$)) from A, forward the request to the simulated $F_{N-MC}^{bc}$ and return whatever is returned to A.
\item[-] Upon receiving (swap, sid, mid, $\mathrm{mid}\mathrm{'}$) from A, forward the request to the simulated $F_{N-MC}^{bc}$ and return whatever is returned to A.
\end{itemize}
Simulation of Network $F_{N-MC}^{tx}$ (over which chains are sent) towards A:
\begin{itemize}
\item[-] Upon receiving (multicast, sid,($m_{{i_{1}}}$, $U_{{i_{1}}}$),...,($m_{{i_{l}}}$, $U_{{i_{l}}}$))with a list of transactions from A on behalf some corrupted $P\in P_{net}$, then do the following:
\begin{enumerate}
\item Submit the transaction(s) to the ledger on behalf of this corrupted party, and receive for each transaction the transaction id txid
\item Forward the request to the internally simulated $F_{N-MC}^{tx}$, which replies for each message with a message-ID mid
\item Remember the association between each mid and the corresponding txid.
\item Provide A with whatever the network outputs.
\end{enumerate}

\item[-] Upon receiving (multicast, sid, m) from A on behalf some corrupted $P$, then execute the corresponding steps 1.
        To 4. above.
\item[-] Upon receiving (fetch, sid) from A on behalf of some corrupted party $P\in P_{net}$, forward the request to the simulated $F_{N-MC}^{tx}$ and return whatever is returned to A.
\item[-] Upon receiving (delays, sid, (${T_{mid}}_{{i_{1}}}$, $mid_{{i_{1}}}$),...,(${T_{mid}}_{{i_{l}}}$, $mid_{{i_{l}}}$)) from A, forward the request to the simulated $F_{N-MC}^{tx}$ and return whatever is returned to A.
\item[-] Upon receiving (swap, sid, mid, $\mathrm{mid}\mathrm{'}$) from A, forward the request to the simulated $F_{N-MC}^{bc}$ and return whatever is returned to A.
\end{itemize}
{\centering \textbf{Simulator }$\boldsymbol{S}_{\boldsymbol{ledg}}$\textbf{ (Part 3 \textendash{}Internal Procedures)}

\par}\textbf{procedure} \textbf{SimulateMaintence}($U_{p}$\textit{, }$t_{now}$)\newline

Simulate the (in the UC interruptible manner) the maintenance procedure of party P as in the protocol at time $t_{now}$ global round $sl=\uptau _{L}$ ($t_{now}$ may change as protocol simulating, need to ask $G_{\textit{AutoClock}}$ multiple times)i.e., run \textbf{LedgerMaintenance }(·) for this simulated party.
\begin{enumerate}[{1:}]
\item If party P gives up activation then 
\item    If party P has reached the instruction FinishRound(P, $\mathrm{t}_{next}^{(U_{p})}$) at time $t_{now}$ then 
\item       Ask $G_{\textit{AutoClock}}$ for current time if $t_{now}< \mathrm{t}_{next}^{(U_{p})}$ return activation to A.
\item    End if 
\item    If party P has completed JoinProc(·) and $\text{updateInitTime}_{P,{\uptau _{L}},}$ is false then 
\item       Execute AdjustTime($U_{P}$\textit{,} $t_{now}$) and then set $\text{updateInitTime}_{P,{\uptau _{L}},}\leftarrow \text{true}$ 
\item    End if 
\item    If party P has reached the instruction SelectChain(·) and $\text{updateState}_{P,{\uptau _{L}},}$ is false then Execute ExtendLedgerState($\uptau _{L},t_{now}$) and then set $\text{updateState}_{P,{\uptau _{L}},}\leftarrow \text{true}$.
\item    End if
\item    If party P has reached the instruction UpdateTime(·) and $\text{updateTime}_{P,{\uptau _{L}}}$ is false then Execute AdjustTime($U_{P}$\textit{,} $t_{now}$) and then set $\text{updateTime}_{P,{\uptau _{L}}}\leftarrow \text{true}$ 
\item    End if
\item    Return activation to A 
\item End if
\end{enumerate} 

\textbf{End procedure}

\textbf{Procedure} ExtendLedgerState(\textit{${\tau}$,} $t_{now}$)
\begin{enumerate}[{1:}]
\item For each synchronized party $U_{p}$ ${\in}$$S_{\textit{alert}}$${\cup}$$S_{\text{syncStalled}}$ of round ${\tau}$ do
\item    Let $\mathrm{C}_{loc}^{(U_{p})}$ be the party's currently stored local chain.
\item    \textcolor{color-2E74B5}{ Let }$\textcolor{color-2E74B5}{\mathrm{C}}_{\textcolor{color-2E74B5}{1}}^{\textcolor{color-2E74B5}{(}\textcolor{color-2E74B5}{U}_{\textcolor{color-2E74B5}{p}}\textcolor{color-2E74B5}{)}}$\textcolor{color-2E74B5}{ ,{\ldots}, }$\textcolor{color-2E74B5}{\mathrm{C}}_{\textcolor{color-2E74B5}{k}}^{\textcolor{color-2E74B5}{(}\textcolor{color-2E74B5}{U}_{\textcolor{color-2E74B5}{p}}\textcolor{color-2E74B5}{)}}$\textcolor{color-2E74B5}{ be the chains contained in the receiver buffer }$\vec{\textcolor{color-2E74B5}{M}}^{\left(\textcolor{color-2E74B5}{U}_{\textcolor{color-2E74B5}{p}}\right)}$\textcolor{color-2E74B5}{ of} $\textcolor{color-2E74B5}{F}_{\textcolor{color-2E74B5}{N}\textcolor{color-2E74B5}{-}\textcolor{color-2E74B5}{M}\textcolor{color-2E74B5}{C}}^{\textcolor{color-2E74B5}{b}\textcolor{color-2E74B5}{c}}$ \textcolor{color-2E74B5}{with delivery time }$\textcolor{color-2E74B5}{D}_{\textcolor{color-2E74B5}{m}\textcolor{color-2E74B5}{i}\textcolor{color-2E74B5}{d}}\textcolor{color-2E74B5}{\leq }\textcolor{color-2E74B5}{t}_{\textcolor{color-2E74B5}{n}\textcolor{color-2E74B5}{o}\textcolor{color-2E74B5}{w}}$\textcolor{color-2E74B5}{.}
\item    Evaluate $C_{{U_{p}}}\leftarrow $ maxvalid-bg($\mathrm{C}_{loc}^{(U_{p})}$, $\mathrm{C}_{1}^{(U_{p})},\ldots , \mathrm{C}_{k}^{(U_{p})}$)
\item End for
\item Let $\overrightarrow{st}$ be the longest state among all such states $\overrightarrow{st}_{{U_{p}}}$, $U_{p}$ ${\in}$$S_{\textit{alert}}$${\cup}$$S_{\text{syncStalled}}$.
\item Compare $\overrightarrow{st}^{\left\lceil k\right\rceil}$ with the current state \textit{state }of the ledger
\item If {\textbar}state{\textbar}{\textgreater}$\overrightarrow{st}^{\left\lceil k\right\rceil}$ then 
\item    Execute AdjustView(state)
\item End if 
\item Define the difference \textit{diff }to be the block sequence s.t. state{\textbar}{\textbar}diff = $\overrightarrow{st}^{\left\lceil k\right\rceil}$.
\item Parse diff := $\text{diff}_{1}${\textbar}{\textbar} \textit{. . . }{\textbar}{\textbar}$\text{diff}_{\mathrm{n}}$
\item For \textit{j }= 1 to \textit{n }do
\item    Map each transaction tx in this block to its unique transaction ID txid.
\item    If a transaction does not yet have a txid, then submit it to the ledger first and receive the corresponding txid from $\mathrm{G}_{\textit{ledger}}$
\item    Let $list_{j}$ = ($txid_{j,1}$\textit{, . . . ,} $txid_{j,{l_{j}}}$) be the corresponding list for this block $\text{diff}_{\mathrm{j}}$
\item    If coinbase $txid_{j,1}$ specifies a party honest at block creation time then
\item       $\text{hFlag}_{\mathrm{j}}$ \textit{${\leftarrow}$ }1
\item    Else
\item       $\text{hFlag}_{\mathrm{j}}$ \textit{${\leftarrow}$ }0
\item    End if
\item    Output (\textit{next-block, }$\text{hFlag}_{\mathrm{j}}$\textit{, }$\text{list}_{\mathrm{j}}$) to \textit{Gledger }(receiving (\textit{next-block, ok}) as an immediate answer)
\item End for
\item If \textit{Fraction of blocks with hFlag = 0 in the recent }k \textit{blocks }{\textgreater} \textit{1 }- ${\mathrm{\mu}}$ then
\item    Abort \textit{simulation: chain quality violation. // Event }$\mathrm{BAD}-\mathrm{CQ}_{\mu ,k}$
\item Else if \textit{State increases less than }k \textit{blocks during the last }$\frac{k}{\tau _{CG}}$ \textit{rounds }then
\item    Abort \textit{simulation: chain growth violation. // Event }$\mathrm{BAD}-\mathrm{CQ}_{{\tau _{CG}},k/{\tau _{CG}}}$
\item End if
\item \textcolor{color-A6A6A6}{// If no bad event occurs, we can adjust pointers into this new state.}
\item Execute AdjustView(state{\textbar}{\textbar}diff)
\end{enumerate}
\textbf{End procedure}

\textbf{Procedure} AdjustTime($U_{P}$\textit{,} $t_{now}$)
\begin{enumerate}[{1:}]
\item Simulate \textbf{UpdateTime}($U_{P},R,\mathrm{C}_{loc}^{(U_{p})},\textcolor{color-2E74B5}{t}_{\textcolor{color-2E74B5}{r}\textcolor{color-2E74B5}{u}\textcolor{color-2E74B5}{n}})$ function for $U_{p}$, using $\mathrm{C}_{loc}^{(U_{p})}$. 
\item Denote current slot number for this party as $sl^{({U_{p}})}$, current round length for this party as $\mathrm{t}_{\textit{round}}^{(U_{p})}$.
\item If $\textit{\textbf{updatecompleted}}_{sl}=0$ then
\item    Set $sl=sl+1$
\item    If $sl mod R=0$ then 
\item       Set $t_{\textit{round}}=\mathrm{t}_{\textit{round}}^{(U_{p})}$
\item    End if
\item    Set $t_{\textit{begin}}=t_{next}$, $t_{next}=t_{\textit{round}}+t_{\textit{begin}}$
\item    Set $\textit{\textbf{updatecompleted}}_{sl}=\mathbf{1}$
\item Else if $sl\neq sl^{\left(U_{p}\right)}$ or $t_{\textit{round}}\neq $ $\mathrm{t}_{\textit{round}}^{(U_{p})}$
\item    If $U_{p}\in S_{\textit{alert}}$ then abort simulation: round-synchrony violation.
\item    End if 
\item End if 
\end{enumerate}
\textbf{End procedure}

\chapter{D proof of stake assumption as a UC wrapper}

{\centering \textbf{Functionality} $W_{OG}^{PoS}(\cdot )$

\par}The wrapper functionality is parameterized by the bounds ${\alpha}$,\textit{${\beta}$ }on the alert and participating stake ratio (see Definition 2 from {[}15{]}).
        The wrapper is assumed to be registered with the global clock $\mathrm{G}_{\text{CLOCK}}$ and is aware of sets of registered parties, and the set of corrupted parties and message delivery ratio of each round.

\textit{General:}
\begin{itemize}
\item[-] Upon receiving any request I from any party $\mathrm{U}_{\mathrm{P}}$ or from A (possibly on behalf of a party $\mathrm{U}_{\mathrm{P}}$ which is corrupted) to a wrapped hybrid functionality, record the request I together with its source and the current time.
\item[-] The wrapper keeps track of the active parties and their relative share to the stake distribution.
\end{itemize}
\textit{Restrictions on obtaining VRF proofs:}
\begin{itemize}
\item[-] Upon receiving (\textit{EvalProve}, sid, ·) to $\mathrm{F}_{VRF}$ from A on behalf of a party $\mathrm{U}_{\mathrm{P}}$ which is corrupted or registered but de-synchronized do the following:
\begin{enumerate}
\item If the fraction of alert stake relative to all active stake in this round \textit{${\tau}$ during }$\textcolor{color-2E74B5}{t}_{\textcolor{color-2E74B5}{r}\textcolor{color-2E74B5}{u}\textcolor{color-2E74B5}{n}}$ \textcolor{color-2E74B5}{period} so far does not satisfy the honest majority condition 4 (of Theorems 1 and 2) then ignore the request.
\item Otherwise, forward the request to $\mathrm{F}_{VRF}$ and return to A whatever $\mathrm{G}_{RO}$ returns
\end{enumerate}

\end{itemize}

\begin{itemize}
\item[-] Upon receiving (\textit{EvalProve}, sid, ·) to $\mathrm{F}_{VRF}$ from an alert party $\mathrm{U}_{\mathrm{P}}$ do the following:
\begin{enumerate}
\item Forward the request to $\mathrm{F}_{VRF}$ and return to \textit{A }whatever $\mathrm{G}_{RO}$ returns.
\item If the minimal fraction (in stake) of participation (of alert parties running procedure SimulateMaintence and in total) as demanded by Theorem 1 (and Theorem 2) has not been reached \textit{during }$\textcolor{color-2E74B5}{t}_{\textcolor{color-2E74B5}{r}\textcolor{color-2E74B5}{u}\textcolor{color-2E74B5}{n}}$ \textcolor{color-2E74B5}{period}, halt and outputs error.
\item If the minimal fraction (in stake) of alert parties message send success ratio $\upeta $ as demanded by Theorem 1 (and Theorem 2) has not been reached at the end of this round ($\textcolor{color-2E74B5}{t}_{\textcolor{color-2E74B5}{n}\textcolor{color-2E74B5}{e}\textcolor{color-2E74B5}{x}\textcolor{color-2E74B5}{t}}$\textcolor{color-2E74B5}{)}, halt and outputs error.
\end{enumerate}

\begin{itemize}
\item[-] Any other request is relayed to the underlying functionality (and recorded by the wrapper) and the corresponding output is given to the destination specified by the underlying functionality.
\end{itemize}

\end{itemize}
\chapter{ Proof of Theorem}

\section{ The Reduction Mapping}

\textbf{Characteristic string under message delay.}

\textbf{Lemma 3. }In the light of ({[}15{]}, Definition 8 \textbf{characteristic string}) $\mathrm{W}={\mathrm{W}_{1}},\ldots ,\mathrm{W}_{\mathrm{r}}$, the characteristic string induced by the protocol \textbf{\textit{Ouroboros-AutoSyn }}in the single-epoch setting over a sequence of \textit{r }slots under message delivery ratio $\eta $, denoted as $W^{r}~ .$ Then we have $W^{r}=\rho _{r}(W)$

\textbf{\textit{Proof.}} Consider the characteristic string of 2 consecutive slots by the protocol Ouroboros-Praos in the single-epoch setting, if message delay happens at the first slot, there are 4 cases of situation,

$\mathbf{00}\colon $ When delay happens, we have either $\mathbf{00}\rightarrow \mathbf{0}\bot $ or $\mathbf{00}\rightarrow \bot \mathbf{0}$, depending on the network situation

$\mathbf{01}\colon $ we have $\mathbf{01}\rightarrow \mathbf{0}\bot $ or $\mathbf{01}\rightarrow \bot \mathbf{1}$, but under the assumption that adversary’s message will not be delayed and faster than honest party, so we get $\mathbf{01}\rightarrow \bot \mathbf{1}$ 

$\mathbf{10}\colon $ we have $\mathbf{10}\rightarrow \mathbf{1}\bot $ or $\mathbf{10}\rightarrow \bot \mathbf{0}$, same as above, so we get $\mathbf{10}\rightarrow \mathbf{1}\bot $ 

$\mathbf{11}\colon $ we have $\mathbf{11}\rightarrow \mathbf{1}\bot $ or $\mathbf{11}\rightarrow \bot \mathbf{1}$, same as above, we get $\mathbf{11}\rightarrow \mathbf{11}$

Specially, for $\mathbf{000}$, if message delay happens at the first two slots, we can conclude that $\mathbf{000}\rightarrow \mathbf{0}\bot \mathbf{0}$ different than case ``00'' or $\mathbf{000}\rightarrow \bot \bot \mathbf{0}$.

And for $\mathbf{0}\bot \ldots \bot $ or $\mathbf{1}\bot \ldots \bot $ the more $\bot $ appears after $\mathbf{1}$ or $\mathbf{0}$ the more likely $\mathbf{1}$ or $\mathbf{0}$ will be accepted by other parties.
        So we have $\mathbf{1}\bot \rightarrow \mathbf{1}\bot $ or $\mathbf{0}\bot \rightarrow \mathbf{0}\bot $ with probability more than $\boldsymbol{\eta }$\textbf{.}

In the light of discussion above we have $\rho _{r}\left(0| | w\mathrm{'}\right)=0| | \rho _{r}\left(w\mathrm{'}\right)$

\textbf{Definition 3 (}$\bot $\textbf{-Reduction mapping)}\textbf{. }We define the function $\rho _{\bot }$\textit{:} $\{0,1,\bot \}^{\mathrm{*}}\rightarrow \{0,1\}^{\mathrm{*}}$ inductively as follows: 

\begin{align*}
\rho _{\bot }\left(\epsilon \right)&=\epsilon , \\
\rho _{\bot }\left(\bot | | w'\right)&=\rho _{\bot }\left(w'\right), 
\end{align*}

\textbf{Lemma 4 (Structure of the induced distribution without boundary conditions)}. ({[}15{]}, Lemma 6).
        Let $\mathrm{W}=W_{1},W_{2},\ldots $  be an infinite sequence of random variables, each taking values in \{${\bot}$,0,1\}, which satisfy the $(f;\upgamma )$-characteristic conditions.
        Let $W^{r}=W_{1}^{r},\ldots $ be a family of variables, taking values in $\{\mathbf{0},\mathbf{1},\bot \}$  satisfying the $W^{r}$= $\rho _{r}$ (W) and let $\mathrm{X}=\rho _{\bot }(W^{r})$.
        Then $\mathrm{X}=X_{1},\ldots $ satisfy the $\gamma \eta $-martingale conditions.

\textit{Proof}: With$\mathrm{W}=W_{1},W_{2},\ldots $  taking values in \{${\bot}$, 0, 1\}, and satisfying the $(f;\upgamma )$-characteristic conditions, we have 
\begin{equation*}
\Pr \left[\mathrm{W}_{\mathrm{k}}=0\right| \mathrm{W}_{1},..., \mathrm{W}_{\mathrm{k}-1},\mathrm{W}_{\mathrm{k}}\neq \bot ]\geq \upgamma 
\end{equation*}
\textbf{And }
\begin{equation*}
\Pr \left[W_{k}^{r}=0\right| W_{i}^{r}=w_{1}^{r}]=\Pr \left[\mathrm{W}_{l}=0\right| \mathrm{W}_{1},..., \mathrm{W}_{l-1},\mathrm{W}_{l}\neq \bot ]\times \Pr [\mathrm{W}_{l}\text{successfully sent}]
\end{equation*}
Thus, it follows that for any fix values $\mathrm{x}_{1},..., \mathrm{x}_{\mathrm{k}-1}$ and $\mathrm{w}_{1},..., \mathrm{w}_{\mathrm{k}-1}$
\begin{equation*}
\Pr \left[\mathrm{X}_{\mathrm{k}}=0\right| \mathrm{X}_{\mathrm{k}}=\mathrm{x}_{\mathrm{k}-1}]\geq \Pr \left[W_{k}^{r}=0\right| W_{k-1}^{r}=w_{k-1}^{r}]\geq \gamma \eta 
\end{equation*}
\textbf{Lemma 5}\textbf{\textit{ (Structure of the induced distribution).}} ({[}15{]}, Lemma 8) \textit{Let }$\textcolor{color-2E74B5}{\boldsymbol{W}}^{\textcolor{color-2E74B5}{\boldsymbol{r}}}\textcolor{color-2E74B5}{=}\textcolor{color-2E74B5}{\boldsymbol{W}}_{\textcolor{color-2E74B5}{\mathbf{1}}}^{\textcolor{color-2E74B5}{\boldsymbol{r}}}\textcolor{color-2E74B5}{,}\textcolor{color-2E74B5}{\ldots }\textcolor{color-2E74B5}{,} \textcolor{color-2E74B5}{\boldsymbol{W}}_{\textcolor{color-2E74B5}{\boldsymbol{r}}}^{\textcolor{color-2E74B5}{\boldsymbol{r}}}$\textit{ be a sequence of random variables induced by the protocol }\textbf{Ouroboros-AutoSyn}\textit{, each taking values in \{${\bot}$, 0, 1\}, let}
\begin{equation*}
X=X_{1},\ldots ,X_{l}=\rho _{\bot }\left(W_{1}^{r},\ldots , W_{R}^{r}\right),l> 2
\end{equation*}
\textit{be the random variables obtained by applying the reduction mapping to W .
        Then there is a sequence of random variables }$Z_{1},Z_{2},\ldots $\textit{  each taking values in }$\{0,1\}$\textit{, so that}
\begin{description}
\item[\textit{(i).}]\textit{the random variables }$Z_{1},\ldots $ \textit{satisfy the} $\upgamma \upeta $\textit{-martingale conditions;}
\item[\textit{(ii).}]$X_{1},\ldots ,X_{l-2}=\uprho _{\bot }({\boldsymbol{W}^{\boldsymbol{r}}})^{\lceil 2}$ \textit{is a prefix of }$Z_{1}Z_{2}\ldots $\textit{.} · · · \textit{.}
\end{description}
\textit{Under the further condition that }$\Pr [\textcolor{color-2E74B5}{\boldsymbol{W}}_{\textcolor{color-2E74B5}{\boldsymbol{i}}}^{\textcolor{color-2E74B5}{\boldsymbol{r}}}\textcolor{color-2E74B5}{,}=\bot | \textcolor{color-2E74B5}{\boldsymbol{W}}_{\textcolor{color-2E74B5}{\mathbf{1}}}^{\textcolor{color-2E74B5}{\boldsymbol{r}}},..., \textcolor{color-2E74B5}{\boldsymbol{W}}_{\textcolor{color-2E74B5}{\boldsymbol{i}}\textcolor{color-2E74B5}{-}\textcolor{color-2E74B5}{\mathbf{1}}}^{\textcolor{color-2E74B5}{\boldsymbol{r}}}]\leq (1-\mathrm{a})$\textbf{\textit{, we also have:}}
\begin{description}
\item[\textit{(iii).}]\textit{the random variable }$l$\textit{ satisfies, for any }$\delta > 0$\textit{,}
\begin{equation*}
\mathit{\Pr } \left[l< \left(1-\delta \right)an\right]\leq exp (-\delta ^{2}a^{2}n/2);
\end{equation*}

\item[\textit{(iv).}]finally, if $\textcolor{color-2E74B5}{\gamma }\textcolor{color-2E74B5}{\eta }> (1+\epsilon )/2$ for some $\epsilon \geq 0$ then
\end{description}

\begin{equation}
\tag{13}
\Pr \left[\# _{0}\left(X\right)< \frac{\textcolor{color-2E74B5}{(}\textcolor{color-2E74B5}{1}\textcolor{color-2E74B5}{+}\textcolor{color-2E74B5}{\upepsilon }\textcolor{color-2E74B5}{)}\textcolor{color-2E74B5}{a}\textcolor{color-2E74B5}{n}}{\textcolor{color-2E74B5}{4}}\textcolor{color-2E74B5}{-}\textcolor{color-2E74B5}{2}\right]\leq \exp \left(-\frac{\mathrm{a}^{2}n}{32}\right)+\exp \left(-\frac{an}{64}\right)\leq 2\exp \left(-\frac{\mathrm{a}^{2}n}{64}\right);
\end{equation}
And
\begin{equation}
\tag{14}
\Pr \left[\# _{0}\left(X\right)-\# _{1}\left(X\right)< \frac{\textcolor{color-2E74B5}{\upepsilon }\textcolor{color-2E74B5}{a}\textcolor{color-2E74B5}{n}}{\textcolor{color-2E74B5}{4}}\textcolor{color-2E74B5}{-}\textcolor{color-2E74B5}{4}\right]\leq \exp \left(-\frac{\mathrm{a}^{2}n}{8}\right)+n\exp \left(-\frac{\upepsilon ^{2}an}{64}\right)\leq (\mathrm{n}+1)\exp \left(-\frac{\upepsilon ^{2}\mathrm{a}^{2}n}{64}\right);
\end{equation}
In the light of lemma 4 and lemma 5, and with theorem 4 from {[}15{]}, we have the following.

\textbf{Theorem 4}.
        Let $\boldsymbol{W}^{\boldsymbol{r}}=\boldsymbol{W}_{\mathbf{1}}^{\boldsymbol{r}},\ldots ,~\boldsymbol{W}_{\boldsymbol{R}}^{\boldsymbol{r}}$ be a family of random variables, taking values in $\{\mathbf{0},\mathbf{1},~ \bot \}$ and satisfying the $\boldsymbol{W}^{\boldsymbol{r}}$= $\boldsymbol{\rho }_{\boldsymbol{r}}$ (W).If $\boldsymbol{\varepsilon }> \mathbf{0}$ and message delivery ratio $\boldsymbol{\eta }$, satisfy $\boldsymbol{\gamma \eta }\geq (\mathbf{1}+\boldsymbol{\varepsilon })/\mathbf{2}$ then 
\begin{equation*}
\Pr \left[div_{\Delta }\left(W^{r}\right)\geq k+2\right]\leq \frac{19R}{\varepsilon ^{4}}\exp (-\varepsilon ^{4}k/18)
\end{equation*}
Proof .We let $X=\rho _{\bot }(W^{r})$ 
\begin{equation*}
div_{\Delta }\left(W^{r}\right)=div_{0}\left(X\right)
\end{equation*}
And with $div_{0}\left(xy\right)\leq div_{0}\left(x\right)+| y| $, we have
\begin{equation*}
div_{0}\left(X\right)\leq div_{0}\left(X^{\lceil 2}\right)+2
\end{equation*}
By applying $\mathrm{Z}_{1},\ldots $from Lemma 8 we have
\begin{equation*}
div_{0}\left(X^{\lceil 2}\right)+2\leq div_{0}\left(\mathrm{Z}_{1},\ldots ,\mathrm{Z}_{\mathrm{R}}\right)+2
\end{equation*}
Applying Theorem 4, together, we have 
\begin{equation*}
\Pr \left[div_{\Delta }\left(W^{r}\right)\geq k+2\right]\leq \frac{19R}{\varepsilon ^{4}}\exp (-\varepsilon ^{4}k/18)
\end{equation*}
E.2 Distribution of Characteristic Strings in a Single Epoch

\textbf{Lemma 6.} \textit{The protocol }\textbf{\textit{Ouroboros-AutoSyn}}\textit{, when executed in the single-epoch setting, induces characteristic strings} $W_{1}^{r},\ldots , W_{R}^{r}$\textit{ (with each }$W_{t}^{r}\in \{0,1,\bot \}$\textit{) satisfying,} ${\upalpha}$ \textit{is a lower-bound on the alert stake ratio over the execution and,}
\begin{equation*}
\Pr [W_{t}^{r}=0| W_{1}^{r},..., W_{t-1}^{r}]=\Pr [\mathrm{W}_{\mathrm{k}}=0| \mathrm{W}_{1},..., \mathrm{W}_{\mathrm{k}-1}]\cdot \eta \geq \textcolor{color-2E74B5}{\upalpha }\textcolor{color-2E74B5}{(}\textcolor{color-2E74B5}{1}\textcolor{color-2E74B5}{-}\textcolor{color-2E74B5}{f}\textcolor{color-2E74B5}{)}^{\textcolor{color-2E74B5}{2}}\textcolor{color-2E74B5}{\eta }
\end{equation*}
\textbf{In the light of {[}Ouroboros Genesis, }Corollary 2{]}\textbf{.}

Active party is consist of adversary, alert party and other honest party.
\begin{equation*}
\Pr [W_{t}^{r}=\bot | W_{1}^{r},..., W_{t-1}^{r}]\leq 1-f\cdot \mathrm{S}^{-}(\mathrm{P}_{\text{alert}}[t])\cdot \eta -f\cdot {\mathrm{S}^{-}}(\mathrm{P}_{\text{other}}[t])\cdot \eta -f\cdot \mathrm{S}^{-}(\mathrm{P}_{\text{adversary}}[t])\leq 1-f\cdot \mathrm{S}^{-}(\mathrm{P}_{\text{alert}}[t])\cdot \eta -f\cdot {\mathrm{S}^{-}}(\mathrm{P}_{\text{other}}[t])\cdot \eta -f\cdot \mathrm{S}^{-}(\mathrm{P}_{\text{adversary}}[t])\cdot \eta =1-f\cdot \mathrm{S}^{-}(\mathrm{P}_{\text{active}}[t])\cdot \eta 
\end{equation*}
And we got
\begin{equation*}
\Pr [W_{t}^{r}=\bot | W_{1}^{r},..., W_{t-1}^{r}]\leq 1-f\cdot \mathrm{S}^{-}(\mathrm{P}_{\text{active}}[t])\textcolor{color-2E74B5}{\cdot }\textcolor{color-2E74B5}{\eta }
\end{equation*}
$\mathrm{P}_{\text{alert}}[t]$ \textit{denotes the set of }alert\textit{ participants at time t}

$\mathrm{P}_{\text{other}}[t]$ \textit{denotes the set of }other honest \textit{participants at time t}

$\mathrm{P}_{\text{adversary}}[t]$ \textit{denotes the set of }adversarial\textit{ participants at time t}

$\mathrm{P}_{\text{active}}[t]$ \textit{denotes the set of }active\textit{ participants at time t}
\begin{equation*}
\mathrm{P}_{\text{active}}=\mathrm{P}_{\text{alert}}\left[t\right]\cup \mathrm{P}_{\text{other}}\left[t\right]\cup \mathrm{P}_{\text{adversary}}[t]
\end{equation*}
Now we can extend our discussion common prefix quality of the protocol execution in a single epoch setting.

\textbf{Corollary 2 (Common prefix). }\textit{Let }$W^{r}=W_{1}^{r},\ldots , W_{R}^{r}$\textit{ denote the characteristic string induced by the }\textbf{\textit{Ouroboros-AutoSyn}}\textit{ protocol in the single-epoch setting over a sequence of r slots} \textit{under message delivery ratio }$\eta $\textit{.
        Assume that }$\varepsilon > 0$\textit{ satisfies }
\begin{equation*}
\upalpha (1-f)^{2}\eta \geq (1+\varepsilon )/2,
\end{equation*}
\textit{where }${\upalpha}$ \textit{is a lower-bound on the alert stake ratio over the execution.
        Then }
\begin{equation*}
\Pr \left[div_{\Delta }\left(W^{r}\right)\geq k+2\right]\leq \frac{19R}{\varepsilon ^{4}}\exp (-\varepsilon ^{4}k/18)
\end{equation*}
\textit{and hence a }k\textit{-common-prefix violation occurs with probability at most}
\begin{equation*}
\overline{\varepsilon }_{CP}(k;r,\Delta =2,\varepsilon )\triangleq \frac{19r}{\varepsilon ^{4}}\exp (2-\varepsilon ^{4}k/18)
\end{equation*}
\textbf{\textit{Proof. }}The statement is a direct consequence of combining Theorem 7 with Lemma 10.

\textbf{Lemma 7. }\textit{Let }$\textcolor{color-2E74B5}{\boldsymbol{W}}^{\textcolor{color-2E74B5}{\boldsymbol{r}}}\textcolor{color-2E74B5}{=}\textcolor{color-2E74B5}{\boldsymbol{W}}_{\textcolor{color-2E74B5}{\mathbf{1}}}^{\textcolor{color-2E74B5}{\boldsymbol{r}}}\textcolor{color-2E74B5}{,}\textcolor{color-2E74B5}{\ldots }\textcolor{color-2E74B5}{,} \textcolor{color-2E74B5}{\boldsymbol{W}}_{\textcolor{color-2E74B5}{\boldsymbol{r}}}^{\textcolor{color-2E74B5}{\boldsymbol{r}}}$\textit{ be a sequence of random variables induced by the protocol }\textbf{Ouroboros-AutoSyn}\textit{ in the single-epoch setting over a sequence of r slots.
        Let }$\alpha , \beta \in [0,1]$ \textit{denote lower bounds on the alert stake ratio and the participating stake ratio over the execution as per Definition 2, and assume that for some }$\varepsilon \in (0,1)$\textit{ the parameter ${\alpha}$ satisfies}
\begin{equation*}
\upalpha (1-f)^{2}\eta > (1+\epsilon )/2
\end{equation*}
\textit{Then }\textbf{HCG, HCQ}\textbf{\textit{,}}\textit{ and }\textbf{${\exists}$}\textbf{CQ}\textit{ are guaranteed with the following parameters:}

\textbf{HCG:}\textit{ For }$\textcolor{color-2E74B5}{\mathrm{s}}\textcolor{color-2E74B5}{\geq }\textcolor{color-2E74B5}{16}\textcolor{color-2E74B5}{/}\textcolor{color-2E74B5}{(}\textcolor{color-2E74B5}{\beta }\textcolor{color-2E74B5}{f}\textcolor{color-2E74B5}{\eta }\textcolor{color-2E74B5}{)}$\textit{ and }$\tau =\textcolor{color-2E74B5}{\beta }\textcolor{color-2E74B5}{f}\textcolor{color-2E74B5}{\eta }\textcolor{color-2E74B5}{/}\textcolor{color-2E74B5}{8}$
\begin{equation*}
\Pr \left[W \textit{admits} a \left(\tau ,s\right)-HCG \textit{violation}\right]\leq \overline{\epsilon }_{HCG}\left(\tau ,s;r\right)\triangleq \textcolor{color-2E74B5}{2}\textcolor{color-2E74B5}{r}^{\textcolor{color-2E74B5}{2}}\textcolor{color-2E74B5}{\exp }\textcolor{color-2E74B5}{ }\left(\textcolor{color-2E74B5}{-}\left(\textcolor{color-2E74B5}{f}\textcolor{color-2E74B5}{\beta }\textcolor{color-2E74B5}{\eta }\right)^{\textcolor{color-2E74B5}{2}}\textcolor{color-2E74B5}{s}\textcolor{color-2E74B5}{/}\textcolor{color-2E74B5}{64}\right).
\end{equation*}
\textit{(Similar proven way)}

\textbf{HCQ:}\textit{ For }$\mathrm{s}\geq \textcolor{color-2E74B5}{32}\textcolor{color-2E74B5}{/}\textcolor{color-2E74B5}{(}\textcolor{color-2E74B5}{\upepsilon }\textcolor{color-2E74B5}{\beta }\textcolor{color-2E74B5}{f}\textcolor{color-2E74B5}{\eta }\textcolor{color-2E74B5}{)}$\textit{ and }$\tau =\textcolor{color-2E74B5}{\beta }\textcolor{color-2E74B5}{f}\textcolor{color-2E74B5}{\eta }\textcolor{color-2E74B5}{/}\textcolor{color-2E74B5}{8}$
\begin{equation*}
\Pr \left[W \textit{admits} a \left(\tau ,s\right)-HCQ \textit{violation}\right]\leq \overline{\epsilon }_{HCQ}\left(\tau ,s;r,\epsilon \right)\triangleq \textcolor{color-2E74B5}{r}^{\textcolor{color-2E74B5}{2}}\left(\textcolor{color-2E74B5}{\mathrm{s}}\textcolor{color-2E74B5}{+}\textcolor{color-2E74B5}{1}\right)\textcolor{color-2E74B5}{\exp }\textcolor{color-2E74B5}{ }\left(\textcolor{color-2E74B5}{-}\left(\textcolor{color-2E74B5}{\epsilon }\textcolor{color-2E74B5}{f}\textcolor{color-2E74B5}{\beta }\textcolor{color-2E74B5}{\eta }\right)^{\textcolor{color-2E74B5}{2}}\textcolor{color-2E74B5}{s}\textcolor{color-2E74B5}{/}\textcolor{color-2E74B5}{64}\right).
\end{equation*}
$\exists $\textbf{CQ:}\textit{ For }$\mathrm{s}\geq \textcolor{color-2E74B5}{24}\textcolor{color-2E74B5}{/}\textcolor{color-2E74B5}{(}\textcolor{color-2E74B5}{\upepsilon }\textcolor{color-2E74B5}{\beta }\textcolor{color-2E74B5}{f}\textcolor{color-2E74B5}{\eta }\textcolor{color-2E74B5}{)}$
\begin{equation*}
\Pr \left[W \textit{admits} a s-\exists CG \textit{violation}\right]\leq \overline{\epsilon }_{\exists CG}\left(s;r,\epsilon \right)\triangleq \textcolor{color-2E74B5}{r}^{\textcolor{color-2E74B5}{2}}\left(\textcolor{color-2E74B5}{\mathrm{s}}\textcolor{color-2E74B5}{+}\textcolor{color-2E74B5}{1}\right)\textcolor{color-2E74B5}{\exp }\textcolor{color-2E74B5}{ }\left(\textcolor{color-2E74B5}{-}\left(\textcolor{color-2E74B5}{\epsilon }\textcolor{color-2E74B5}{f}\textcolor{color-2E74B5}{\beta }\textcolor{color-2E74B5}{\eta }\right)^{\textcolor{color-2E74B5}{2}}\textcolor{color-2E74B5}{s}\textcolor{color-2E74B5}{/}\textcolor{color-2E74B5}{64}\right).
\end{equation*}
For convenience, let us call a slot i good if $\textcolor{color-2E74B5}{\mathbf{W}}_{\textcolor{color-2E74B5}{\mathbf{i}}}^{\textcolor{color-2E74B5}{\mathbf{r}}}\textcolor{color-2E74B5}{=}\textcolor{color-2E74B5}{0}$, and bad if $\textcolor{color-2E74B5}{\mathbf{W}}_{\textcolor{color-2E74B5}{\mathbf{i}}}^{\textcolor{color-2E74B5}{\mathbf{r}}}\textcolor{color-2E74B5}{=}\textcolor{color-2E74B5}{1}$\textcolor{color-2E74B5}{ and we override the notion W (induced by }\textbf{Ouroboros-Praos}\textcolor{color-2E74B5}{) used to prove HCG,}\textbf{ HCQ} $\exists $\textbf{CG with }$\textcolor{color-2E74B5}{\mathbf{W}}^{\textcolor{color-2E74B5}{\mathbf{r}}}$\textcolor{color-2E74B5}{ .
        And using similar method we can get the result.}

In the light of Lemma 7 and Lemma 10 from {[}15{]}, we have chain growth, and chain quality property under 2-rounds listening time.

\textbf{Corollary 3 (Chain Growth) }\textit{Let }$\textcolor{color-2E74B5}{\boldsymbol{W}}^{\textcolor{color-2E74B5}{\boldsymbol{r}}}\textcolor{color-2E74B5}{=}\textcolor{color-2E74B5}{\boldsymbol{W}}_{\textcolor{color-2E74B5}{\mathbf{1}}}^{\textcolor{color-2E74B5}{\boldsymbol{r}}}\textcolor{color-2E74B5}{,}\textcolor{color-2E74B5}{\ldots }\textcolor{color-2E74B5}{,} \textcolor{color-2E74B5}{\boldsymbol{W}}_{\textcolor{color-2E74B5}{\boldsymbol{r}}}^{\textcolor{color-2E74B5}{\boldsymbol{r}}}$\textit{  denote the characteristic string induced by the protocol }\textbf{Ouroboros-AutoSyn}\textit{ in the single-epoch setting over a sequence of r slots} \textit{under message delivery ratio }$\eta $\textit{.
        Let }$\alpha , \beta \in [0,1]$ \textit{denote lower bounds on the alert stake ratio and the participating stake ratio over the execution as per Definition 2, and assume that for some }$\varepsilon \in (0,1)$\textit{ the parameter ${\alpha}$ satisfies}
\begin{equation*}
\upalpha (1-f)^{2}\eta > (1+\epsilon )/2
\end{equation*}
\textbf{\textit{Then for}}

{\centering $\mathrm{s}=\textcolor{color-2E74B5}{96}\textcolor{color-2E74B5}{/}\textcolor{color-2E74B5}{(}\textcolor{color-2E74B5}{\upepsilon }\textcolor{color-2E74B5}{\beta }\textcolor{color-2E74B5}{f}\textcolor{color-2E74B5}{\eta }\textcolor{color-2E74B5}{)}$ \textit{and} $\tau =\textcolor{color-2E74B5}{\beta }\textcolor{color-2E74B5}{f}\textcolor{color-2E74B5}{\eta }\textcolor{color-2E74B5}{/}\textcolor{color-2E74B5}{16}$

\par}\textbf{\textit{we have }}
\begin{equation*}
\Pr \left[W \textit{admits} a \left(\tau ,s\right)-CG \textit{violation}\right]\leq \overline{\epsilon }_{CG}\left(\tau ,s;r,\epsilon \right)\triangleq \frac{\textcolor{color-2E74B5}{1}}{\textcolor{color-2E74B5}{2}}\textcolor{color-2E74B5}{s}\textcolor{color-2E74B5}{r}^{\textcolor{color-2E74B5}{2}}\textcolor{color-2E74B5}{\exp }\textcolor{color-2E74B5}{ }\left(\textcolor{color-2E74B5}{-}\left(\epsilon \textcolor{color-2E74B5}{f}\textcolor{color-2E74B5}{\beta }\textcolor{color-2E74B5}{\eta }\right)^{\textcolor{color-2E74B5}{2}}\textcolor{color-2E74B5}{s}\textcolor{color-2E74B5}{/}\textcolor{color-2E74B5}{256}\right).
\end{equation*}
\textbf{Corollary 4 (Chain Quality) }\textit{Let }$\textcolor{color-2E74B5}{\boldsymbol{W}}^{\textcolor{color-2E74B5}{\boldsymbol{r}}}\textcolor{color-2E74B5}{=}\textcolor{color-2E74B5}{\boldsymbol{W}}_{\textcolor{color-2E74B5}{\mathbf{1}}}^{\textcolor{color-2E74B5}{\boldsymbol{r}}}\textcolor{color-2E74B5}{,}\textcolor{color-2E74B5}{\ldots }\textcolor{color-2E74B5}{,} \textcolor{color-2E74B5}{\boldsymbol{W}}_{\textcolor{color-2E74B5}{\boldsymbol{r}}}^{\textcolor{color-2E74B5}{\boldsymbol{r}}}$\textit{  denote the characteristic string induced by the protocol }\textbf{Ouroboros-AutoSyn}\textit{ in the single-epoch setting over a sequence of r slots} \textit{under message delivery ratio }$\eta $\textit{.
        Let }$\alpha , \beta \in [0,1]$ \textit{denote lower bounds on the alert stake ratio and the participating stake ratio over the execution as per Definition 2, and assume that for some }$\varepsilon \in (0,1)$\textit{ the parameter ${\alpha}$ satisfies}
\begin{equation*}
\upalpha (1-f)^{2}\eta > (1+\epsilon )/2
\end{equation*}
\textbf{\textit{Then for}}

{\centering $\mathrm{k}=\textcolor{color-2E74B5}{96}\textcolor{color-2E74B5}{/}\textcolor{color-2E74B5}{(}\textcolor{color-2E74B5}{\upepsilon }\textcolor{color-2E74B5}{\beta }\textcolor{color-2E74B5}{f}\textcolor{color-2E74B5}{\eta }\textcolor{color-2E74B5}{)}$ \textit{and} $\mu =\epsilon \textcolor{color-2E74B5}{\beta }\textcolor{color-2E74B5}{f}\textcolor{color-2E74B5}{\eta }\textcolor{color-2E74B5}{/}\textcolor{color-2E74B5}{16}$

\par}\textbf{\textit{we have }}
\begin{equation*}
\Pr \left[W \textit{admits} a \left(\mu ,k\right)-CQ \textit{violation}\right]\leq \overline{\epsilon }_{CQ}\left(\mu ,k;r,\epsilon \right)\triangleq \frac{\textcolor{color-2E74B5}{1}}{\textcolor{color-2E74B5}{2}}\textcolor{color-2E74B5}{k}\textcolor{color-2E74B5}{r}^{\textcolor{color-2E74B5}{2}}\textcolor{color-2E74B5}{\exp }\textcolor{color-2E74B5}{ }\left(\textcolor{color-2E74B5}{-}\left(\epsilon \textcolor{color-2E74B5}{f}\textcolor{color-2E74B5}{\beta }\textcolor{color-2E74B5}{\eta }\right)^{\textcolor{color-2E74B5}{2}}\textcolor{color-2E74B5}{k}\textcolor{color-2E74B5}{/}\textcolor{color-2E74B5}{256}\right).
\end{equation*}
\textbf{E.3 Lifting to Multiple Epochs}

\textbf{Theorem 5.} \textit{Consider the execution of Ouroboros-AutoSyn with adversary A and environment Z in the setting with static }$F_{N-MC}$\textit{ registration.
        Let f be the active-slot coefficient, let }$\textcolor{color-2E74B5}{\eta }$\textit{ be the lower bound on message success ratio.
        Let} $\alpha , \beta \in [0,1]$\textit{ denote a lower bound on the alert and participating stake ratios throughout the whole execution, respectively.
        Let R and L denote the epoch length and the total lifetime of the system (in slots), and let Q be the total number of queries issued to }$G_{RO}$\textit{.
        If for some }$\varepsilon \in (0,1)$\textit{ we have}
\begin{equation*}
\upalpha (1-f)^{2}\eta > (1+\epsilon )/2
\end{equation*}
\textbf{Then Ouroboros-AutoSyn achieves the same guarantees for common prefix (resp. chain growth, chain quality, existential chain quality) as given in Corollary 3,4,5 and Lemma 11 except with an additional error probability of }
\begin{equation*}
QL\cdot (2\overline{\epsilon }_{CG}(\tau ,R/3;R,\epsilon )+2\overline{\epsilon }_{CP}(\tau R/3;R,2,\epsilon )+2\overline{\epsilon }_{\exists CQ}(R/3;R,\epsilon )),
\end{equation*}
\textit{where }$\tau =\beta f\eta /16$\textit{.
        If }$R\geq 288/\epsilon \beta f$\textit{ then this term can be upper-bounded by}
\begin{equation*}
\upvarepsilon _{\text{lift}}\triangleq QL\cdot \left[R^{3}\cdot \exp \left(-\frac{\left(\epsilon f\beta \eta \right)^{2}R}{768}\right)+\frac{38R}{\varepsilon ^{4}}\cdot \exp \left(2-\frac{\varepsilon ^{4}f\beta \eta R}{864}\right) \right]
\end{equation*}
Proof. 

In addition to stake distribution updates and randomness updates, we need also establish slot length adjusting. 

Consider the last R/2 slots of each epoch, if we have common prefix during the last R/3 of each epoch then we must have every alert part agree on the same chain for the first R/2 slots at the end of the epoch.
        Hence round length for alert party will be the same.  
\end{appendices}
\end{document}